\documentclass[ip,twocolumn]{jpsj3}
\pdfoutput=1

\usepackage{bm}
\usepackage{color}
\usepackage{txfonts}
\usepackage{braket}
\usepackage{url}

\title{
Sparse Modeling in Quantum Many-Body Problems
}

\author{%
Junya Otsuki$^1$,
Masayuki Ohzeki$^2$,
Hiroshi Shinaoka$^3$, and
Kazuyoshi Yoshimi$^4$
}

\inst{%
$^1$Research Institute for Interdisciplinary Science, Okayama University, Okayama 700-8530, Japan\\
$^2$Graduate School of Information Sciences, Tohoku University, Sendai 980-8579, Japan\\
$^3$Department of Physics, Saitama University, Saitama 338-8570, Japan\\
$^4$Institute for Solid State Physics, University of Tokyo, Chiba 277-8581, Japan
}

\recdate{\today}

\abst{%
This review paper describes the basic concept and technical details of sparse modeling and its applications to quantum many-body problems.
Sparse modeling refers to methodologies for finding a small number of relevant parameters that well explain a given dataset.
This concept reminds us physics, where the goal is to find a small number of physical laws that are hidden behind complicated phenomena.
Sparse modeling extends the target of physics from natural phenomena to data, and may be interpreted as ``physics for data''.
The first half of this review introduces sparse modeling for physicists.
It is assumed that readers have physics background but no expertise in data science.
The second half reviews applications.
Matsubara Green's function, which plays a central role in descriptions of correlated systems, has been found to be sparse, meaning that it contains little information.
This leads to 
(i) a new method for solving the ill-conditioned inverse problem for analytical continuation, and
(ii) a highly compact representation of Matsubara Green's function, which enables efficient calculations for quantum many-body systems.
}

\newcommand{\bx}{\bm{x}}
\newcommand{\by}{\bm{y}}
\newcommand{\bz}{\bm{z}}
\newcommand{\bv}{\bm{v}}
\newcommand{\bu}{\bm{u}}
\newcommand{\bnu}{\bm{\nu}}
\newcommand{\bS}{S}
\newcommand{\bU}{U}
\newcommand{\bV}{V}
\newcommand{\bK}{K}
\newcommand{\brho}{\bm{\rho}}
\newcommand{\bG}{\bm{G}}

\newcommand{\br}{\bm{r}}
\newcommand{\bk}{\bm{k}}

\newcommand{\wmax}{\ensuremath{{\omega_\mathrm{max}}}}

\definecolor{mycolor}{rgb}{0, 0.5, 0}

\DeclareMathOperator*{\argmin}{arg\,min}

\begin{document}
\maketitle


\section{Introduction}
\label{sec:intro}

A small number of physical laws exist behind apparently complicated behaviors: This is the basic notion of physics.
In condensed matter physics, we expect the existence of simple laws that approximately explain the behavior of ensembles in some parameter region. 
In this context, finding simple laws means finding an effective model that well explains behaviors of physical quantities with simple, hopefully mean-field-level, calculations.

In the field of data science, one of the main goals is to find \emph{features} that discriminate different kinds of data efficiently.
The use of fewer features makes it easier to understand what is happening.
Moreover, a small set of features is more flexible for describing a wide range of data than a large set of features designed to fit a specific dataset.
Sparse modeling offers methodologies for this purpose.
Solving an optimization problem, one can find essential parameters (cause) from a complicated dataset (effect) just as physicists find relevant physical laws from complicated natural phenomena.

Technically, sparse modeling treats inverse problems.
Consider a well-defined mapping rule $\bx \mapsto \by$, where $\by$ is known.
The inverse problem is to derive $\bx$ for a given $\by$.
In practical situations, however, this inverse transformation is often difficult to perform
because the observed data $\by$ may be disturbed by noise or the inversion may, in principle, not be uniquely defined.
A standard strategy for \emph{inferring} a seemingly correct solution relies on \emph{prior knowledge}, which compensates for the lack of information for the inversion.

Which prior knowledge leads to a plausible solution?
The maximum entropy method (MEM) uses the prior knowledge that $\bx$ should be close to an ``ideal solution'' that carries all the desired features.
On the other hand, sparse modeling takes advantage of \emph{sparsity}; that is, it assumes that the solution $\bx$ has only a small number of non-zero components.
Therefore, $\by$ is fitted with a small number of components in $\bx$ even if agreement with $\by$ is sacrificed.

A brilliant success of the sparsity criterion has been demonstrated in the applications to MRI~\cite{Candes06a,Candes06b,Donoho06,Lustig07,Lustig08}.
With this criterion, incomplete signals measured in the Fourier domain can be stably transformed into a real-space image that looks as if complete signals had been used.
A recent observation of a shadow of a supermassive black hole relied on data analysis that included sparse modeling~\cite{EventHorizonTelescope19a,EventHorizonTelescope19b,EventHorizonTelescope19c,EventHorizonTelescope19d,EventHorizonTelescope19e,EventHorizonTelescope19f}.
These successful applications hint at even wider applicability of sparse modeling beyond data analysis of measurements.

In the second half of this paper, we present the application of sparse modeling to quantum many-body problems.
The first question from an sparse modeling point of view is: What is sparse in quantum many-body theory?
Recent investigations have shown that the information handled by the imaginary-time ($\tau\equiv it$) framework is sparse.
More precisely, the information that the imaginary-time (Matsubara) Green's function $G(\tau)$ can carry is quite limited.
Analytically, the imaginary-time representation $G(\tau)$ and real-frequency representation $G_\mathrm{R}(\omega)$ are equivalent in the sense that the transformation from one to the other preserves information.
In practice, however, $G(\tau)$ is more fragile to noise, meaning that the numerically computed $G(\tau)$ has lost a large part of the information on $G_\mathrm{R}(\omega)$.
Therefore, $G(\tau)$ is sparse.

Given that the numerically computed $G(\tau)$ contains less information on real-frequency dynamics, we now turn our attention to how to extract the relevant information.
This can be achieved using a new basis set, which is placed as an intermediate representation (IR) between the imaginary and real frequencies.
With the IR basis, the sparse-modeling technique leads to a new algorithm for conversion (analytical continuation) from $G(\tau)$ to $G_\mathrm{R}(\omega)$ for efficient computation in quantum many-body theories.

The rest of this review paper is organized as follows.
In Sect.~\ref{sec:spm}, the fundamentals of sparse modeling are presented.
Section~\ref{sec:algorithm} focuses on the technical details of the sparse modeling.
Readers may skip this section, if numerical calculations are not of interest.
Selected applications are reviewed in Sect.~\ref{sec:applications}, with particular focus on condensed matter physics research.
The applications to quantum many-body problems are presented in Sects.~\ref{sec:green_function}--\ref{sec:compressed_sampling}.
Section~\ref{sec:green_function} focuses on the ``sparsity'' of Matsubara Green's function and introduces a proper basis.
This basis is utilized for a new method for analytical continuation in Sect.~\ref{sec:analytical_continuation} and efficient calculations of many-body problems in Sect.~\ref{sec:compressed_sampling}.
Finally, this review is closed with a summary in Sect.~\ref{sec:summary}.

\section{Inverse Problem Revisited}
\label{sec:spm}
\subsection{Inverse problem of underdetermined systems}
In this section, we present the concept of sparse modeling, which opens a new paradigm for solving inverse problems.
Starting from a simple problem, we try unveiling the essence of the sparse modeling.
To this end, we consider a simple inverse problem, namely, a linear equation of the form
\begin{equation}
\by = A\bx,
\label{eq:y_Ax}
\end{equation}
where $\bx$ and $\by$ are vectors of $N$- and $M$-dimensions, respectively, and $A$ is an $M\times N$ matrix.
Let us suppose that we want to obtain $\bx$ from known variables $\by$ and $A$.
Obviously, the equation can be solved immediately, if $M=N$ and the inverse of $A$ exists:
\begin{equation}
\bx = A^{-1}\by.
\end{equation}
However, if $M<N$, the number of equations is insufficient for $\bx$ to be determined uniquely.
We consider such systems, called \emph{underdetermined systems}, in the rest of this section.

Even for underdetermined systems, there are cases where the equation can be solved \emph{exactly}.
Here, ``sparsity'' plays an essential role.
The vector $\bx$ is called sparse if most of its components are zero.
Let $n$ be the number of non-zero components in $\bx$.
If we can find the positions of zeros in $\bx$ and remove these zeros from the set of equations, then the number of unknown components will be reduced from $N$ to $n$. 
Thus, equations that belong to underdetermined systems are solvable if $M>n$.

\subsection{Methods for finding sparse solutions}\label{subsec:methods}

The problem now is whether and how to find the positions of the non-zero components in $\bx$.
One of the simplest methods is $L_0$-norm minimization.
The $L_0$-norm, represented by $\left\|\bx\right\|_0$, counts the number of non-zero components in $\bx$.
By selecting the solution that minimizes $\left\|\bx\right\|_0$ from the set of solutions of the underdetermined system, the most sparse solution is obtained.
This statement is formulated as 
\begin{equation}
\min_{\bx} \left\|\bx\right\|_0
\quad \text{subject to} \quad
\by = A\bx.
\end{equation}
However, $L_0$-norm minimization is a combinatorial optimization problem, which requires exponential cost of computation.
Therefore, an alternative formulation is required from a practical point of view.

\begin{figure}[tb]
\centering
\includegraphics[width=5cm]{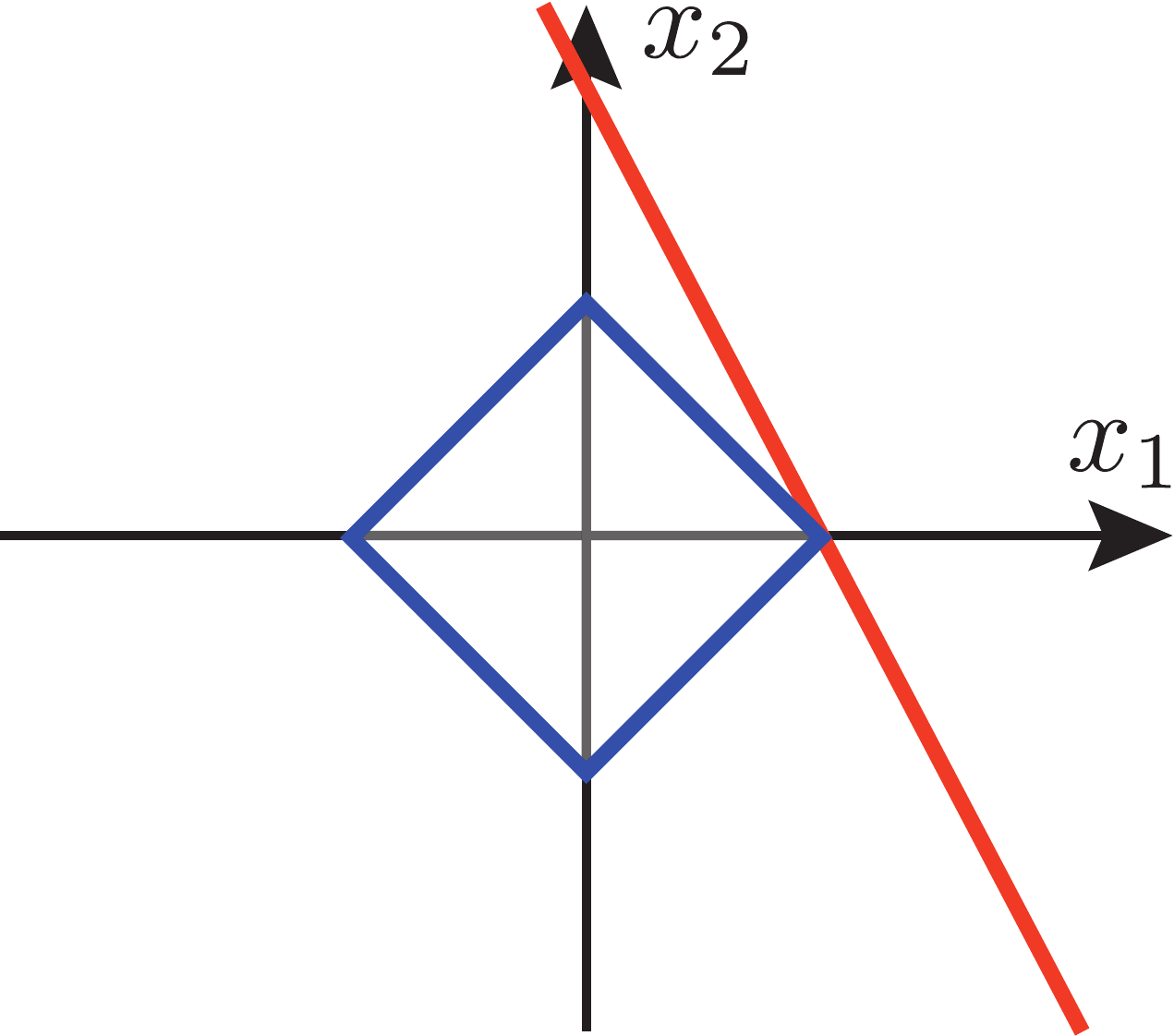}
\caption{
(Color online) Graphical solution for $L_1$-norm minimization, Eq.~(\ref{eq:min_L1}), with $(N, M)=(2, 1)$.
The solution is given by the intersection between the solid line, which represents the set of solutions for $\by=A\bx$, and the rhombus describing the contour of the $L_1$ norm.
}
\label{fig:L1norm}
\end{figure}

A feasible approach for simultaneously handling the sparsity requirement and computational cost is to relax the norm from $L_0$ to $L_1$, which leads to
\begin{equation}
\min_{\bx} \left\|\bx\right\|_1
\quad \text{subject to} \quad
\by = A\bx.
\label{eq:min_L1}
\end{equation}
Here, $\left\|\bx\right\|_1$ denotes the $L_1$ norm defined by
\begin{equation}
\left\| \bx \right\|_1 \equiv \sum_{k=1}^N |x_k|.
\end{equation}
This problem can be solved with moderate computational complexity using the interior point method which is an optimization technique.
The solution of Eq.~(\ref{eq:min_L1}) is sparse, as shown below.
Let us consider the simplest case, with $N=2$ and $M=1$:
One linear equation is given between two unknown variables $x_1$ and $x_2$.
The set of solutions forms a straight line on the $x_1$--$x_2$ plane as shown in Fig.~\ref{fig:L1norm}.
The $L_1$-norm is represented by $\left\| \bx \right\|_1 = |x_1|+|x_2|$. Its contour is a rhombus. 
The the optimization problem in Eq.~(\ref{eq:min_L1}) is now interpreted as follows.
The rhombus must intersect the straight line and its size should be as small as possible.
The solution that satisfies this statement is, for the case in Fig.~\ref{fig:L1norm}, located on the $x_1$-axis because of the cuspidal nature of the $L_1$-norm.
Thus, Eq.~(\ref{eq:min_L1}) yields a unique solution in which some components tend to be zero, that is, a sparse solution.
Sparsity and non-exponential computational cost are therefore compatible for $L_1$-norm minimization.

\subsection{Conventional method: $L_2$-norm minimization}
\begin{figure}[tb]
\centering
\includegraphics[width=5cm]{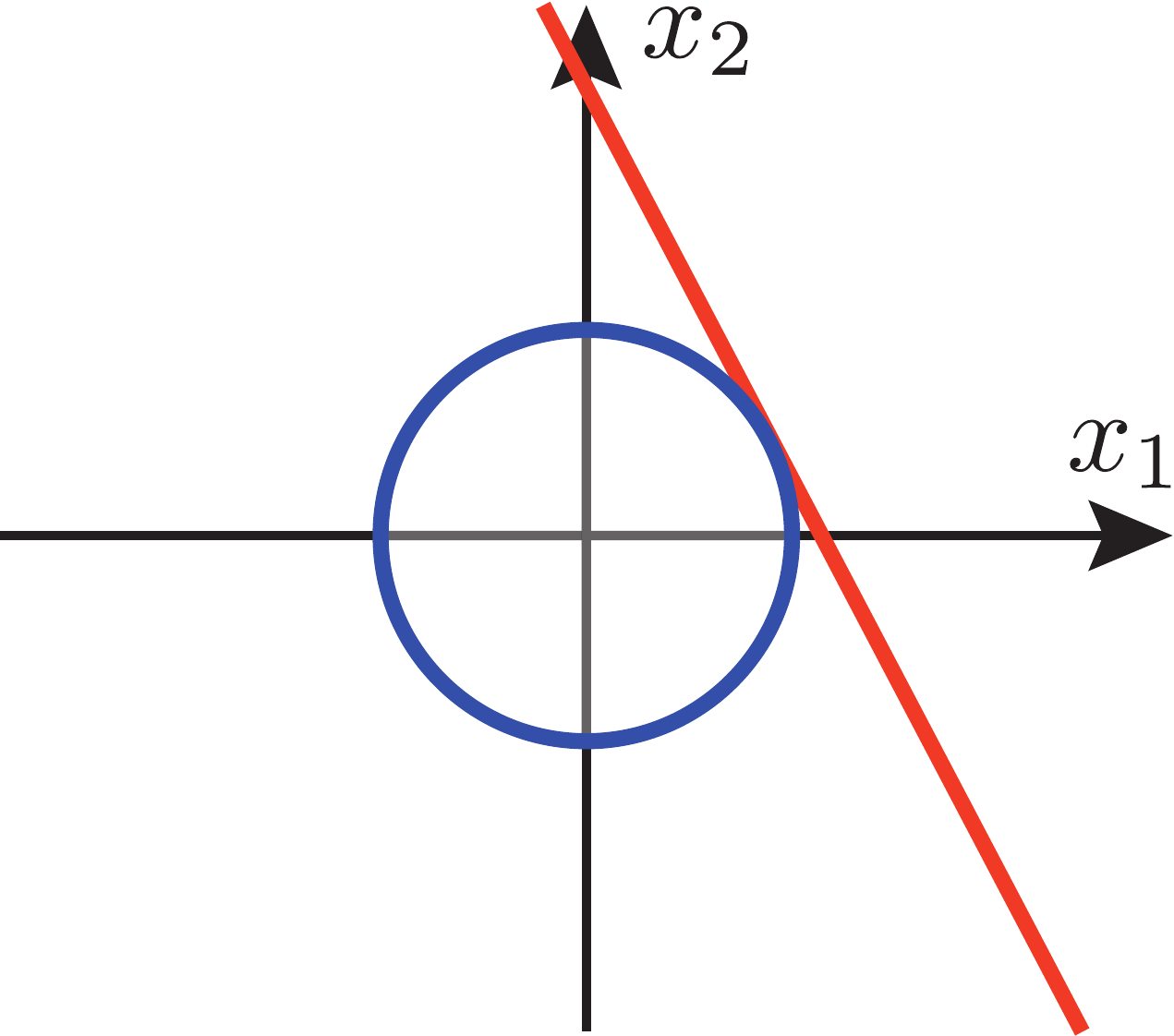}
\caption{
(Color online) Graphical solution for $L_2$-norm minimization, Eq.~(\ref{eq:min_L2}).
See the caption of Fig.~\ref{fig:L1norm} for the explanation.
}
\label{fig:L2norm}
\end{figure}

Traditionally, $L_2$-norm has been used in determining a solution of underdetermined systems.
$L_2$-norm is also referred to as the Euclidean norm, that is, the ``ordinary'' norm defined by
\begin{equation}
\left\|\bx\right\|_2 = \sum_{k=1}^N x_k^2.
\end{equation}
$L_2$-norm minimization is thus written as
\begin{equation}
\min_{\bx} \left\|\bx\right\|_2
\quad \text{subject to} \quad
\by = A\bx.
\label{eq:min_L2}
\end{equation}
The solution of Eq.~(\ref{eq:min_L2}) corresponds to the intersection between the straight line and the circle as shown in Fig.~\ref{fig:L2norm}.
We note that both $x_1$ and $x_2$ are finite unlike the case in $L_1$-norm minimization.

The advantage of $L_2$-norm minimization is that the solution can be evaluated analytically.
Using the Lagrange multiplier method, Eq.~(\ref{eq:min_L2}) is rewritten as
\begin{equation}
\min_{\bx} \left\{ \left\|\bx\right\|_2 + \bm{\lambda}^{\rm T}\left(\by - A\bx\right) \right\},
\end{equation}
where $\bm{\lambda}$ is the Lagrange multiplier.
Taking the derivatives and assuming the underdetermined condition $M<N$, we obtain the solution $\bx^*$of this optimization problem as
\begin{equation}
\bx^* = A^{+}\by,
\end{equation}
where $A^{+}=A^{\rm T}(AA^{\rm T})^{-1}$ is called the Moore-Penrose pseudo-inverse matrix\cite{footnote_minL2}.

\subsection{Finding the true solution}
\label{subsec:true_solution}
As mentioned, a unique solution can be determined for underdetermined systems, if an additional condition is provided.
Then, it is important to judge whether the obtained solution is reasonable.
$L_1$-norm-minimized solution is sparse.
Therefore, if the true solution is also sparse, there is a chance that $L_1$-norm-minimized solution coincides with the true solution.
It has been proven, for a specific model, that $L_1$-norm minimization indeed yields the exact solution~\cite{Donoho05,Donoho06b,Kabashima09,Donoho09}.

In contrast, an $L_2$-norm-minimized solution does not have this feature.
For underdetermined problems, the $L_2$-norm-minimized solution exactly coincides with the true solution only when $N=M$, and thus the additional condition does not make sense for finding the true solution.
One can only avoid obviously unreasonable results that involve infinitely large components.

\begin{figure}[tb]
\centering
\includegraphics[width=\linewidth]{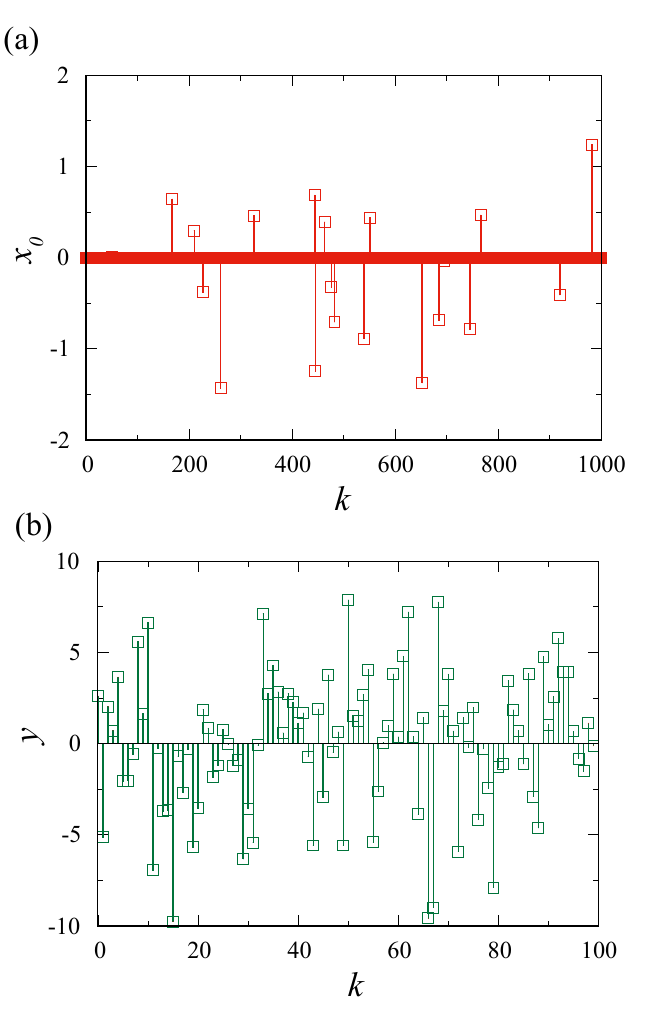}
\caption{(Color online) Sample linear inverse problem, Eq.~(\ref{eq:y_Ax}). (a) Correct solution $\bx_0$, which has $n=20$ non-zero components among $N=1000$. (b) Input vector $\by=A \bx_0$ with dimension $M=100$.}
\label{fig:example_input}
\end{figure}

\begin{figure}[tb]
\centering
\includegraphics[width=\linewidth]{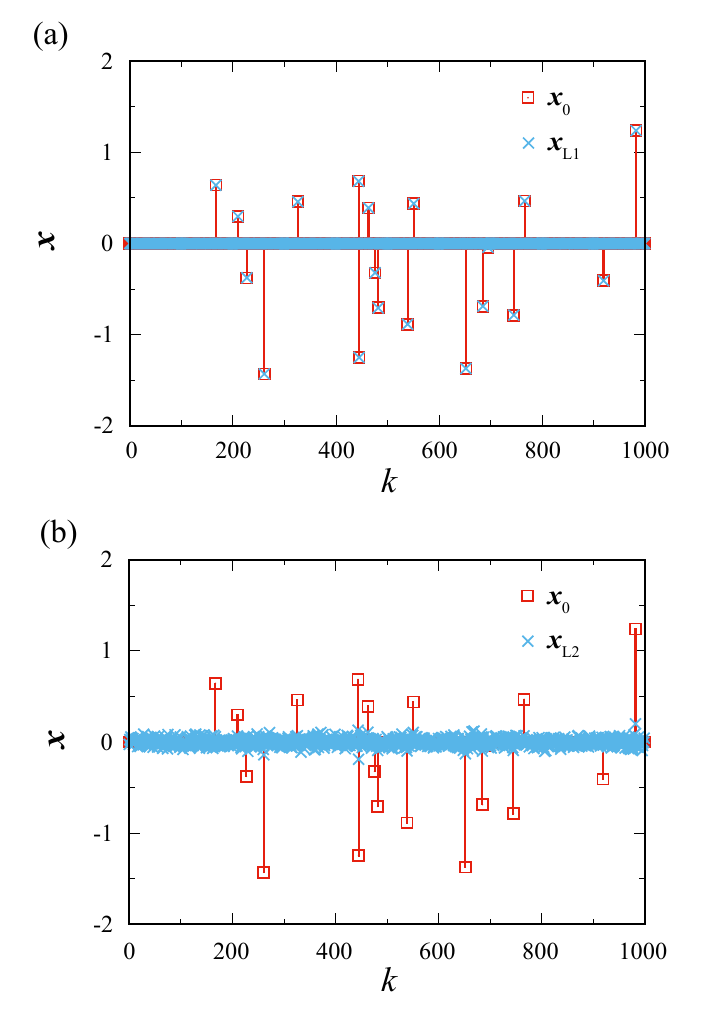}
\caption{(Color online) Comparison between solution $\bx_0$ and vector $\bx$ reconstructed using (a) $L_1$-norm minimization, Eq.~(\ref{eq:min_L1}), and (b) $L_2$-norm minimization, Eq,~(\ref{eq:min_L2}).}
\label{fig:example_result}
\end{figure}

We illustrate the difference between $L_1$-norm and $L_2$-norm minimizations by considering a random matrix $A^\mathrm{rand}$ as an example~\cite{footnote_sample_script}.
We suppose that the true solution $\bx_0$ has only $n=20$ non-zero components among $N=1000$ [Fig.~\ref{fig:example_input}(a)].
This vector is converted to $\by=A^\mathrm{rand} \bx_0$, which has $M=100$ components [Fig.~\ref{fig:example_input}(b)].
This problem satisfies the condition $n<M<N$, meaning that the equations are underdetermined but solvable if the zero components are properly eliminated.
Figures~\ref{fig:example_result}(a) and \ref{fig:example_result}(b) show $\bx$ reconstructed using $L_1$-norm minimization [Eq.~(\ref{eq:min_L1})] and $L_2$-norm minimization [Eq.~(\ref{eq:min_L2})], respectively.
As shown, $L_1$-norm minimization perfectly recovers the true solution $\bx_0$,
whereas $L_2$-norm minimization fails to do so because the weights are distributed among all components.

\begin{figure}[tb]
\centering
\includegraphics[width=\linewidth]{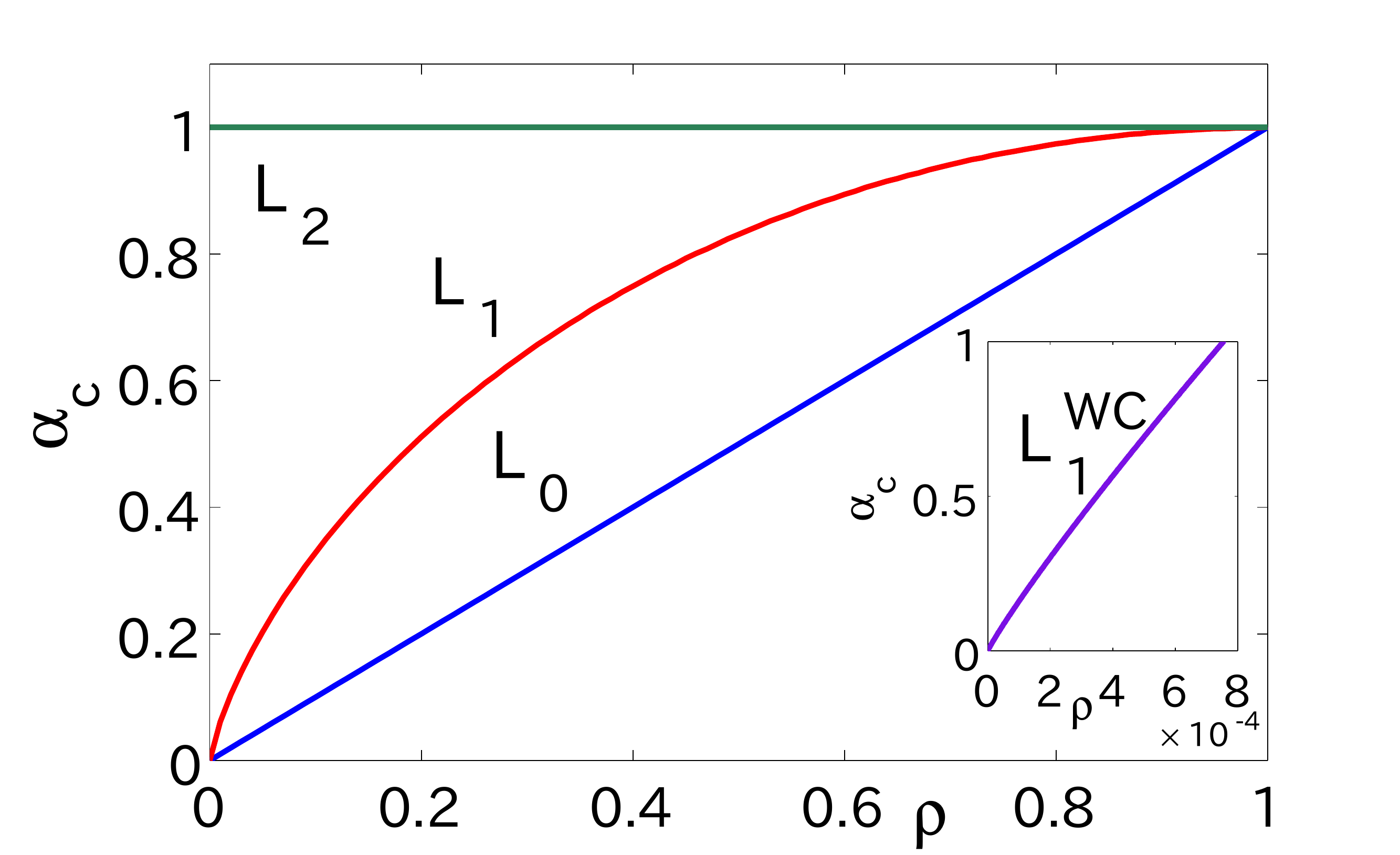}
\caption{Boundary between the region where the exact solution $\bx_0$ is reconstructable ($\alpha>\alpha_\mathrm{c}$) and the region where reconstruction is impossible ($\alpha>\alpha_\mathrm{c}$). The variables $\alpha$ and $\rho$ are defined by $\alpha \equiv M/N$ and $\rho \equiv n/N$, respectively.
Reprinted with permission from Ref.~\citen{Kabashima09} \copyright 2009 IOP Publishing.}
\label{fig:kabashima}
\end{figure}

The property demonstrated above has been proven analytically using integral geometry\cite{Donoho05,Donoho06b} and information statistics\cite{Kabashima09,Donoho09}.
Let us define $\alpha\equiv M/N$ (the ratio between the dimensions of $\by$ and $\bx$) and $\rho=n/N$ (the ratio of non-zero components in $\bx$ to all components), and consider the continuous limit $N\to\infty$.
Then, there exists a critical value $\alpha_\mathrm{c}$ above which the exact solution is reconstructable.
As shown in Fig.~\ref{fig:kabashima}, $L_1$-norm regularization yields a finite region where $\alpha_\mathrm{c} \leq \alpha \leq 1$ is satisfied, while with $L_2$-norm regularization, $\alpha_\mathrm{c}=1$, meaning that the exact solution can be obtained only when the complete information of $\by$ is available.

\subsection{Handling noise in inverse problem}
We have so far assumed that $\bx$ follows the equation exactly.
In practical situations, however, the input to the equation, $\by$, contains noise and the equation does not need to be satisfied rigorously.
This situation is represented as follows:
\begin{equation}
\by = A\bx_0 + \bm{\eta},
\label{eq:y_Ax_noise}
\end{equation}
where $\bx_0$ is the true solution and $\bm{\eta}$ is an $M$-dimensional noise vector.
Assuming a Gaussian distribution with a zero mean, the maximum likelihood estimation for $\bx$ corresponds to the minimization of the function
\begin{equation}
F_0(\bx) = \frac{1}{2} \left\|\by-A\bx\right\|_2^2,
\label{eq:min_mean_square_error}
\end{equation}
which is called the minimum mean square error estimation.
We use the notation $F$ analogous to the free energy in statistical physics.
The subscript 0 indicates that no extra term is introduced besides the quadratic term.
The minimum point of $F_0(\bx)$ can be analytically expressed as
\begin{equation}
\bx^* = (A^{\rm T}A)^{-1}A^{\rm T}\by.
\label{eq:x_min_mean_square}
\end{equation}
However, for the underdetermined condition $M<N$, this solution suffers from numerical instability (division by zero) because the $N\times N$ matrix $A^{\rm T}A$ is rank deficient.

A converged solution can be obtained by granting an additional term to $F_0(x)$ in Eq.~(\ref{eq:min_mean_square_error}).
This approach is called \emph{regularization} and the additional term is called a \emph{regularizer}.
If we adopt the $L_2$ term as a regularizer, we obtain
\begin{equation}
F_\mathrm{Ridge}(\bx) = \frac{1}{2} \left\|\by-A\bx\right\|_2^2 + \lambda \left\| \bx \right\|_2^2,
\label{eq:ridge}
\end{equation}
where $\lambda$ is a small constant.
This regularization is referred to as \emph{Ridge regression}.
The $L_2$ term replace $A^\mathrm{T}A$ with $(A^{\rm T}A + \lambda I)$ in Eq.~(\ref{eq:x_min_mean_square}) to yield
\begin{equation}
\bx^* = (A^{\rm T}A + \lambda I)^{-1}A^{\rm T}\by.
\end{equation}
Here, $I$ denotes the unit matrix.
Because of $\lambda$, the inverse always exists and the solution is well-defined.
However, the $L_2$ term tends to make the solution featureless as shown in Fig.~\ref{fig:example_result}, and thus coincidence with the true solution is not expected.

It is natural to replace $L_2$ with $L_1$ in Eq.~(\ref{eq:ridge}) to select out a sparse solution.
Then, the function $F(\bx)$ to be minimized is
\begin{equation}
F_\mathrm{LASSO}(\bx) = \frac{1}{2} \left\|\by-A\bx\right\|_2^2 + \lambda \left\| \bx \right\|_1.
\label{eq:lasso}
\end{equation}
Here, $\lambda$ is a parameter that controls the sparsity.
The minimization problem in the form of Eq.~(\ref{eq:lasso}) is called the least absolute shrinkage and selection operator (LASSO)~\cite{Tibshirani96}.
The LASSO is a type of convex optimization problem, which guarantees convergence of the iterative update procedure to the unique solution.

The parameter $\lambda$ is often called a \emph{hyperparameter} to distinguish it from the ordinary parameters that specify a model.
The value of $\lambda$ should be determined so that the effect of regularization is moderate. The inverse problem is unstable if $\lambda$ is too small, and the result becomes artificial if $\lambda$ is too large.
Techniques for automatically fixing the value of $\lambda$ are presented in Sect.~\ref{subsec:lambda}.

\subsection{Maximum entropy method}
As mentioned, an additional condition (regularization) helps to determining a unique solution of underdetermined equations.
The MEM is also one of such methods~\cite{MaxEnt}.
As a regularizer, the MEM employs a distance with an ``ideal'' solution called the default model $\bm{m}$.
The default model is prepared so that it fulfills all prior knowledge for an expected solution.
A solution obtained is thus not far from $\bm{m}$ as expected.

The function $F(\bx)$ to be minimized in the MEM is given by~\cite{MaxEnt}
\begin{equation}
F_\mathrm{MEM}(\bx) = \frac{1}{2}\left\| \by - A\bx\right\|_2^2 - \lambda S(\bm{m},\bx),
\end{equation}
where $S(\bm{m},\bx)$ is defined as
\begin{equation}
S(\bm{m},\bx) = \sum_{k=1}^N\left[ x_k - m_k -x_k \log\left(\frac{x_k}{m_k}\right)\right].
\end{equation}
Here, $S(\bm{m},\bx)$  quantifies a ``distance'' between $\bm{m}$ and $\bm{x}$,  referred to as \emph{information entropy} or \emph{Kullback--Leibler divergence}~\cite{Kullback-book}.

\begin{figure}
\centering
\includegraphics[width=7cm]{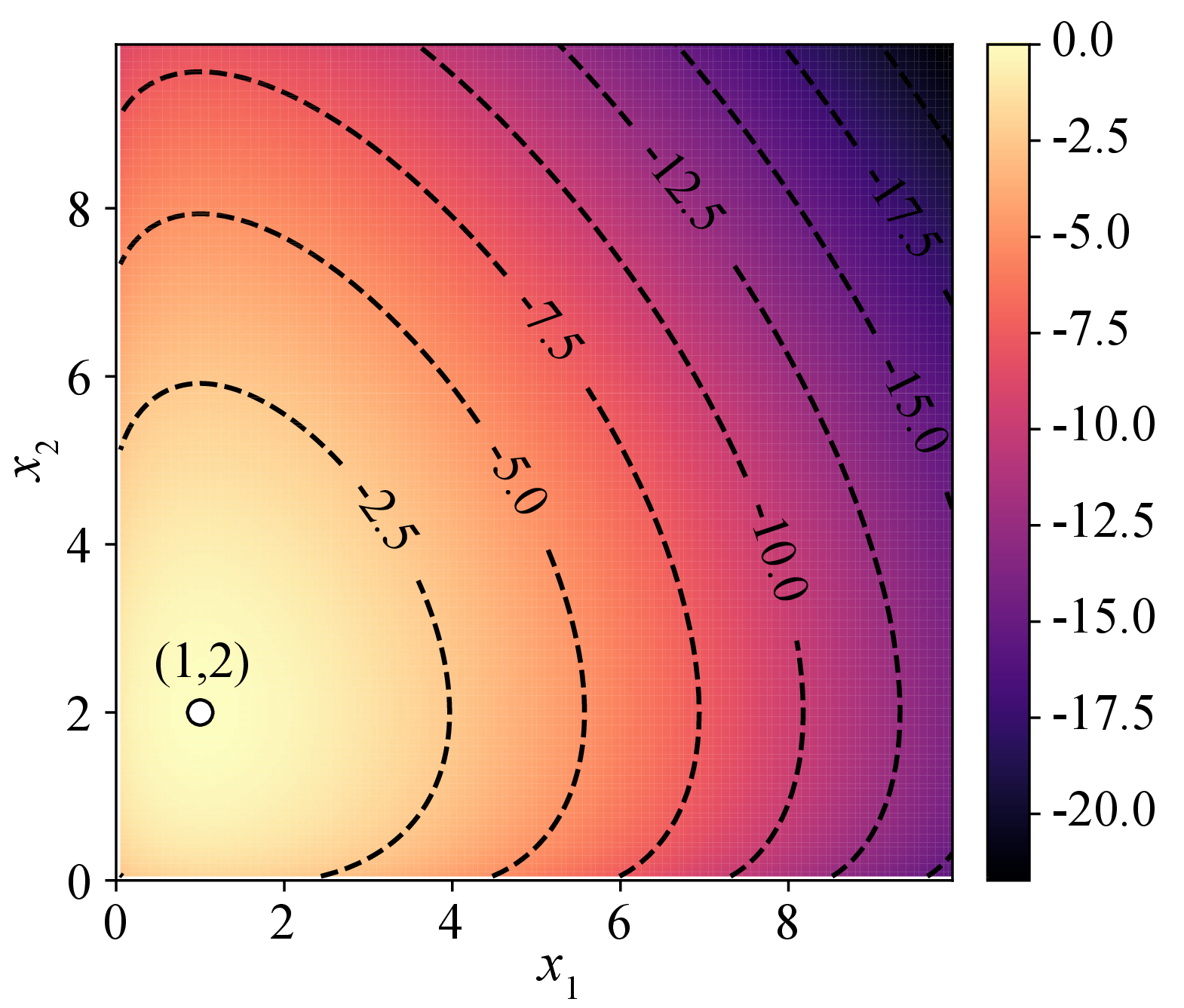}
\caption{(Color online) Contour lines of $S(\bm{m},\bx)$ for $N=2$ and $\bm{m}=(1, 2)$.}
\label{fig:MaxEnt}
\end{figure}

Let us consider the effect of the regularizer $S(\bm{m},\bx)$, as done for $L_1$- and $L_2$-norm regularization in Figs.~\ref{fig:L1norm} and \ref{fig:L2norm}.
Figure~\ref{fig:MaxEnt} shows contour lines of $S(\bm{m},\bx)$ in the $x_1$--$x_2$ plane.
As noted earlier, the solution is given by the intersection between the straight line, which shows the set of solutions for $\by=A\bx$, and one of the contour lines.
In this view, the solution exhibits no particular characteristics except that it is close to $\bm{m}$.
There is no reason to expect coincidence between the MEM solution and the true solution because the former depends on the choice of default model $\bm{m}$.

The MEM has been applied in various fields.
Examples include the calculation of spectral functions, which we discuss in Sect.~\ref{sec:analytical_continuation}, and astronomical image analysis~\cite{Gull-Skilling84, Carcamo18}.
Practically, the ambiguity due to the default model may be less severe in the case of astronomy because a typical default model is available from the observed data.
We have to keep in mind, however, that we should not design the default model elaborately to avoid biased analysis.

\subsection{Is the true solution really sparse?}
\label{subsec:really_sparse}
The success of $L_1$-norm regularization strongly relies on the sparsity of the true solution.
One may think that the true solution in realistic problems is not always sparse and that therefore $L_1$-norm regularization would not work in these cases.
We emphasize that sparsity is basis-dependent.
In other words, we can manifest the potential sparsity of the data by transforming the basis.

Let us suppose that $\bx$ represents images.
Then, the spatial variation of $\bx$ is expected to be smooth in an extensive region and abrupt at the edge of some object.
This contrast indicates a possible sparsity of the data in the representation $D\bx$, where $D$ is a matrix obtained by taking the difference between adjacent elements of $\bx$, namely, $(D\bx)_{k} = x_{k+1} - x_{k}$ for one dimension.

We can generalize the LASSO defined in Eq.~(\ref{eq:lasso}) so that the basis transformation of $\bx$ is taken into account. The general form of the LASSO is thus 
\begin{equation}
F_\mathrm{LASSO}(\bx) = \frac{1}{2} \left\|\by-A\bx\right\|_2^2 + \lambda \left\| B \bx \right\|_1,
\tag{\ref{eq:lasso}${}^\prime$}
\label{eq:lasso_gen}
\end{equation}
where $B$ is an $M'\times N$ matrix, which transforms $\bx$ to a sparse representation $\bx' \equiv B\bx$ with dimension $M'$.
It is crucial to properly choose the basis $\bx'$ in applying the LASSO to a problem.
In Sect.~\ref{sec:green_function}, we present an example in which the use of LASSO leads to a new basis that compactly represents $\bx$.

\subsection{Relation to machine learning}
We close this section by discussing the relation between sparse modeling and machine learning.
In applications of machine learning, we are interested in the result for prediction, classification, etc, but not in \emph{how} machines predict. Optimized parameters inside machines are typically not analyzed or are difficult to analyze because of the huge number of parameters and complexity of nonlinear transformations.
In sparse modeling, in contrast, we are interested in the processes used for prediction. For this aim, it is crucial to find relevant parameters (or descriptors) among all parameters. The sparsity criterion plays a central role in this task.

Although the main concepts of sparse modeling and machine learning are different, these methods share some technical details.
$L_1$-norm regularization is utilized in a learning process to remove redundant parameters, which cause overfitting and thus make prediction unreliable.
As the number of retained parameters is reduced, it becomes possible to examine the optimized parameter set and determine which parameters mostly control the prediction.

Finally, we will mention a direct application of sparse modeling to machine learning. We have so far assumed that the matrix $A$ is given in Eq.~(\ref{eq:y_Ax}) and a sparse solution for $\bx$ is pursued.
There is another class of optimization problems that searches for $A$ as well as $\bx$ for a given dataset $\{ \by_i \}$. These problems, called dictionary learning, are presented in more detail in Sec.~\ref{sec:dictionary_learning}.

\section{Algorithms for Solving Inverse Problems}
\label{sec:algorithm}

The previous section demonstrated that the sparsity constraint can be implemented with $L_1$-norm regularization.
In this section, we discuss how to solve a minimization problem that includes the $L_1$-norm term.
Two classes of method are introduced, namely the iterative shrinkage thresholding algorithm (ISTA) in subsection \ref{subsec:ista} and the alternating direction method of multipliers (ADMM) in subsection \ref{subsec:admm}.
Then, subsection~\ref{subsec:lambda} describes methods for determining an optimal value of $\lambda$, which controls the sparsity of the solution.

In this section, we consider how to solve the optimization problem
\begin{equation}
\bx^* = \argmin_{\bx} F(\bx),
\label{eq:argmin_Fx}
\end{equation}
where $\argmin_{\bx}$ is an operator that returns $\bx$ that minimizes the operand.
We suppose that the target function $F(\bx)$ is represented by the following general form:
\begin{equation}
F(\bx) = f(\bx) + g(\bx),
\label{eq:fx_gx}
\end{equation}
where $f(\bx)$ and $g(\bx)$ are the differentiable and non-differentiable functions, respectively.
For the LASSO in Eq.~(\ref{eq:lasso}), $f(\bx)$ and $g(\bx)$ are given by
\begin{align}
f(\bx) &= \frac{1}{2} \left\| \by - A \bx \right\|_2^2,
\label{eq:fx}
\\
g(\bx) &= \lambda \left\| \bx \right\|_1.
\label{eq:gx}
\end{align}
Recall that the dimensions of $\bx$ and $\by$ are $N$ and $M$, respectively, and the underdetermined condition, $N>M$, is supposed.

\subsection{Soft threshold function}
\begin{figure}[tb]
\centering
\includegraphics[width=6cm]{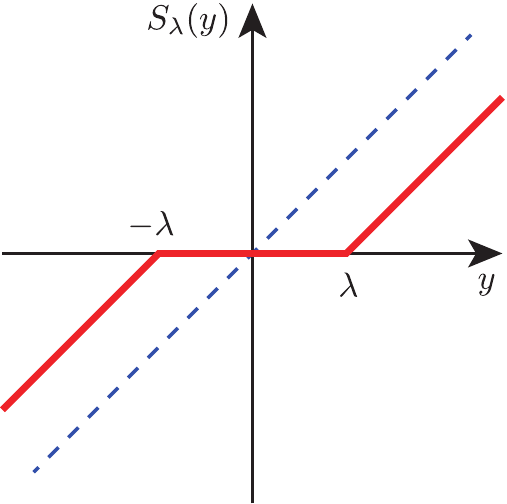}
\caption{
(Color online) Soft threshold function $x=S_{\lambda}(y)$ as a function of $y$.
The dashed line shows $x=S_{0}(y)=y$.
}
\label{fig:soft_threshold}
\end{figure}

It is instructive to begin with the one-dimensional case, $N=M=1$, and see the effect of the $L_1$-norm term.
The LASSO in Eq.~(\ref{eq:lasso}) is rewritten as
\begin{equation}
F(x) = \frac{1}{2} (y-x)^2 + \lambda | x |.
\label{eq:lasso_1d}
\end{equation}
Minimization of this function can be performed by considering $x>0$ and $x<0$ separately.
The solution is called the \emph{soft threshold function} $S_{\lambda}(y)$, and is given by
\begin{equation}
S_{\lambda}(y) \equiv
\begin{cases}
y - \lambda & (y > \lambda) \\
0 & (-\lambda \leq y \leq \lambda) \\
y + \lambda & (y < -\lambda)
\end{cases},
\label{eq:soft_threshold}
\end{equation}
Figure~\ref{fig:soft_threshold} shows the solution $x=S_{\lambda}(y)$ compared with $x=y$, i.e., the solution in the case without the $L_1$ term ($\lambda=0$).
It turns out that for a given $y$, its absolute value is reduced by $\lambda$ and approaches zero as a lower limit.
A small input is thus discarded and a sparse solution is generated.

\subsection{Method I: ISTA}
\label{subsec:ista}
Gradient descent is a simple and fundamental method for finding the solution of inverse problems.
However, because the derivative of $g(\bx)=\lambda \left\| \bx \right\|$ is discontinuous, it cannot be naively applied to the LASSO.
In the following, starting from the gradient descent, we consider how to treat the non-differentiable function $g(\bx)$ and derive an alternative update formula that is applicable to the LASSO.

\subsubsection{Majorization-minimization (MM)}
In the gradient descent, the vector $\bx$ is updated iteratively using the derivative of $f(\bx)$ as
\begin{equation}
\bx_{t+1} = \bx_{t} - \eta \nabla f(\bx_{t}),
\label{eq:gradient_descent}
\end{equation}
where $\eta$ is a small quantity.
The exact solution is obtained as long as $f(\bx)$ is a convex differentiable function in the region considered.
This update formula can be expressed as
\begin{equation}
\bx_{t+1} = \argmin_{\bx} \tilde{f}_{1/\eta}(\bx, \bx_{t}).
\end{equation}
The function $\tilde{f}_{1/\eta}(\bx, \bx_{t})$ is defined by 
\begin{equation}
\tilde{f}_{1/\eta}(\bx, \bx_{t}) \equiv f(\bx_{t}) + [\nabla f(\bx_{t})]^\mathrm{T} (\bx-\bx_{t}) + \frac{1}{2\eta}\left\| \bx -\bx_{t} \right\|^2_2,
\end{equation}
which is called the \emph{majorizer} of the function $f(\bx)$.
The majorizer approximates $f(\bx)$ around $\bx=\bx_{t}$ with a quadratic function as a Taylor expansion, but the quadratic term is simplified with an isotropic form [no dependence on $\nabla f(\bx_{t})$].
The most important feature of the majorizer is that, if $f(\bx)$ is sufficiently smooth around $\bx_{t}$ and if $\eta$ is sufficiently small\cite{footnote_smooth}, $\tilde{f}_{1/\eta}(\bx, \bx_{t})$ satisfies the following inequality against the original function $f(\bx)$~\cite{Lange-book}:
\begin{equation}
f(\bx) \le \tilde{f}_{1/\eta}(\bx, \bx_{t}).
\label{eq:majorizer_ineq}
\end{equation}
The equality is satisfied at $\bx=\bx_{t}$ if $\nabla f(\bx_{t})=0$, namely, when the temporary solution $\bx_{t}$ reaches the exact solution.
This inequality indicates that one may use the majorizer $\tilde{f}_{1/\eta}(\bx, \bx_{t})$ instead of $f(\bx)$ to find the minimum of $f(\bx)$.
The gradient descent can be regarded as a successive minimization of $\tilde{f}_{1/\eta}(\bx, \bx_{t})$.

Now, we exploit the inequality~(\ref{eq:majorizer_ineq}) for establishing an algorithm for solving minimization problems that include a non-differential function.
Adding $g(\bx)$ to both sides of Eq.~(\ref{eq:majorizer_ineq}), we obtain
\begin{equation}
F(\bx) \le \tilde{F}_{1/\eta}(\bx, \bx_{t}),
\end{equation}
where
\begin{equation}
\tilde{F}_{1/\eta}(\bx, \bx_{t}) \equiv f(\bx_{t}) + [\nabla f(\bx_{t})]^\mathrm{T} (\bx-\bx_{t}) + \frac{1}{2\eta}\left\| \bx -\bx_{t} \right\|^2_2 + g(\bx).
\end{equation}
This majorizer defines an alternative minimization problem that does not include $A\bx$ and hence is solvable in contrast to the original $F(\bx)$.
The solution attained through minimizing $\tilde{F}_{1/\eta}(\bx, \bx_{t})$ is expressed as
\begin{equation}
\bx_{t+1} = \argmin_{\bx} \left\{ \frac{1}{2}\left\| \bx -\bv\right\|^2_2 + \eta g(\bx) \right\},
\label{eq:MM}
\end{equation}
where
\begin{equation}
\bv = \bx_{t} - \eta \nabla f(\bx_{t}).
\label{eq:MM_v}
\end{equation}
The update formula, Eq.~(\ref{eq:MM}), is very generic and can be used even if $g(\bx)$ is not a differentiable function.
This method is called the majorization-minimization (MM) algorithm.

\subsubsection{Application of MM to LASSO}
Let us apply the MM algorithm to the LASSO.
The update formula is obtained by substituting $g(\bx)$ in Eq.~(\ref{eq:MM}) with Eq.~(\ref{eq:gx}).
Then, the minimization is performed for each element separately because all elements are independent and have the form of Eq.~(\ref{eq:lasso_1d}).
The solution is hence the soft-threshold function defined in Eq.~(\ref{eq:soft_threshold}):
\begin{equation}
[\bx_{t+1}]_k = S_{\lambda\eta}([\bv]_k),
\end{equation}
where $[\cdot]_k$ denotes the $k$-th element of the vector.
We represent the above equation simply by
\begin{equation}
\bx_{t+1} = S_{\lambda\eta}(\bv).
\label{eq:MM_lasso}
\end{equation}
Here, $S_{\lambda\eta}(\bv)$ is regarded as an element-wise soft threshold function.
An explicit expression for $\bv$ is obtained from Eq.~(\ref{eq:MM_v}) as
\begin{equation}
\bv = \bx_{t} + \eta A^\mathrm{T} (\by - A \bx_{t}).
\end{equation}
We set $\eta=1/\| A^\mathrm{T}A \|_2$ to satisfy the condition of the majorizer~\cite{Nocedal-Wright-book}.
The update in Eq.~(\ref{eq:MM_lasso}) is repeated until $\bx_{t}$ converges.
This iterative algorithm based on the MM method is called ISTA.
Based on Nesterov's acceleration~\cite{NAG}, a faster version of ISTA (called FISTA) has been proposed~\cite{FISTA}.

\subsection{Method II: ADMM}
\label{subsec:admm}

We now describe the ADMM, a flexible method developed by Boyd \textit{et al.}~\cite{Boyd11}
As discussed in Sect.~\ref{subsec:really_sparse}, the choice of basis is crucial in applications of the LASSO.
Hence, we need to consider the $L_1$ term of the form $g(\bx)=\lambda \left\| B\bx \right\|_1$, where $B$ is a transformation matrix to a sparse basis.
For this case, ISTA does not work well without introducing complications~\cite{Beck2009} but the ADMM does~\cite{Wahlberg2012}.
The only possible difficulty of the ADMM is the required computation of an inverse of a $N\times N$ matrix (shown later).
If the inverse matrix can be obtained, the ADMM should be the first choice in practice.

\subsubsection{Augmented Lagrange multiplier method}

We first represent the minimization problem, $\min_{\bx} F(\bx)$, of the function in Eq.~(\ref{eq:fx_gx}) as
\begin{equation}
\min_{\bx, \bz} \left\{ f(\bx)+g(\bz) \right\}
\quad \text{subject to} \quad
\bm{h}(\bx, \bz) \equiv B\bx-\bz=0.
\label{eq:fx_gz}
\end{equation}
The single minimization problem is split into two problems with an additional constraint.
A minimization is performed for $\bm{x}$ and $\bm{z}$ separately, and the constraint is then imposed gradually.
This treatment leads to fast convergence and allows another constraints to be flexibly handled.

Normally, a constraint is treated using the Lagrange multiplier method.
However, this method is not a good choice in the present situation, because the Lagrange multiplier term enforces the constraint rigorously in each iteration step, and the two minimization problems are strongly coupled.
To relax the constraint, we apply the \emph{augmented} Lagrange multiplier method~\cite{Bertsekas1996},
and formulate Eq~(\ref{eq:fx_gz}) as
\begin{equation}
F(\bx, \bz; \bnu)
= f(\bx) + g(\bz)
+ \bnu^\mathrm{T} \bm{h}(\bx, \bz)
+ \frac{\mu}{2} \left\| \bm{h}(\bx, \bz) \right\|_2^2,
\label{eq:aux_Lagrange}
\end{equation}
where $\bnu$ is the Lagrange multiplier and $\mu$ is the coefficient of the penalty term, $\left\| \bm{h}(\bx, \bz) \right\|_2^2$.
The penalty term pushes the solution to follow the constraint, $\bm{h}(\bx, \bz)=0$, rather gradually.
Although the penalty term itself does not enforce the constraint rigorously, the Lagrange multiplier term instead enforces the constraint in a converged solution.
The augmented Lagrange multiplier method thus takes advantage of the two methods; it achieves fast convergence and a rigorous implementation of the constraint.
Minimization of $F(\bx, \bz; \bnu)$ with respect to $\bx$, $\bz$, and $\bnu$ is performed iteratively for a fixed value of $\mu$.
The vectors $\bx$ and $\bz$ are updated based on the solution of the individual minimization problems at each iteration step.
The solutions $\bx^\mathrm{new}$ and $\bz^\mathrm{new}$ are written symbolically as
\begin{align}
\bx^\mathrm{new} &= \argmin_{\bx} \left\{ f(\bx) + \frac{\mu}{2} \left\| \bm{h}(\bx, \bz) + \frac{\bnu}{\mu} \right\|_2^2 \right\},
\label{Ap1}
\\
\bz^\mathrm{new} &= \argmin_{\bz} \left\{ g(\bz) + \frac{\mu}{2} \left\| \bm{h}(\bx, \bz) + \frac{\bnu}{\mu} \right\|_2^2 \right\}.
\label{Ap2}
\end{align}
Explicit solutions depend on the form of $f(\bz)$ and $g(\bx)$.
The multiplier $\bnu$ is updated using the rule~\cite{Bertsekas1996}
\begin{equation}
\bnu^\mathrm{new} = \bnu + \mu \bm{h}(\bx, \bz).
\label{Ap3}
\end{equation}
The update in Eqs.~(\ref{Ap1})--(\ref{Ap3}) are repeated until convergence is reached.

\subsubsection{Application of ADMM to LASSO}
Let us apply the discussion above to the LASSO.
Replacing $f(\bx)$ and $g(\bx)$ with Eqs.~(\ref{eq:fx}) and (\ref{eq:gx}), respectively,
we rewrite Eqs.~(\ref{Ap1}) and (\ref{Ap2}) as
\begin{align}
\bx^\mathrm{new} &= \argmin_{\bx} \left\{ \frac{1}{2\lambda} \left\| \by - A\bx \right\|_2^2 + \frac{\mu}{2} \left\| B\bx-\bz + \bu \right\|_2^2 \right\},
\label{Ap1_lasso}
\\
\bz^\mathrm{new} &= \argmin_{\bz} \left\{ \left\| \bz \right\|_1 + \frac{\mu}{2} \left\| B\bx-\bz + \bu \right\|_2^2 \right\}.
\label{Ap2_lasso}
\end{align}
Here, we changed the variable $\bnu$ into $\bu \equiv \bnu/\mu$ to simplify the notation.
The minimization of the quadratic form in Eq.~(\ref{Ap1_lasso}) can be done analytically to yield
\begin{equation}
\bx^\mathrm{new} = \left( \mu B^\mathrm{T}B + \frac{1}{\lambda} A^\mathrm{T}A \right)^{-1}
\left( \frac{1}{\lambda} A^\mathrm{T} \by + \mu B^\mathrm{T}(\bz - \bu) \right).
\label{eq:x_new}
\end{equation}
The minimization problem in Eq.~(\ref{Ap2_lasso}) is independent among vector elements.
Therefore, the solution in the one-dimensional case applies to each element, and $\bz^\mathrm{new}$ is given by
\begin{equation}
\bz^\mathrm{new} = S_{1/\mu}(B\bx + \bu),
\label{eq:z_new}
\end{equation}
where $S_{1/\mu}(B\bx + \bu)$ is the element-wise soft threshold function defined in Eq.~(\ref{eq:soft_threshold}).
The update of $\bu$ is obtained from Eq.~(\ref{Ap3}) as
\begin{equation}
\bu^\mathrm{new} = \bu + B\bx - \bz.
\label{eq:u_new}
\end{equation}

The computation procedure is summarized as follows.
For given $\lambda$ and $\mu$, we begin with initial vectors $\bx=0$, $\bz=0$, and $\bnu=0$.
The updates, Eqs.~(\ref{eq:x_new})--(\ref{eq:u_new}), are repeated until convergence is reached.
We note that, in Eq.~(\ref{eq:x_new}), the inverse of a matrix of size $N\times N$ is performed before the iteration starts.
Then, the updates include only matrix-vector products.

\subsection{How to fix the hyperparameter $\lambda$}
\label{subsec:lambda}

The optimization problem in the LASSO [Eq.~(\ref{eq:fx_gx})] includes the hyperparameter $\lambda$, and its solution $\bx^*(\lambda)$ thus depends on $\lambda$.
We then need to fix the value of $\lambda$ to select the best solution among $\bx^*(\lambda)$.
We introduce two methods that are often employed in the literature.

\begin{figure}[tb]
\centering
\includegraphics[width=\linewidth]{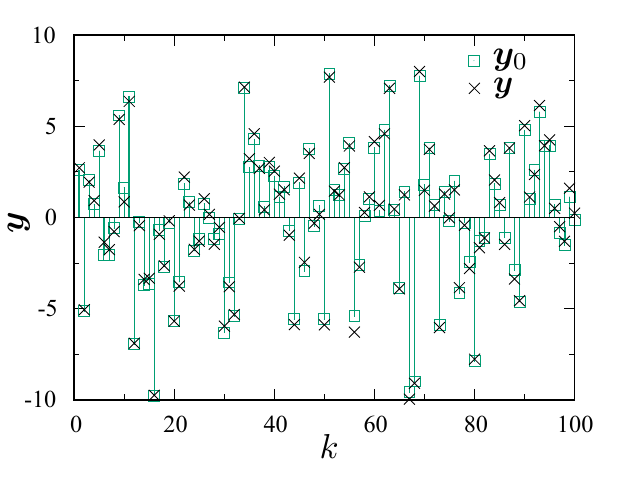}
\caption{(Color online) Input vector $\by = A\bx_0 + \bm{\eta}$ with dimension $M=100$ (black crosses) compared with $A\bx_0 \equiv \by_0$ (green squares). Here, $\bm{\eta}$ is Gaussian noise with amplitude $0.1||\by_{0}||_1/M$ and standard deviation $10^{-1}$.}
\label{fig:lasso_example_input}
\end{figure}

For a better description, we again consider the explicit problem of the random matrix $A^\mathrm{rand}$ introduced in Sec.~\ref{subsec:true_solution}.\cite{footnote_sample_script}
The true solution $\bx_0$ has only $n=20$ non-zero components out of $N=1000$, as depicted in Fig.~\ref{fig:example_input}(a).
The input $\by$ with dimension $M=100$ is now disturbed by a Gaussian noise $\bm{\eta}$ according to Eq.~(\ref{eq:y_Ax_noise}).
Figure~\ref{fig:lasso_example_input} shows comparison between $\by$ and $A\bx_0 \equiv \by_0$.

\subsubsection{``Elbow'' method}
\label{subsubsec:elbow}
Let us begin by observing the influence of $\lambda$ on the solution $\bx^*(\lambda)$.
To this end, we introduce a score $s(\lambda)$ that quantifies
the extent to which the original equation, $\by=A\bx$, is satisfied.
The \emph{coefficient of determination}, $R^2$, is used as a function for the score, which in the present case, is defined as
\begin{equation}
s(\lambda) = R^2(\by, A\bx^*(\lambda))
\equiv 1-\frac{\| \by - A\bx^*(\lambda)\|_2^2}{\| \by - \langle \by \rangle \|_2^2},
\label{eq:score}
\end{equation}
where $\langle \by \rangle$ denotes the mean value of $\by$.
In this definition, the squared error $\| \by - A\bx^*(\lambda)\|_2^2$ is normalized by the variance of $\by$ so that it can be compared between different dataset of $\by$.
The score $s(\lambda)$ yields a maximum of $1$ when the equation $\by=A\bx$ is exactly satisfied; it decreases (even becomes negative) as the deviation between $\by$ and $A\bx^*(\lambda)$ increases.

\begin{figure}[tb]
\centering
\includegraphics[width=\linewidth]{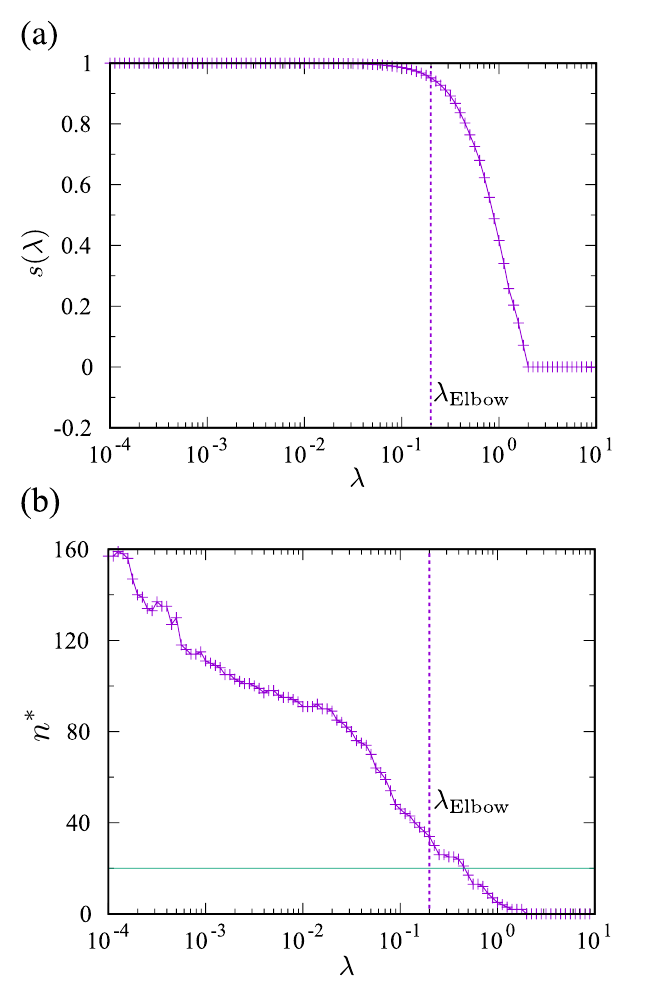}
\caption{(Color online) (a) $\lambda$-dependence of the score $s(\lambda)$. The vertical dashed line indicates the optimal value, $\lambda_\mathrm{Elbow}=0.2$, determined using the elbow method. (b) Number of non-zero components, $n^*$, retained in the solution $\bx^*(\lambda)$. The horizontal line indicates $n=20$ (the number of non-zero components in the true solution $\bx_0$).}
\label{fig:lasso_example_score}
\end{figure}

\begin{figure*}[tb]
    \centering
    \includegraphics[width=\linewidth]{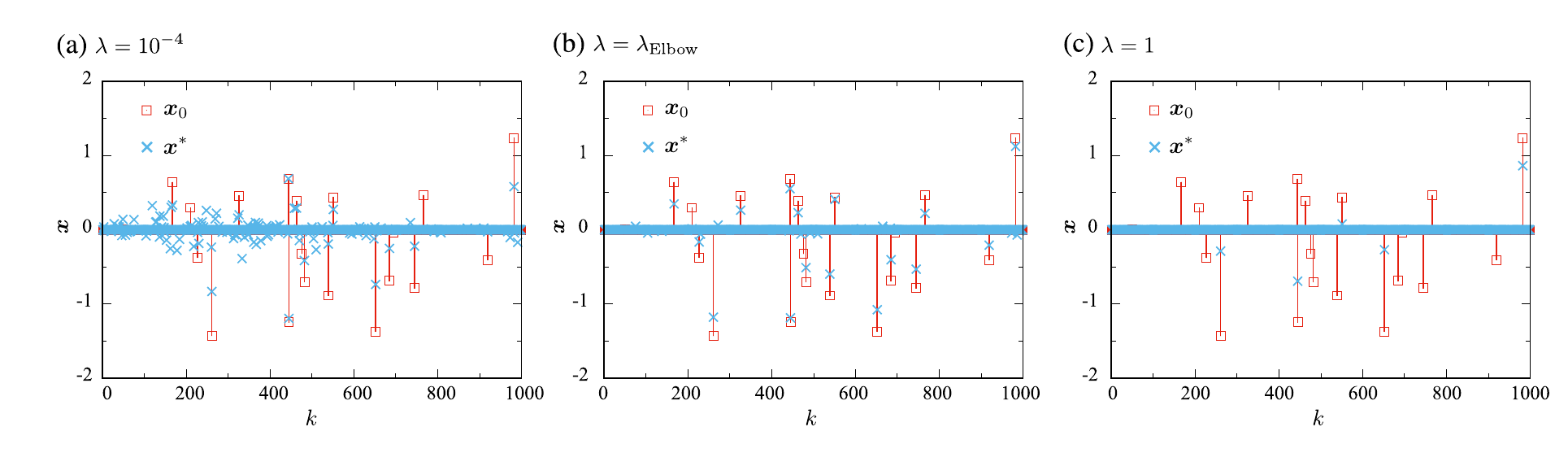}
    \caption{(Color online) Comparison between the solution $\bx^*(\lambda)$ reconstructed using the $L_1$ regularization (blue crosses) and the true solution $\bx_0$ (red squares). (a) $\lambda=10^{-4}$, (b) $\lambda=\lambda_\mathrm{Elbow}=0.2$, and (c) $\lambda=1$.
    }
    \label{fig:lasso_example_result}
\end{figure*}

Figure~\ref{fig:lasso_example_score}(a) shows the numerical result for $s(\lambda)$ in the random matrix problem.
This graph represents the typical behavior of $s(\lambda)$.
When $\lambda$ is sufficiently small, the score reaches a maximum value of $s(\lambda) \fallingdotseq 1$, because the fitting to the equation $\by=A\bx$ takes priority over a reduction of bases by $L_1$-norm regularization.
This high score, however, does not indicate agreement between the obtained solution $\bx^*(\lambda)$ and the true solution $\bx_0$, but instead implies that $\bx^*(\lambda)$ is highly affected by the noise (overfitting).
Figure~\ref{fig:lasso_example_score}(b) shows that the number of non-zero components, $n^*$, in $\bx^*(\lambda)$ is much higher than the actual number $n=20$.
The value of $s(\lambda)$ is insensitive to the variation of $\lambda$ as long as $n^*$ is far above $n=20$, namely, in the region $\lambda \lesssim 10^{-1}$.
Once $\lambda$ exceeds this region and $n^*$ passes below $n=20$, 
$s(\lambda)$ drops drastically and falls below $0.5$, where the original equation $\by=A\bx$ is not respected anymore.
For $\lambda \gtrsim 2\times10^1$, no component is retained, i.e., $n^*=0$ (underfitting).

From this behavior, we expect a reasonable solution around the ``elbow'' (bending point) of $s(\lambda)$, where two effects, i.e., fitting to the original equation and $L_1$-norm regularization, are competing.
We choose $\lambda_\mathrm{Elbow} \equiv 0.2$ by rotating the graph by 45 degree and find the maximum.
This simple strategy has been utilized in clustering analysis~\cite{Thorndike53} and interaction network analysis~\cite{Decelle14,Yamanaka15}.

The $\lambda$-dependent solution $\bx^*(\lambda)$ is shown for three values of $\lambda$ in Fig.~\ref{fig:lasso_example_result}.
The optimal solution $\bx^*(\lambda_\mathrm{Elbow})$ in Fig.~\ref{fig:lasso_example_result}(b) turns out to hit the correct non-zero components, although these absolute values non-negligibly deviate from the exact values.
Recall that an exact recovery of $\bx_0$ as discussed in Sect.~\ref{subsec:true_solution}, is possible only in the absence of noise.
When $\lambda$ is too small (large), the solution $\bx^*(\lambda)$ contains too many (few) non-zero components and is clearly inconsistent with the exact solution $\bx_0$.

\subsubsection{Cross-validation method}
\label{subsubsec:CV}

An alternative, but more sophisticated estimation can be carried out using the cross-validation (CV) method in statistics~\cite{Kohavi95}.
We first describe the concept of the CV method and then discuss practical implementations.

The input dataset $\by$ is divided into two subsets, namely training dataset $\by_\mathrm{T}$ and validation dataset $\by_\mathrm{V}$.
An optimization problem is set up with $\by_\mathrm{T}$, with
the LASSO function in Eq.~(\ref{eq:lasso}) rewritten as 
\begin{equation}
F_\mathrm{LASSO}(\bx) = \frac{1}{2} \left\|\by_\mathrm{T}-\mathcal{P}_\mathrm{T}A\bx\right\|_2^2 + \lambda \left\| \bx \right\|_1,
\end{equation}
where $\mathcal{P}_\mathrm{T}$ is a projection operator into the subspace defined by $\by_\mathrm{T}$.
After the optimization problem is solved, the solution $\bx^*(\lambda)$ is validated with $\by_\mathrm{V}$ by evaluating the score defined by
\begin{equation}
s_\mathrm{CV}(\lambda) = \mathcal{S}(\by_\mathrm{V}, \mathcal{P}_\mathrm{V} A \bx^*(\lambda) ),
\label{eq:score_cv}
\end{equation}
where the operator $\mathcal{P}_\mathrm{V}=1-\mathcal{P}_\mathrm{T}$ projects $A\bx^*(\lambda)$ into the subspace with $\by_\mathrm{V}$.
The important point here is that the training (fitting) and the validation are preformed with different datasets.
Because of this, $s_\mathrm{CV}(\lambda)$ does not approach 1 even in the limit $\lambda\to0$ in contrast to $s(\lambda)$ in Eq.~(\ref{eq:score}).
In the presence of noise, a solution that perfectly fits a specific dataset does not fit a different dataset, resulting in a reduction of $s_\mathrm{CV}(\lambda)$ for $\lambda \to0$.
The highest score is achieved when the input $\by_\mathrm{T}$ is moderately fitted to avoid the influence of noise.

In practice, the division into $\by_\mathrm{T}$ and $\by_\mathrm{V}$ should be repeated to obtain a reliable estimation of $s_\mathrm{CV}(\lambda)$.
$K$-fold CV can be used to systematically generate multiple divisions of the dataset.
We split the input vector $\by$ into $K$ groups.
One of them is assigned as $\by_\mathrm{V}$ and kept for validation, and the other $K-1$ groups are assigned as $\by_\mathrm{T}$. 
A set of training and validation is performed $K$ times, i.e., for all possible configurations.
The final score is obtained by averaging the $K$ estimations of $s_\mathrm{CV}(\lambda)$.

\begin{figure}[tb]
\centering
\includegraphics[width=\linewidth]{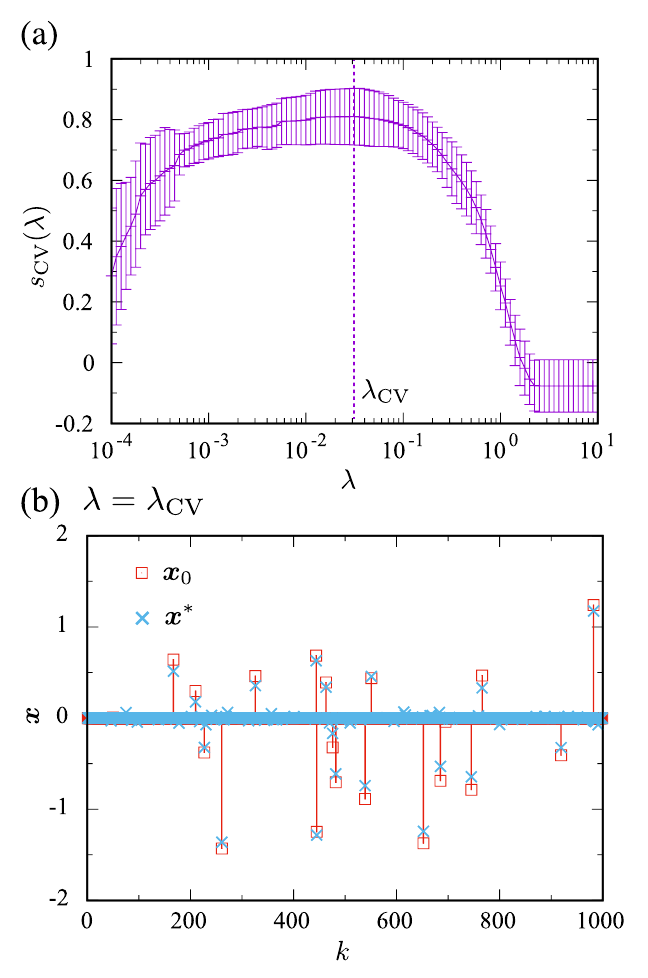}
\caption{(Color online) (a) CV score $s_\mathrm{CV}(\lambda)$ computed with a 5-fold CV calculation.
The vertical dashed line indicates the peak position, $\lambda_\mathrm{CV}=0.03$.
(b) Comparison between $\bx^*(\lambda)$ and $\bx_0$ at $\lambda=\lambda_\mathrm{CV}$.}
\label{fig:lasso_example_cv}
\end{figure}

Figure~\ref{fig:lasso_example_cv}(a) shows $s_\mathrm{CV}(\lambda)$ computed by a $5$-fold CV calculation.
The maximum in $s_\mathrm{CV}(\lambda)$ yields the optimal value $\lambda_\mathrm{CV}\equiv0.03$, which is smaller than the previous estimation $\lambda_\mathrm{Elbow}=0.2$ by order of 10.
The solution $\bx^*(\lambda_\mathrm{CV})$ is plotted in Fig.~\ref{fig:lasso_example_cv}(b) for comparison with $\bx^*(\lambda_\mathrm{Elbow})$ in Fig.~\ref{fig:lasso_example_result}(b).
The non-zero components are better fitted compared with $\bx^*(\lambda_\mathrm{Elbow})$ in Fig.~\ref{fig:lasso_example_result}(b), while at the same time, more redundant components remain finite in $\bx^*(\lambda_\mathrm{CV})$ though their absolute values are small.
Which method yields better estimation depends on the problem.
Nevertheless, the CV method is useful for removing ambiguity due to hyperparameters without bias.

The error bars in Fig.~\ref{fig:lasso_example_cv} can be reduced by increasing the division number $K$.
In particular, CV with the limit $K=M$ (the dimension of $\by$) is referred to as leave-one-out CV (LOO-CV).
An LOO-CV calculation produces a rather smooth curve of $s_\mathrm{CV}(\lambda)$, but has a large computational cost for solving the LASSO problem $M$ times repeatedly.
To mitigate this problem, Obuchi and Kabashima~\cite{Obuchi16} derived approximate formulas for LOO-CV.
The formulas can be evaluated easily using the solution $\bx^*(\lambda)$ computed once for the full dataset of $\by$, meaning that no repeated computation is necessary for different set of $\{ \by_\mathrm{V}, \by_\mathrm{T} \}$.
For details, we refer readers to the original paper, Ref.~\citen{Obuchi16}.

\section{Applications and Further Developments}
\label{sec:applications}

In this section, we introduce applications of sparse modeling based on $L_1$-norm regularization.
The first two subsections describe applications to measurement and experimental data analysis.
Another class of applications, namely the construction of an effective model from first-principles and experimental data, is presented in Sect.~\ref{subsec:model_selections}.
Section~\ref{subsec:develop} introduces further developments and extended theories related to sparse modeling.
Applications to quantum many-body theories are presented in separate sections.

\subsection{Compressed sensing}
\label{subsec:compressed_sensing}
The ability of $L_1$-norm regularization to find the true sparse solution has led to innovation in measurement methods.
An early work in the 1970's pointed out the potential utility of $L_1$ regularization in the context of geophysics~\cite{Claerbout73}.
Theories by Cand\'es \textit{et al.}~\cite{Candes06a,Candes06b} and Donoho~\cite{Donoho06} and their application to MRI by Lustig \textit{et al.}~\cite{Lustig07,Lustig08} have revolutionized measurement.
This technology is called \emph{compressed sensing} (also compressive sensing or compressive sampling)\cite{Eldar-book,Candes08,Elad-book}.

Let us suppose that there are measurements in which the quantities of interest, $\bx$, are connected to measurable quantities, $\by$, through an inverse problem, $\by=A\bx$.
A typical example is the Fourier transform, which is represented by
\begin{equation}
f(\bk_i) = \int_{V} d\br \rho(\br) e^{-i \bk_i \cdot \br}.
\label{eq:CS-DFT}
\end{equation}
In MRI measurement, for instance, $\rho(\br)$ denotes the density distribution of H$_2$O molecules and its Fourier component $f(\bk_i)$ is measured.
Discretizing $V$ into $N$ blocks with equivalent volumes and introducing the notation
\begin{equation}
f(\bk_i) \to y_i,
\quad
\rho(\br_j) (V/N) \to x_j,
\quad
e^{-i \bk_i \cdot \br_j} \to A_{ij},
\end{equation}
we obtain the equation $\by=A\bx$.
In practical situations, $\by$ may be incomplete because the measurement might be fundamentally restricted or the number of sampling points is intentionally reduced. In practice, $\bk_i$ points are randomly sampled to avoid an artificial structure in the reconstructed data. In all cases, solving the equation $\by=A\bx$ with respect to $\bx$ is an underdetermined inverse problem.
To express this situation explicitly, we represent measured data by $\by'=\mathcal{P}\by$, where $\mathcal{P}$ is a projection operator onto the subset of $\by$ with dimension $M$. Then, the LASSO function in Eq.~(\ref{eq:lasso_gen}) is expressed as
\begin{equation}
F_\mathrm{LASSO}(\bx) =  \frac{1}{2} \| \by' - \mathcal{P}A\bx \|_2^2 + \lambda \| B \bx \|_1.
\end{equation}
The basis transformation $B$ is, for example, the finite difference operator $D$, which transforms MRI images into a sparse representation.
If $n$, the number of relevant components of $B\bx$, is sufficiently small compared to $M$, $L_1$ regularization reconstructs a clear image from the incomplete data $\by'$.
One could further reduce $M$, and thus achieves reduction of measurement cost and time.

Compressed sensing can be applied to any experiment that uses the Fourier transform in the analysis procedure.
X-ray (neutron) diffraction experiments measure the structure factor $f(\bk_i)$ as the Fourier transform of the electron (nuclear) density $\rho(\br)$. The MEM has typically been used for the inversion~\cite{Sakata90,Sakata93}, but compressed sensing is now an alternative~\cite{Tanaka19}.
Compressed sensing has also been applied to NMR spectroscopy for studying molecular dynamics~\cite{Kazimierczuk11,Holland11},
scanning tunneling spectroscopy (STS)/scanning tunneling microscopy (STM) for investigating the $\bk$-space electronic properties from real-space measurements~\cite{Nakanishi16},
inverse X-ray fluorescence holography (IXFH) for deriving a three-dimensional image of atomic positions~\cite{Matsushita16}, and
X-ray absorption fine structure (XAFS) for elucidating atomic properties in solids~\cite{Akai18}.
The recent observation of a black hole shadow utilized the sparse-modeling technique because signals are received simultaneously at multiple observatories distributed worldwide and hence the sampling data are inevitably incomplete~\cite{Honma14,EventHorizonTelescope19a,EventHorizonTelescope19b,EventHorizonTelescope19c,EventHorizonTelescope19d,EventHorizonTelescope19e,EventHorizonTelescope19f}.

\subsection{Advanced experimental data analysis}

\subsubsection{Phase retrieval}

X-ray and neutron diffraction experiments measure the intensity $I(\bk_i)$, which is related to the structure factor $f(\bk_i)$ as $I(\bk_i) \propto |f(\bk_i)|^2$. Hence, experiments only provide the information of $|f(\bk_i)|$. That is, in the expression
\begin{equation}
f(\bk_i) = |f(\bk_i)| e^{i\theta(\bk_i)},
\end{equation}
the information of the phase $\theta(\bk_i)$ is missing.
Methods have been established to retrieve the phase factor, such as iterative Fourier transform methods based on the error reduction algorithm~\cite{Fienup78} and the hybrid input-output algorithm~\cite{Fienup82}.
To achieve robust and efficient phase retrieval, compressed sensing has been extended to the situation where only  $( |y_1|, |y_2|, \cdots, |y_M| )$ are known in the equation $\by=A\bx$~\cite{Matthew07}.
A similar approach has been applied to terahertz imaging\cite{Chan08} and
coherent X-ray diffraction imaging (CDI)\cite{Newton12,Yokoyama19}.

\subsubsection{Peak identification}
\begin{figure}[tb]
\centering
\includegraphics[width=6cm]{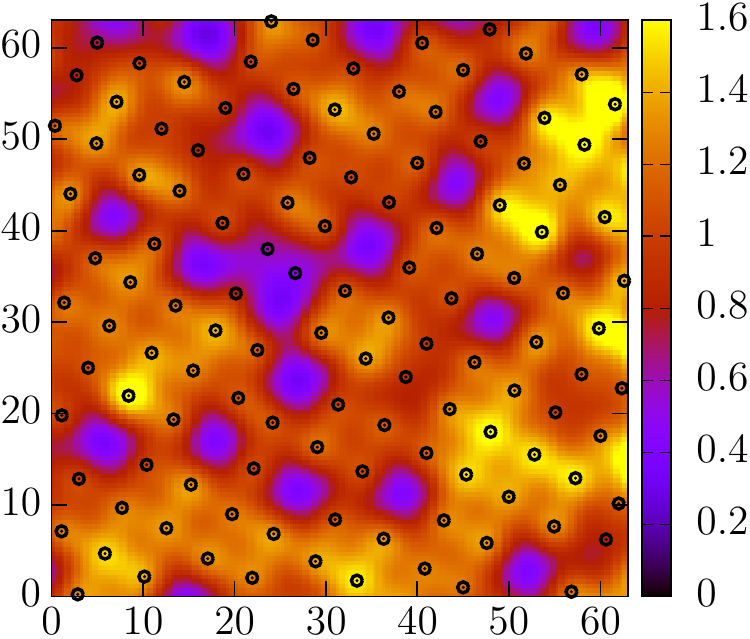}
\caption{
(Color online) STM topography image for SrVO$_3$(100) thin film. The open black circles indicate peak positions inferred using the sparse-modeling approach. The image size is 6.25 nm$^2$ with $64\times64$ pixels.
Reprinted with permission from Ref.~\citen{Miyama18}.
}
\label{fig:miyama}
\end{figure}

There is another interesting application of $L_1$ regularization to data analysis for STM experiments.
The intensity plot in Fig.~\ref{fig:miyama} shows an STM topography image measured for SrVO$_3$ thin film.
High intensities indicate the existence of atoms.
One can see the periodic alignment of atoms, some lattice defects, and a lattice dislocation. The identification of such features partly relies on experience.
Definite criteria are desired for an unbiased analysis of topography images.

Miyama and Hukushima proposed an algorithm to identify the peak position in the topography image~\cite{Miyama18}.
In this problem, $\by$ is a one-dimensional representation of the image and $x_k$ represents the weight of a peak at position $\br_k$.
Here, $\bx$ consists of a large number of peaks (as many as the number of pixels).
Using $L_1$-norm regularization, they retained relevant peaks and succeeded in identifying peak positions (open black circles  in Fig.~\ref{fig:miyama}) without any tuning parameters. Thus, their method offers unbiased peak identification for topography images.
A related approach for one-dimensional peak deconvolution has been proposed based on Bayesian inference~\cite{Nagata12,Igarashi16,Tokuda17}.

\subsection{Model selection}
\label{subsec:model_selections}
In many situations, it is useful to construct an effective model that describes the numerical data obtained from numerical simulations or experiments.
This model is sometimes used to make further accurate simulations while avoiding expensive calculations.
Using experiments, one may want to get insight into a microscopic mechanism from the observed experimental data.
For both simulations and experiments, sparse modeling (compressed sensing) is useful, as we will show in this subsection.

\subsubsection{Acceleration of simulations through construction of effective model}
Typical applications include structure optimizations and molecular dynamics based on first-principles calculations.
First-principles calculations based on density functional theory (DFT) are a powerful tool for investigating and predicting the structure of a solid.
However, fully first-principles calculations may be too expensive because the electronic structure must be determined for each atomic configuration.

Nelson \textit{et al.} applied compressed sensing to cluster expansion (CE).~\cite{Nelson13a}
They illustrated the use of their method on two CE models of configurational energetics for Ag--Pt alloys and protein folding in Ref.~\citen{Nelson13a}.
The CE method constructs an energy model for describing some configurations of a crystal.
The total energy can be expressed as  
\begin{align}
E(\sigma) &= E_0 + \sum_f \bar{\Pi}_f(\sigma) J_f,\label{eq:CE}
\end{align}
where $f$ represents symmetrically distinct clusters of lattice sites (points, pairs, triplets, etc.).
The symbol $\sigma$ denotes the atomic configuration, which may be a collection of pseudo spins specifying the type of atom at each site.
The matrix elements in $\bar{\Pi}_f(\sigma)$ are obtained as symmetrized averages of the products of these pseudo spins.
What we want to determine is $J_f$, the effective cluster interactions (ECIs).
Once ECIs are determined from expensive first-principles data (for many atomic configurations),
one can compute an approximate value for the energy of any atomic configuration.

The ECIs may be considered as the extension of exchange couplings for Ising spins
but contain vastly more complicated clusters than nearest neighbor pairs.
Using $L_1$-norm regularization, they fitted first-principles data with a small number of $J_f$ in Eq.~(\ref{eq:CE}), and succeeded in constructing a model with high predictive power.
They further extended their method by using a Bayesian implementation of compressed sensing, which removes an adjustable parameter in the original implementation.~\cite{Nelson13b}
The importance of the selection of descriptors has been discussed in the context of describing the ground-state structure of binary compounds~\cite{Ghiringhelli15,Ghiringhelli17}.

\begin{figure}[tb]
	\centering
	\includegraphics[width=0.9\linewidth]{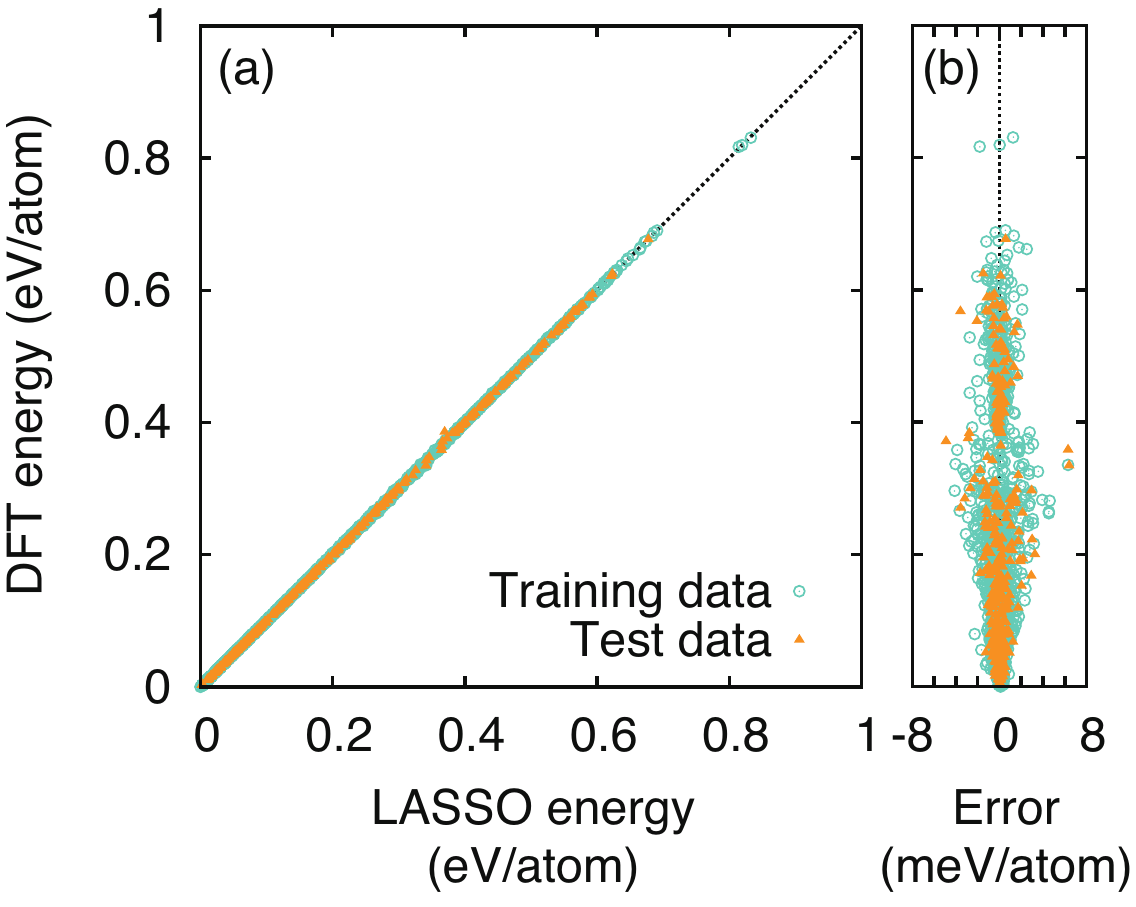}
	\caption{
		(Color online) (a) Comparison of energies predicted using effective model and DFT for Mg.
		Energy is measured from the energy of the ideal hexagonal close-packed structure.
        (b) Prediction errors for the data shown in (a).
		Reprinted with permission from Ref.~\citen{Seko14} \copyright 2014 the American Physical Society.
	}
	\label{fig:seko}
\end{figure}

The potential energy surface (PES) plays a key role in molecular dynamics calculations.
A PES describes the relationship between the energy and atomic configurations, and can be constructed from training data obtained using DFT calculations.
The key factor for success is the choice of descriptors, which are basis functions for representing atomic configurations.
Seko \textit{et al.} developed an algorithm for automatically optimizing and selecting important descriptors and demonstrated the efficiency of their method for elemental metals.~\cite{Seko14,Seko15}
Their candidate descriptors are powers of various types of pair functions of the distance between two atoms.
Using linear regression based on $L_1$-norm regularization, they showed that the energy can be expressed by a linear model with simple basis functions with a small number of coefficients.
Figure~\ref{fig:seko} shows the accuracy of their model constructed for Mg with 95 basis functions.
The prediction errors are typically smaller than 1 meV/atom.
\begin{figure}[tb]
	\centering
	\includegraphics[width=0.9\linewidth]{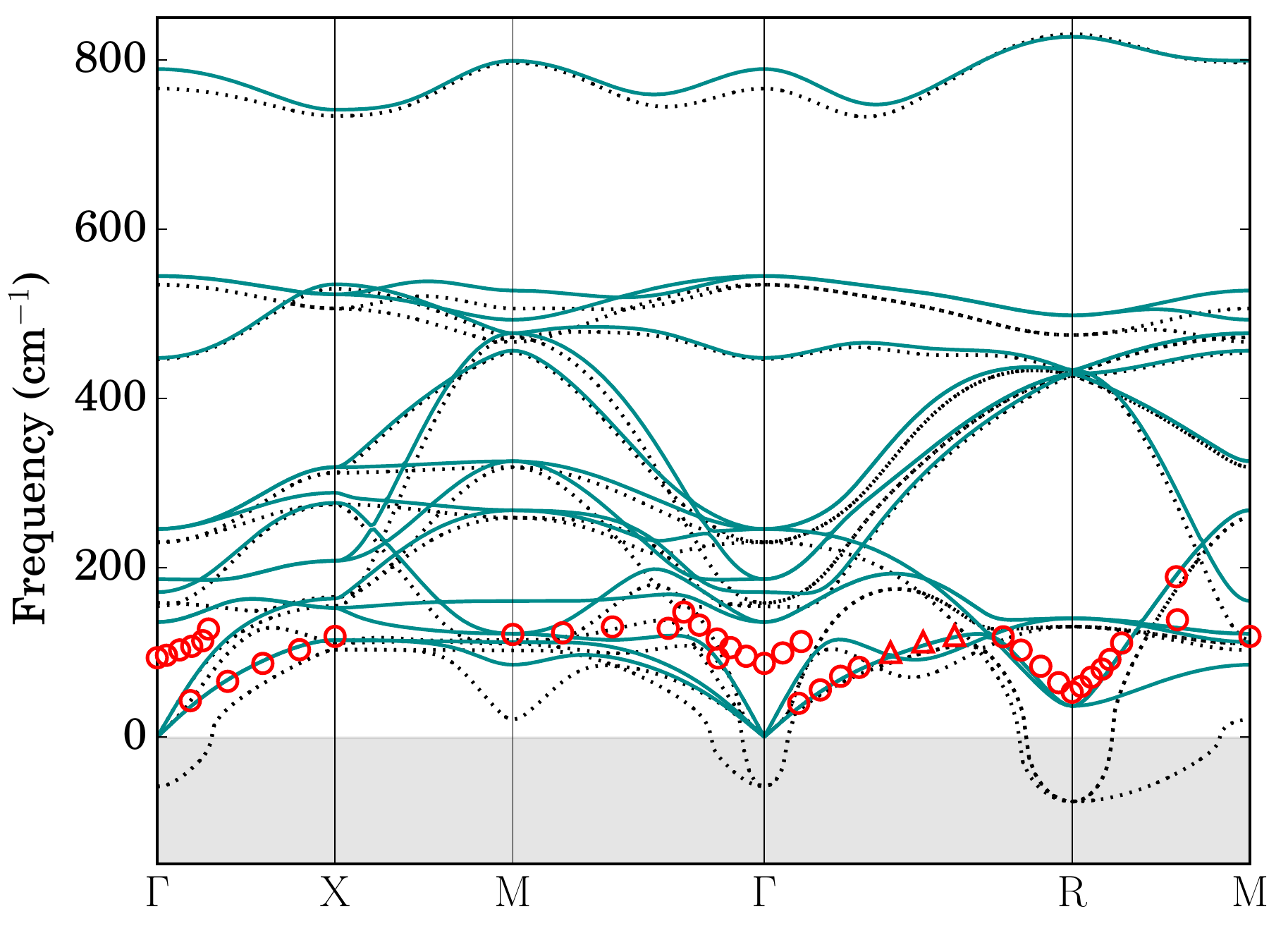}
	\caption{
		(Color online) Anharmonic phonon dispersion of cubic STO at 300 K.
		The dotted lines denote harmonic phonon dispersion and the open symbols are experimental values.
		Reprinted with permission from Ref.~\citen{Tadano15} \copyright 2015 the American Physical Society.
	}
	\label{fig:tadano}
\end{figure}

Compressed sensing has also been used to construct effective models of phonons for strongly anharmonic crystals.
Examples include the prediction of lattice thermal conductivity for compounds such as Cu$_{12}$Sb$_4$S$_{13}$\cite{Zhou14} and anharmonic phonon frequency and phonon lifetime for cubic SrTiO$_3$ (STO)\cite{Tadano15}.
Figure~\ref{fig:tadano} shows the anharmonic phonon dispersion of cubic STO computed using the effective model; good agreement with experimental data can be seen.

\subsubsection{Construction of minimal model for describing theoretical or experimental data}
The sparse modeling can be used to construct a minimal effective model for describing theoretical or experimental data.
For instance, one can construct a spatially localized model for reproducing the eigenfunctions and eigenvalues of a given system, such as localized Wannier functions.~\cite{Ozolins:2013ig}
This approach was extended and used to search for the localized Wannier functions of topological band structures.~\cite{Budich14}

The sparse modeling can be used to bridge experiments and theories.
Tamura and Hukushima proposed the use of the sparse-modeling technique for estimating relevant microscopic parameters in an effective spin model from magnetization curves which can be measured in experiments.~\cite{Tamura17}
Similarly, Mototake \textit{et al.} proposed a method for analyzing core-level X-ray photoelectron spectroscopy (XPS) spectra based on the Bayesian model selection framework, where relevant parameters are automatically selected.~\cite{Mototake19}
Another interesting theoretical approach is the construction of effective models from numerical data for Hubbard-type models~\cite{Fujita18}.
It was shown that the resultant effective models reproduce the results of conventional approaches, such as perturbation theories.
This approach is more general and may be applicable to analysis of experimental data.

\subsection{Further developments}
\label{subsec:develop}

\subsubsection{Diagnosis of compressed sensing results}

Compressed sensing yields a result even from an incomplete dataset.
We should, however, keep in mind that the result may fail to capture the correct features in the exact solution if the dataset is excessively incomplete.
The example of the random-matrix model in Sect.~\ref{subsec:true_solution} revealed the existence of a boundary that separates success and failure.
To obtain a successful result, the sampling number $M$ (the dimension of $\by$) should be sufficiently large to satisfy $\alpha\equiv M/N>\alpha_\mathrm{c}$ (Fig.~\ref{fig:kabashima}).
An important and practical question is whether it is possible to know that the number of samples is sufficient for compressed sampling without knowing the correct solution in the limit of $M\to N$.

\begin{figure}[tb]
\centering
\includegraphics[width=\linewidth]{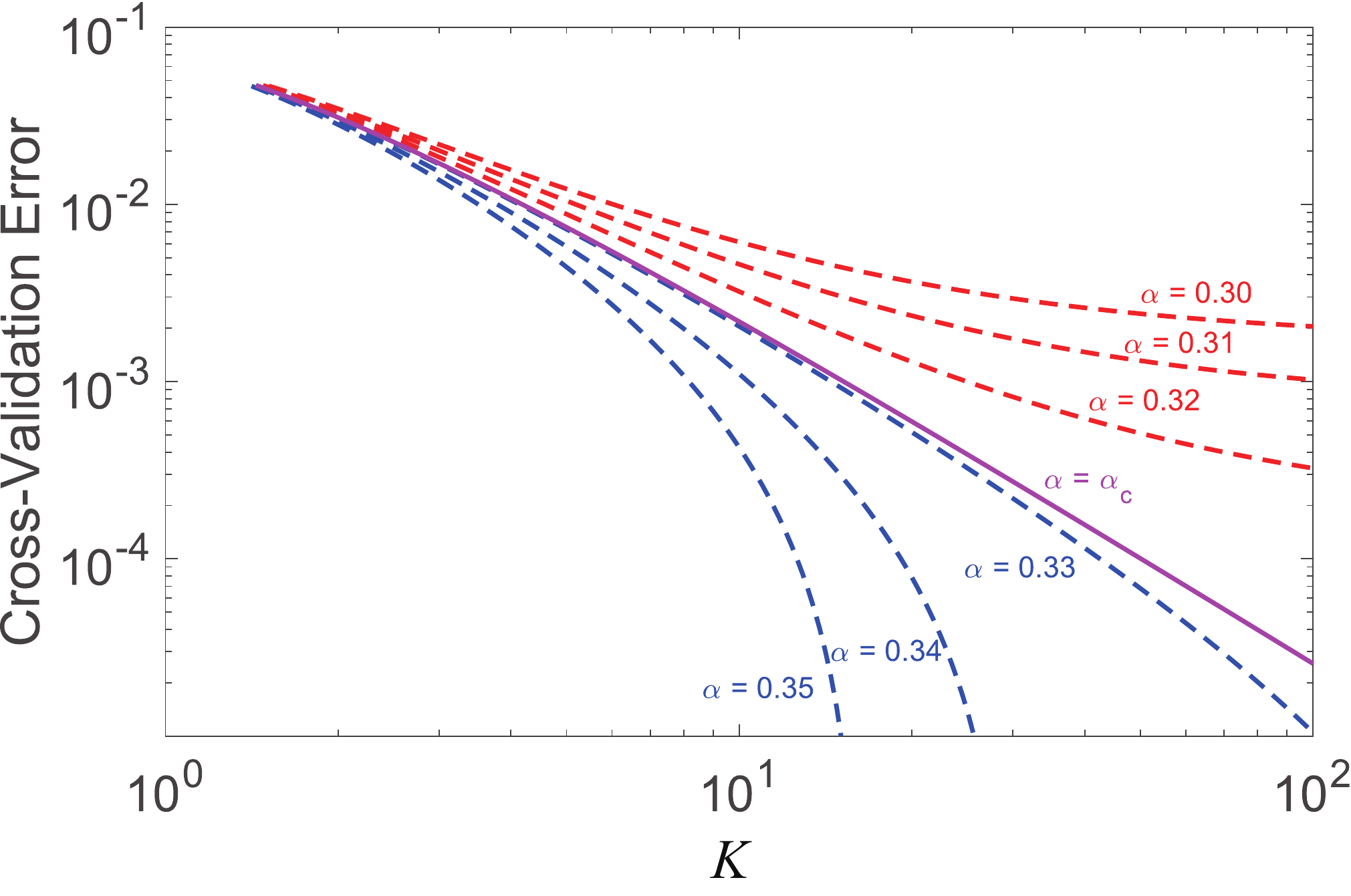}
\caption{
(Color online) CV error [Eq.~(\ref{eq:CVE})] as a function of $K$ in the $L_1$-norm minimization of the random-matrix model. $\rho\equiv n/N$ is fixed at $\rho=0.1$ and $\alpha\equiv M/N$ is varied. $\alpha_\mathrm{c} \fallingdotseq 0.3288$.
Reprinted with permission from Ref.~\citen{Nakanishi18b}.
}
\label{fig:nakanishi}
\end{figure}

Nakanishi and Hukushima proposed a method for diagnosing the results of compressed sensing as success or failure~\cite{Nakanishi18b,Nakanishi18a}.
They applied $K$-fold CV (Sect.~\ref{subsubsec:CV}) and computed the CV error (CVE) defined by
\begin{equation}
\text{CVE} = \frac{1}{2M_\mathrm{V}} \| \by_\mathrm{V} - \mathcal{P}_\mathrm{V} A \bx_\mathrm{T} \|_2^2,
\label{eq:CVE}
\end{equation}
where $M_\mathrm{V}=M/K$ is the size of $\by_\mathrm{V}$, $\bx_\mathrm{T}$ is the solution computed for a given training dataset $\by_\mathrm{T}$, and $\by_\mathrm{V}$ and $\mathcal{P}_\mathrm{V}$ are explained in Eq.~(\ref{eq:score_cv}).
Figure~\ref{fig:nakanishi} shows a log-log plot of CVE versus $K$ for several values of $\alpha$.
The CVE vanishes to zero for $\alpha>\alpha_\mathrm{c}$ (success), converges to a finite value for $\alpha<\alpha_\mathrm{c}$ (failure), and exhibits a power-law decay at $\alpha=\alpha_\mathrm{c}$.
This result indicates that it is possible to judge the success or failure of compressed sensing using only a given dataset.
If this diagnosis concludes success, then one can safely cease further measurement to reduce the measurement time and cost.

\subsubsection{Dictionary learning}
\label{sec:dictionary_learning}
The performance of compressed sensing strongly depends on the sparsity of the vector $\bx$ to be reconstructed.
Therefore, it is crucially important to choose a proper basis (representation) such that $\bx$ is expected to be sparse.
However, there are situations where sparse representations are not known in advance.
This fact motivates us to consider another problem of finding a sparse \emph{representation} as well as a sparse solution for a given representation.

The above problem is formulated as follows~\cite{Chen98,Elad06}.
Let us suppose that in the equation $\by=A\bx$, only the vector $\by$ is known and that a plural number of datasets, $\{ \by_i \}$, are given.
We want to find a representation of $A$ that makes $\bx$ sparse for all $\{ \by_i \}$.
This statement is expressed as
\begin{equation}
\min_{\bx_i, A}\left\{ \frac{1}{2}\left\|\by_i - A\bx_i \right\|_2^2 + \lambda \left\| \bx_i \right\|_0 \right\}
\quad
\text{subject to}
\quad
\left\| A \right\|_2 = 1.
\label{eq:dictionary_learning}
\end{equation}
We stress that, unlike in the optimization problem considered so far [see, e.g., Eq.~(\ref{eq:argmin_Fx})], the matrix $A$ is not given in this equation but is determined so that $\bx_i$ becomes as sparse as possible.
The constraint is to remove arbitrariness from the solution for $A$.
In this context, the matrix $A$ is called a dictionary and the problem defined in Eq.~(\ref{eq:dictionary_learning}) is called \emph{dictionary learning} or \emph{sparse coding}, which is a kind of machine learning.
The standard method for approximately solving this optimization problem is K-SVD (SVD stands for singular value decomposition)~\cite{Elad06,Aharon06,Mairal08}.

In Eq.~(\ref{eq:dictionary_learning}), we may replace the $L_0$ norm with the $L_1$ norm to avoid computational difficulty, as in the derivation of the LASSO in Eq.~(\ref{eq:lasso}).
For this LASSO-type dictionary learning, Mairal \textit{et al.} proposed an efficient algorithm for online learning that allows iterative updating of the dictionary as new data of $\{ \by_i \}$ become available incrementally~\cite{Mairal10}.

\subsubsection{Low-rank matrix completion}
So far, we have considered the target quantity $\bx$ to be a vector and focused on its sparsity.
There is a related problem that takes advantage of the sparsity of a \emph{matrix} instead of a vector.
Let $i$ and $j$ represent indices of customers and products, respectively, and $X_{ij}$ be the correlation between $i$ and $j$ ($X_{ij}$ is large if customer $i$ has checked or bought product $j$).
Given partial data of the matrix $X$, we want to complete $X$ to predict the extent of interest that customer $i'$ has in product $j'$.
This problem is called \emph{matrix completion}.

For this problem, the sparsity of $X$ has been shown to play a relevant role.
Cand\'es and Recht demonstrated that, if $X$ is low-rank, then $X$ can be reconstructed perfectly from incomplete data of $X$ (denoted by $A$) by solving the optimization problem~\cite{Candes09}
\begin{equation}
\min_{X} {\rm rank}(X)
\quad
\text{subject to}
\quad
X_{ij} = A_{ij} 
\quad
\text{for}
\
\forall i,j \in \mathcal{E}, \label{eq:matrix_completion}
\end{equation}
where $\mathcal{E}$ is the set of sampling components in $A$
and $\mathrm{rank}(X)$ is defined as the number of non-zero singular values $s_l$ of the matrix $X$.
The rank of a matrix corresponds to the $L_0$ norm of a vector.
Therefore, the computational complexity of solving the optimization problem (\ref{eq:matrix_completion}) is NP-hard.
In analogy with the LASSO, it is natural to replace $\mathrm{rank}(X)$ with the matrix counterpart of the $L_1$-norm, that is, the \emph{nuclear norm} $\left\|X\right\|_{\ast}$ defined by 
\begin{equation}
\left\|X\right\|_{\ast} \equiv \sum_l s_l,
\end{equation}
where $s_l$ is the singular value of $X$.
An alternative optimization problem can thus be written as
\begin{equation}
\min_{X} \left\|X\right\|_{\ast}
\quad
\text{subject to}
\quad
X_{ij} = A_{ij} 
\quad
\text{for}
\
\forall i,j \in \mathcal{E}, \label{eq:matrix_completion2}
\end{equation}
which is in the class of convex relaxation.
A practical method for solving the convex relaxation problem is singular value thresholding algorithm~\cite{Cai10}.
The influence of noise on matrix completion was discussed in Ref.~\citen{Candes10}.

\section{Sparsity of Many-Body Green's Functions}
\label{sec:green_function}
This and the following sections discuss quantum many-body physics.
In connection with the previous sections, we begin with the equation $\by=A\bx$ and see where it appears in quantum many-body problems.
From this consideration, we will find sparsity hidden in many-body Green's functions.

\subsection{$\by=A\bx$ in quantum many-body problems}
The quantity we want to know, $\bx$, is the spectral function $\rho(\omega)$,
such as the single-particle excitation spectrum measured in angle-resolved photoemission spectroscopy experiments or the magnetic excitation spectrum measured in inelastic neutron scattering experiments.
One of the main tasks of theoretical investigations is to compute $\rho(\omega)$ starting from a microscopic Hamiltonian.
However, because it is difficult to treat $\rho(\omega)$ directly, we introduce imaginary time $it \equiv \tau$ and consider correlation functions $G(\tau)$ in the imaginary-time domain~\cite{AGD}.
This is called an imaginary-time or Matsubara Green's function.
We assign $G(\tau)$ to $\by$.
Imaginary-time descriptions enable sophisticated treatments of interactions, such as perturbative expansions using the Feynman diagram and quantum Monte Carlo (QMC) simulations.

After $G(\tau)$ is evaluated with the aid of some analytical or numerical methods, the imaginary time $\tau$ should be transformed back to real time to derive $\rho(\omega)$.
This procedure is the analytical continuation.
In practical calculations, one may use the exact relation between $G(\tau)$ and $\rho(\omega)$, expressed as
\begin{align}
\bG = \bK \brho,
\label{eq:G_K_rho}
\end{align}
where $\bG$ and $\brho$ denote vector representations of $G(\tau)$ and $\rho(\omega)$, respectively.
An equation of the form $\by=A\bx$ thus appears.
The problem of analytical continuation can be interpreted as the inverse problem of evaluating $\brho$ for a given $\bG$.

\subsection{Exact relations}\label{sec:exact_relations}

In this subsection, we complement Eq.~(\ref{eq:G_K_rho}) with a rigorous derivation and some remarks.
$G(\tau)$ and $\rho(\omega)$ are related to each other through the Lehmann representation, or spectral representation, of the form\cite{AGD}
\begin{align}
G^{\alpha}(\tau) = \int_{-\infty}^{\infty} d\omega K^{\alpha}(\tau, \omega) \rho^{\alpha}(\omega),
\label{eq:Lehmann}
\end{align}
where $\alpha$ specifies either the fermionic statistics ($\alpha=\mathrm{F}$) or the bosonic statistics ($\alpha=\mathrm{B}$).
The variable $\tau$ ranges from 0 to $\beta$.
The kernel function $K^{\alpha}(\tau,\omega)$ is given by
\begin{align}
K^{\alpha}(\tau, \omega)=
\begin{cases}
\displaystyle \frac{e^{-\tau\omega}}{1 + e^{-\beta\omega}} & (\alpha=\mathrm{F})\\
\displaystyle \frac{\omega e^{-\tau\omega}}{1 - e^{-\beta\omega}} & (\alpha=\mathrm{B})
\end{cases}.
\label{eq:kernel}
\end{align}
The spectral function $\rho^{\alpha}(\omega)$ is related to the retarded Green's function $G_\mathrm{R}(\omega)$ by
\begin{align}
\rho^{\alpha}(\omega) =
\begin{cases}
\displaystyle \frac{1}{\pi} \mathrm{Im} G_\mathrm{R}(\omega) & (\alpha=\mathrm{F})\\
\displaystyle \frac{1}{\pi \omega} \mathrm{Im} G_\mathrm{R}(\omega) & (\alpha=\mathrm{B})
\end{cases}.
\end{align}
In the bosonic case, the kernel is defined with an extra $\omega$ to cancel the divergence of the Bose distribution function, $1/(1-e^{-\beta\omega}) \sim 1/\beta\omega$, around $\omega=0$.\cite{MaxEnt}
Hereafter, we will omit the index $\alpha$ for simplicity when a particular distinction is unnecessary.

Now, we transform the integral equation (\ref{eq:Lehmann}) into a matrix-vector representation.
To this end, we first introduce a cutoff $\wmax$ for the infinite integral over $\omega$.
The variables $\omega$ and $\tau$ are then discretized into $N$ and $M$ slices, respectively, with equal intervals.
Defining
$G_i \equiv G(\tau_i)$ and
$\rho_j \equiv \rho(\omega_j) \Delta \omega$,
we obtain Eq.~(\ref{eq:G_K_rho}).
Here, $K$ is an $(M\times N)$ matrix defined by $K_{ij} = K(\tau_i, \omega_j)$.
We note that the discretization error can be reduced by increasing $N$ and $M$ toward the continuous limit, $N, M \to \infty$.
In contrast, we cannot take the limit $\wmax\to\infty$, meaning that the cutoff $\wmax$ is essential.
We will see that $\beta \wmax \equiv \Lambda$ is the relevant parameter that controls the bases presented below.
A detailed description is given in Sect.~\ref{sec:compressed_sampling}.

\subsection{``Intermediate representation'' (IR) of Green's functions}
Let us consider the inverse problem in Eq.~(\ref{eq:G_K_rho}) from the sparse-modeling point of view.
The central question in this respect is which representation (basis set) makes $\brho$ sparse because sparsity relies on representation, as emphasized in Sect.~\ref{subsec:really_sparse}.
We address this question by looking into the nature of the kernel $K$.

\begin{figure}
	\centering
	\includegraphics[width=0.8\linewidth]{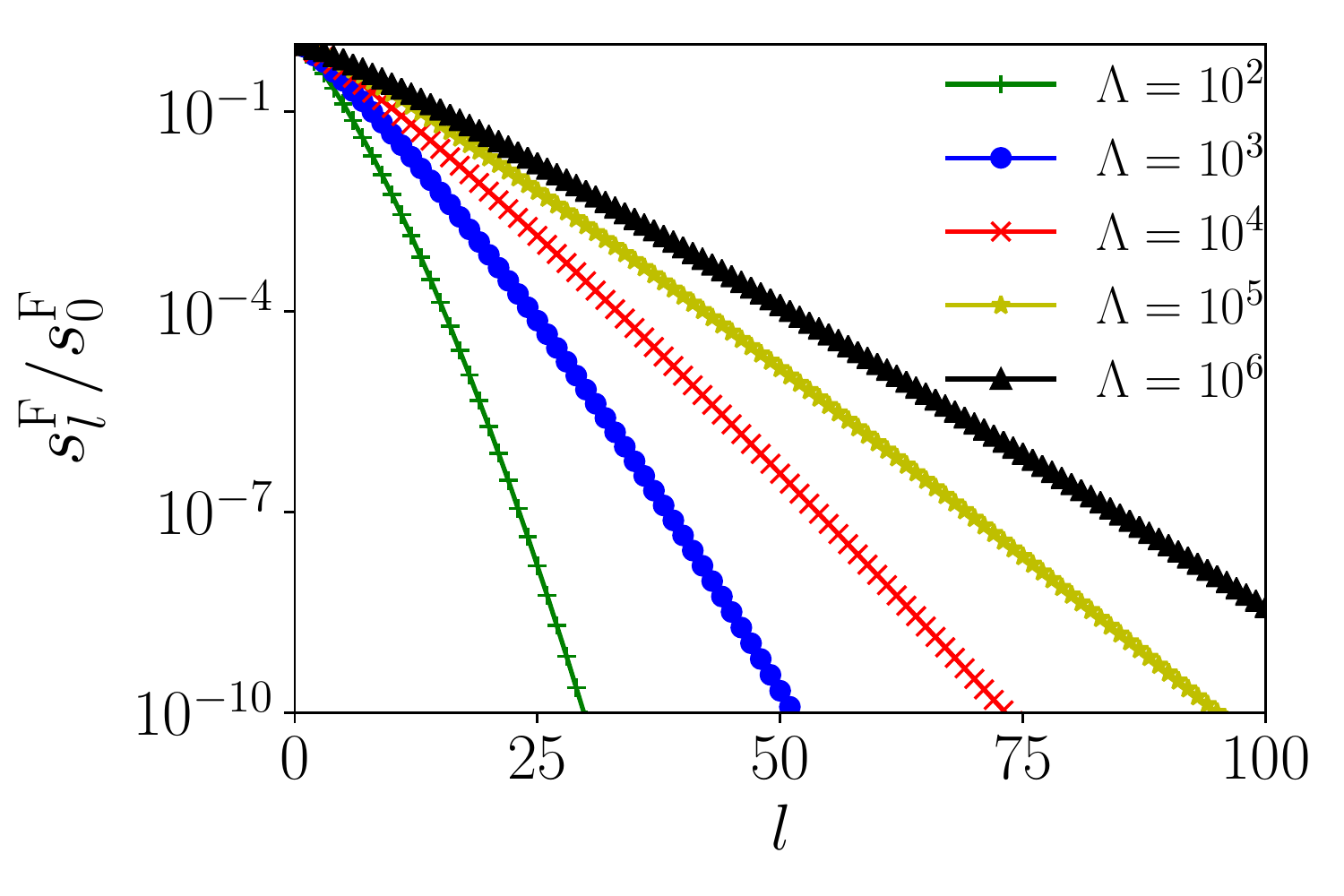}
	\includegraphics[width=0.8\linewidth]{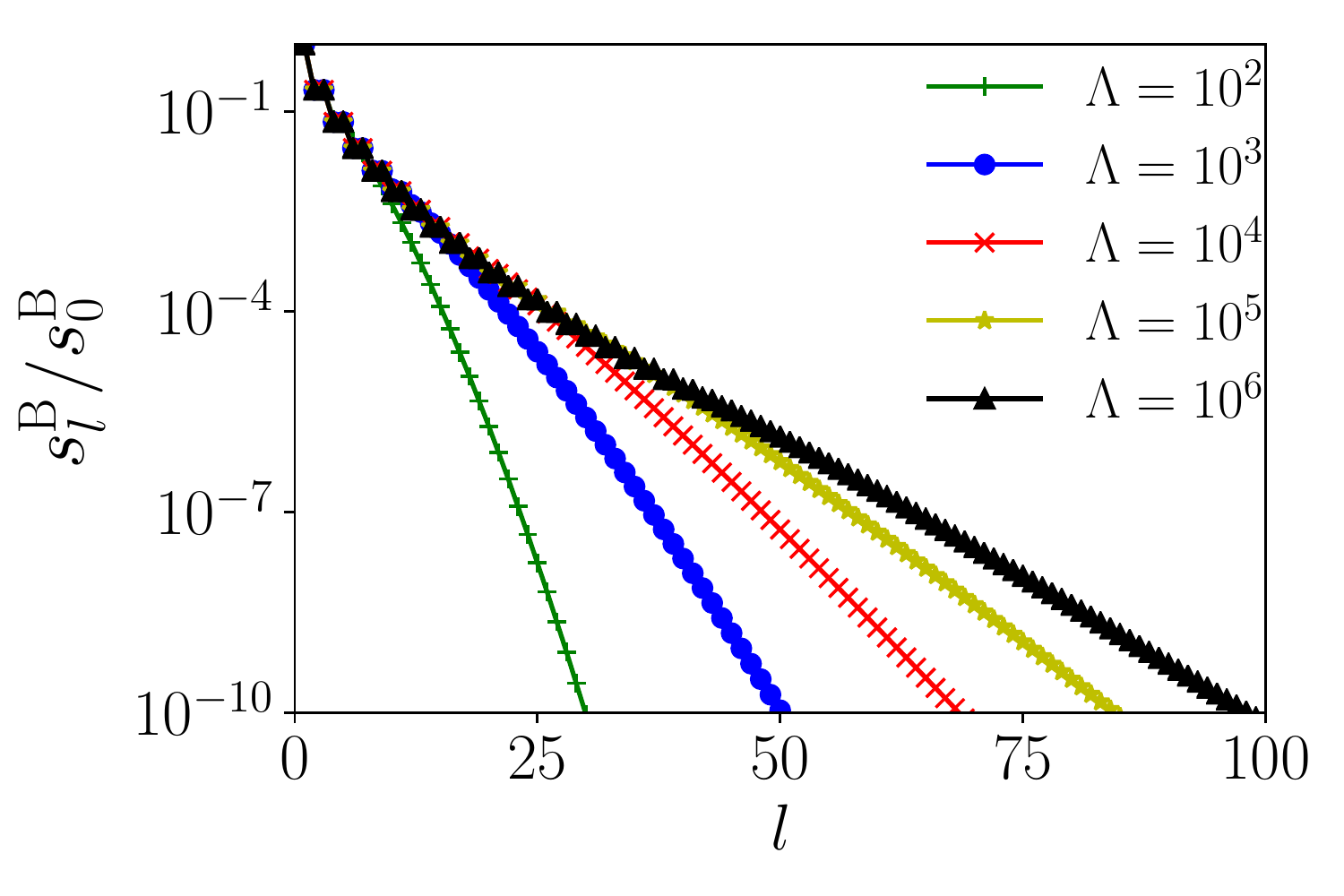}
	\caption{(Color online) Singular values $s_l$ of the kernel matrix for various values of $\Lambda= \beta\wmax$.
	The upper and lower panels are for fermions and bosons, respectively.
	Reprinted with permission from Ref.~\citen{Chikano18PRB} \copyright 2018 the American Physical Society.
	}
	\label{fig:sv}
\end{figure}

Using singular value decomposition (SVD), the real non-square matrix $K$ can be decomposed as
\begin{align}
\bK = \bU \bS \bV^\mathrm{T},
\label{eq:SVD}
\end{align}
where the superscript $\mathrm{T}$ indicates the transpose of a matrix.
$U$ and $V$ are $(M\times M)$ and $(N\times N)$ orthogonal matrices, respectively.
$S$ is an $(M\times N)$ matrix that contains singular values $s_l$ [$l=0, 1, 2, \cdots$, $\min(N,M)-1$] at the diagonal.
$s_l$ contains non-negative real numbers aligned in descending order.
An important observation is that $s_l$ decreases exponentially, or ever faster, as shown in Fig.~\ref{fig:sv}.

\begin{figure}
\centering
\includegraphics[width=\linewidth]{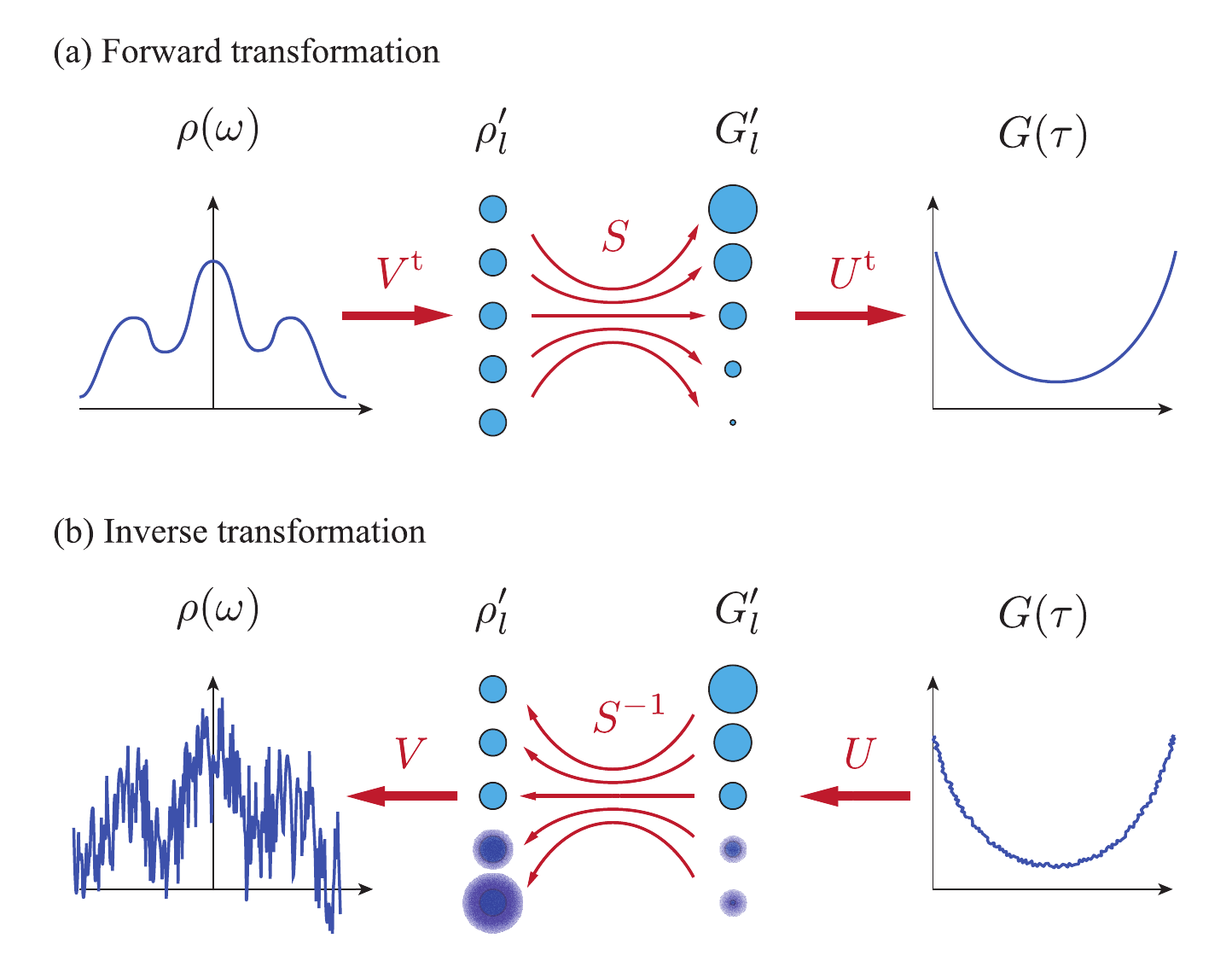}
\caption{(Color online) (a) Schematic diagrams of (a) forward transformation $\rho(\omega) \to G(\tau)$ and (b) inverse transformation $G(\tau) \to \rho(\omega)$. The transformation matrix $K$ is decomposed into three matrices, namely $V$, $S$, and $U$ as in Eq.~(\ref{eq:SVD}). Noise contained in $G(\tau)$ is amplified in the inverse transformation.}
\label{fig:schematic}
\end{figure}

The meaning of the singular values can be explained as follows.
We first transform the vectors $\bG$ and $\brho$ using the orthogonal matrices $U$ and $V$ as
\begin{align}
\bG^\prime \equiv \bU^\mathrm{T} \bG,
\quad
\brho^\prime \equiv \bV^\mathrm{T}\brho.
\end{align}
Substituting Eq.~(\ref{eq:SVD}) into Eq.~(\ref{eq:G_K_rho}), we obtain
\begin{align}
\label{eq:G_S_rho}
\bG^\prime = \bS \brho^\prime.
\end{align}
Because $S$ is diagonal, this equation is reduced to an element-wise expression as
\begin{align}
G'_l = s_l \rho'_l.
\label{eq:G_S_rho-element}
\end{align}
This equation explains the role of singular values.
The transformation from $\brho$ to $\bG$ consists of three procedures: basis transformation using matrix $V$, weighting by $s_l$, and transformation to the original basis using matrix $U$.
The exponential decay of $s_l$ indicates that $\bG$ contains all pieces of information of $\brho$ in extremely different weights.
This situation is schematically shown in Fig.~\ref{fig:schematic}(a).

Mathematically, the difficulty of treating a matrix on computers is quantified by the condition number $C$ defined by $C\equiv s_\mathrm{max}/s_\mathrm{min}$, where $s_\mathrm{max}$ and $s_\mathrm{min}$ denote the maximum and minimum singular values, respectively.
Unitary matrices yield the smallest value, $C=1$, which corresponds to the most well-conditioned case.
On the other hand, when $C\gg1$, the matrix is said to be \emph{ill-conditioned}.
As shown in Fig.~\ref{fig:sv}, $s_l$ of the kernel $K(\tau,\omega)$ exhibits an exponential dependence, and hence the condition number $C$ exceeds the machine precision, namely $C=\infty$ in practice.
Thus, $K$ is in the class of extremely ill-conditioned matrices and Eq.~(\ref{eq:G_K_rho}) is an ill-conditioned inverse problem.

\begin{figure}
	\centering
	\includegraphics[width=.6\linewidth]{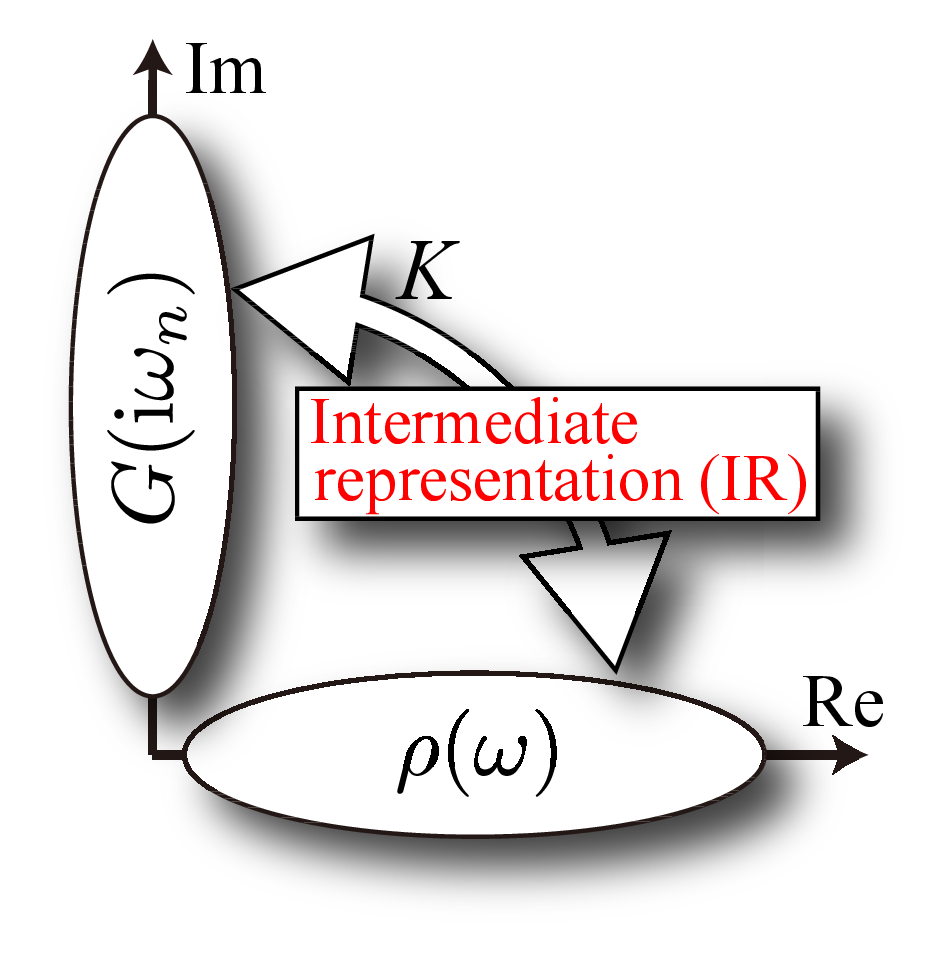}
	\caption{(Color online) Schematic diagram of IR defined between real- and imaginary-frequency domains.
	Reprinted with permission from Ref.~\citen{Shinaoka17} \copyright 2017 the American Physical Society.}
	\label{fig:IR}
\end{figure}

Recall that the kernel $K$ depends only on $\beta$.
This means that the two unitary matrices, $U$ and $V$, constitutes \emph{model-independent} basis sets.
Because these bases connect real- and imaginary-frequency representations [Eq.~(\ref{eq:G_S_rho-element})], they are called intermediate representations (IRs)\cite{Shinaoka17}.
Figure~\ref{fig:IR} shows a schematic diagram of an IR.
We will show below a striking conclusion that results from the ill-conditioned nature of $K$.

\subsection{Example}

\begin{figure*}[t]
    \centering
    \includegraphics[width=0.9\linewidth]{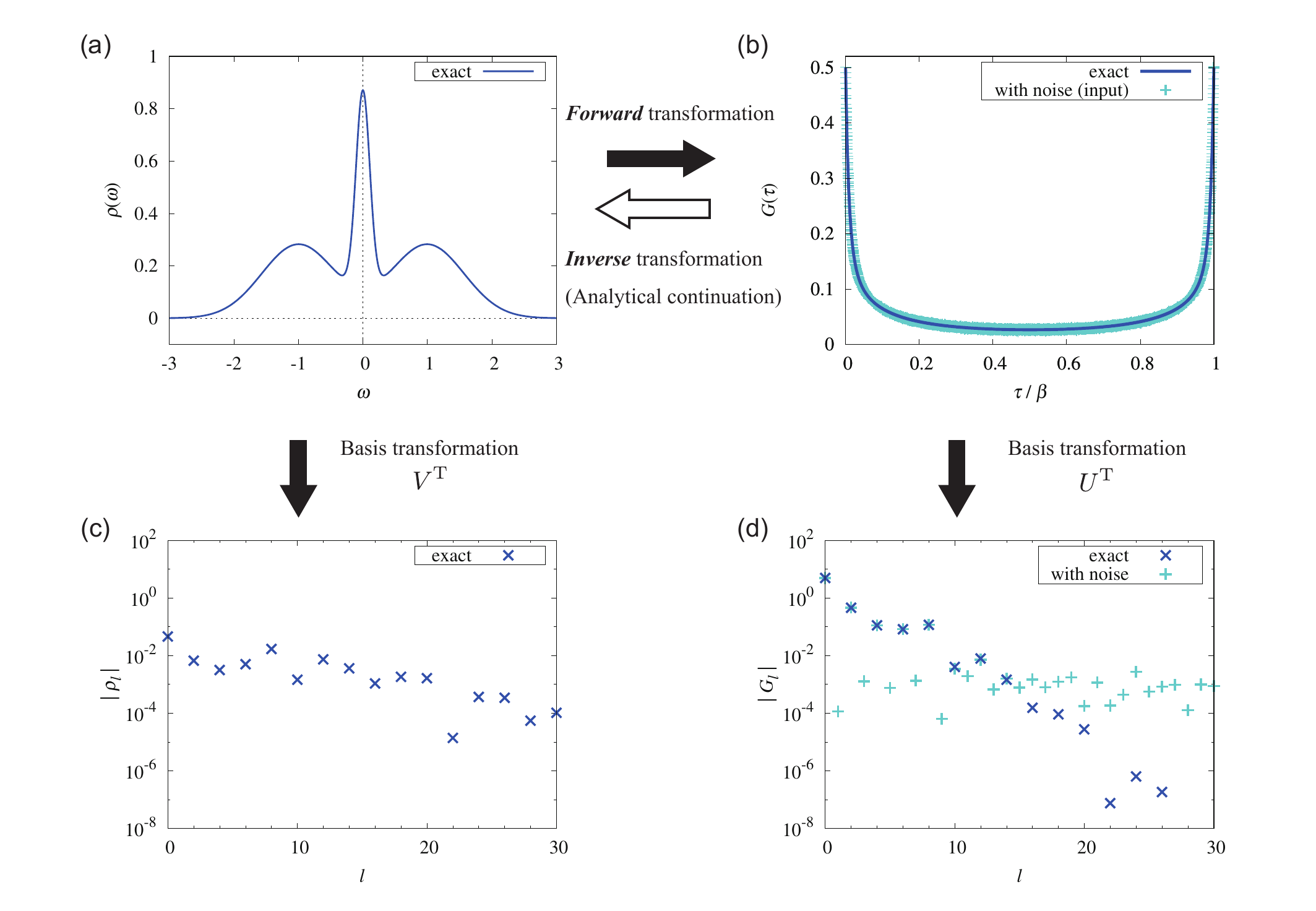}
    \caption{(Color online) (a) Test spectrum $\rho(\omega)$. (b) Imaginary-time Green's function $G(\tau)$ computed using Eq.~(\ref{eq:Lehmann}). Expansion coefficients (c) $\rho'_l$ and (d) $G'_l$. The $+$ points (light blue) in (b) and (c) are data with added Gauss noise.
    Data are taken from Ref.~\citen{Otsuki17}.
    }
    \label{fig:transform}
\end{figure*}

We consider the model spectrum $\rho(\omega)$ shown in Fig.~\ref{fig:transform}(a).
It is transformed into $G(\tau)$, as plotted in Fig.~\ref{fig:transform}(b), by performing numerical integration in Eq.~(\ref{eq:Lehmann}).
The inverse transformation from $G(\tau)$ to $\rho(\omega)$ is the main subject (analytical continuation) in the next section.
Here, we focus on the basis transformation of each quantity.

Figures~\ref{fig:transform}(c) and \ref{fig:transform}(d) [blue crosses] show the expansion coefficients $\rho_l$ and $G_l$, respectively.
Here, only the even-number components of $\rho_l$ and $G_l$ are shown because the odd-number components vanish due to the symmetric condition, $\rho(\omega)=\rho(-\omega)$.
Both quantities exhibit exponential decay as $l$ increases.
In particular, we stress that $G_l$ decays much faster than $\rho_l$ as a result of the exponential behavior of $s_l$ [see Eq.~(\ref{eq:G_S_rho-element}) and Fig.~\ref{fig:sv}].
It should be emphasized that the fast decay of $G_l$ does not depend on a particular model but is an intrinsic feature of $G_l$ because it purely relies on the nature of the kernel $K(\tau,\omega)$.

\subsection{Consequences of the exponential decay of $G'_l$}
What consequences are implied by the exponential decay of $G_l$?
To this end, we consider a situation where the input $G(\tau)$ has errors, such as the statistical errors in QMC simulations.
We simulate this situation by adding Gauss noise with width $\sigma=10^{-3}$ onto the exact data in Fig.~\ref{fig:transform}(b).
The data with noise and the exact data are transformed into $G'_l$ in Fig.~\ref{fig:transform}(d).
The influence of errors is apparent in this expression.
Because of the exponential decay of the exact data, the impact of noise increases as $l$ increases, and finally dominates the exact value for $l \gtrsim 14 \equiv l_0$.
The high-order components contain only noise.

This implies two contrasting consequences depending on what $G(\tau)$ is used for.
Let us first suppose that one wants to know $\rho(\omega)$.
Then, the exponential decay of $G'_l$ indicates that \emph{the numerically computed $G(\tau)$ does not contain sufficient information of $\rho(\omega)$}.
In other words, the problem of analytical continuation is essentially an underdetermined problem, where the information required for solving the equation is lacking.
Furthermore, in the next section we will show that the noise in the high-order components makes the inverse transformation unstable.
Sparsity is thus a good precondition in the problem of analytical continuation.

Next, we suppose that one is interested in $G(\tau)$ itself rather than $\rho(\omega)$.
In this case, the exponential decay of $G'_l$ leads to a positive consequence, allowing one to represent $G(\tau)$ using only a few bases without loss of meaningful information.
For the data in Fig.~\ref{fig:transform}, the 4000 points of $G(\tau)$ can be represented by only 7-8 components exactly within numerical accuracy.
Using this representation, we can perform efficient computations of many-body theories such as QMC simulations.
The details are discussed in Sect.~\ref{sec:compressed_sampling}.

\section{SpM Analytical Continuation}
\label{sec:analytical_continuation}
In this section, we discuss how analytical continuation is performed using sparse modeling.
The purpose here is to compute $\rho(\omega)$ for a give $G(\tau)$, which in general contains noise.
This procedure is formulated as an inversion problem of the linear equation in Eq.~(\ref{eq:G_K_rho}).
It involves considerable difficulty as discussed in Sect.~\ref{sec:green_function}.
The matrix $K$ is ill-conditioned and therefore the equation is an inevitably underdetermined system, in which the input $\bG$ has a little meaningful information.
Furthermore, the inverse of the ill-conditioned matrix amplifies noise exponentially as shown below.

\subsection{Historical review}
Let us first review the problem of analytical continuation with a brief overview of related approaches.
Matsubara Green's function $G(i\omega_n)$ is defined as the Fourier transform of the imaginary-time Green's function $G(\tau)$ introduced in Sect.~\ref{sec:green_function},
$G(i\omega_n) = \int_0^{\beta} d\tau G(\tau) e^{i\omega_n \tau}$,
where $\omega_n$ is the Matsubara frequency defined by
$\omega_n = (2n+1)\pi\beta$ for fermionic systems and $\omega_n = 2n\pi\beta$ for bosonic systems, where $n$ is an integer.
When an analytical expression for $G(i\omega_n)$ is known, the spectrum $\rho(\omega)$ is readily obtained by replacing
$i\omega_n$ with $\omega+i0$, namely, $\rho(\omega)=\mathrm{Im}G(\omega+i0)/\pi$~\cite{AGD}.

\begin{figure}
    \centering
    \includegraphics[width=\linewidth]{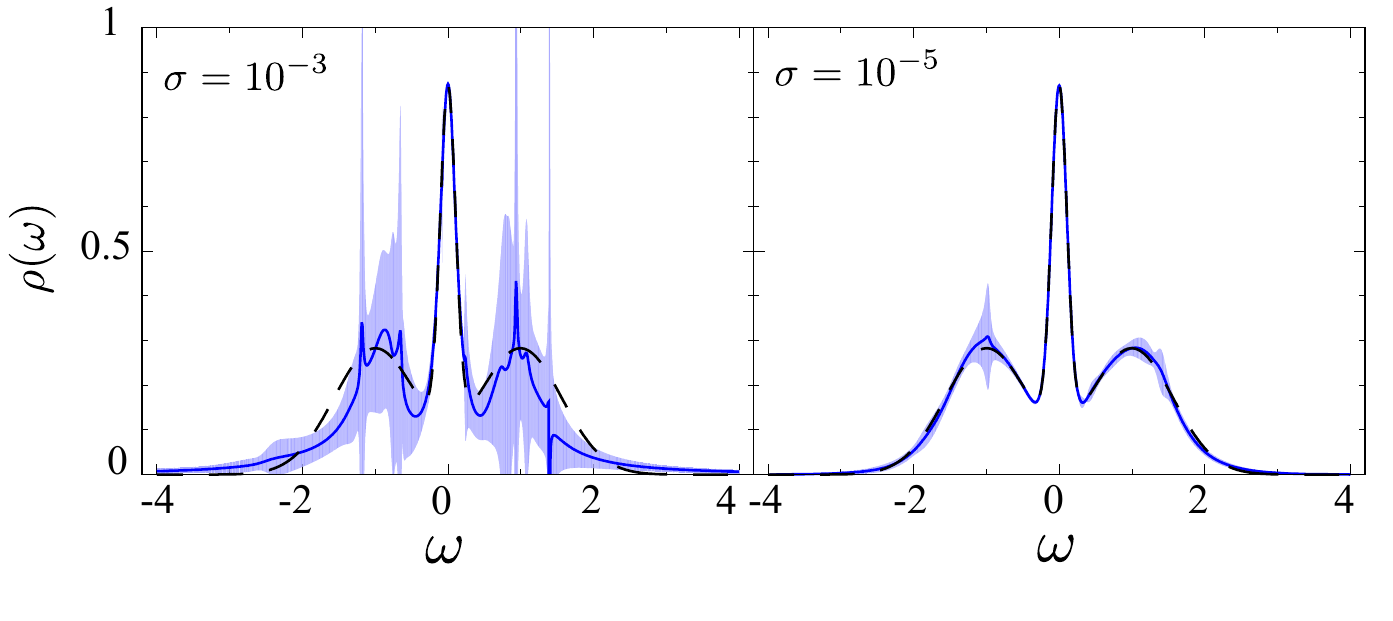}
    \caption{(Color online) Spectrum $\rho(\omega)$ computed using Pad\'e approximation. The black dashed line shows the original model spectrum. The solid blue line and the shaded region respectively show the mean value and the standard deviations of the Pad\'e results evaluated from 30 datasets of $G(i\omega_n)$. The width of the Gaussian noise on $G(i\omega_n)$ is (a) $\sigma=10^{-3}$ and (b) $\sigma=10^{-5}$.
    Reprinted with permission from Ref.~\citen{Yoshimi19} \copyright 2019 Elsevier.}
    \label{fig:results_pade}
\end{figure}

If only numerical values of $G(i\omega_n)$ are given, then the analytical continuation, $i\omega_n \to z$, needs to be performed numerically.
In other words, we infer an analytical expression $\tilde{G}(z)$ in the whole complex frequency plane from the values $G(i\omega_n)$ given at the discrete imaginary frequency points.
Pad\'e approximation~\cite{Vidberg77} uses the rational function
$\tilde{G}(z)=(a_0 + a_1 z + \cdots)/(b_0 + b_1 z + \cdots)$ to fit $G(i\omega_n)$, and then extrapolates it into arbitrary complex frequency $z$.
This method has been widely applied in combinations with, for example, diagrammatic perturbation theories.
However, it is known that Pad\'e approximation is quite sensitive to non-systematic errors in $G(i\omega_n)$ as shown in Fig. \ref{fig:results_pade}.
In particular, there is no guarantee that $\rho(\omega)$ satisfies the fundamental properties such as nonnegativity and the sum rule.
These disadvantages exclude the use of Pad\'e approximation for the analytical continuation of QMC data.

Another method for computing $\rho(\omega)$ is to solve the inverse problem of the linear equation, $\bG=K\brho$, in Eq.~(\ref{eq:G_K_rho}).
We can thus use sophisticated techniques in the field of statistics or data science to address the problem of noise sensitivity.
Various approaches have been developed, including MEM~\cite{MaxEnt,Gunnarsson10b,Bergeron16,Levy17,Sim-arXiv}, which is the \textit{de facto} standard, the stochastic method~\cite{Sandvik98,Mishchenko00,Fuchs10,Beach-arXiv,Sandvik16,Bao16,Krivenko-arXiv},
machine learning approaches~\cite{Arsenault17,Yoon-arXiv},
and others~\cite{Beach00,Oestlin12,Dirks13,Schoett16,Bertaina16,Xuping-arXiv}.

The remaining critical question is how much can we reconstruct $\rho(\omega)$ from noisy $G(i\omega_n)$ in principle.
Recent work by Gaulko \textit{et al.} tackled this problem using a trial-and-error approach~\cite{Goulko17}.
Another approach was proposed based on the sparse modeling techniques, which reveals how much information $G(i\omega_n)$ possesses regarding $\rho(\omega)$\cite{Otsuki17}.

\subsection{Explosion of errors}

Equations~(\ref{eq:G_S_rho}) and (\ref{eq:G_S_rho-element}) show that $\rho(\omega)$ and $G(\tau)$ are directly connected to each other in the IR bases.
Therefore, one could naively evaluate $\rho(\omega)$ by converting the input $G(\tau)$ into $G'_l$ and using the relation
\begin{align}
\rho'_l = G'_l/s_l.
\label{eq:rho_l}
\end{align}
This, however, does not work.
Even tiny noise in the input $G(\tau)$ will cause serious influence on the results for $\rho(\omega)$.

To understand the influence of noise on analytical continuation, let us consider the data in Fig.~\ref{fig:transform}(d) explicitly.
Because the exact value of $G'_l$ decays exponentially ($\times$ symbols), large-$l$ components basically contain only noise, namely $|G'_l| \sim \sigma=10^{-3}$ ($+$ symbols) for $l \simeq l_0 =14$.
Equation~(\ref{eq:rho_l}) then yields $|\rho'_l| \sim \sigma/s_l$. 
This result shows that noise is amplified exponentially at large $l$.
It also indicates that the influence of noise can be removed by truncating the high-order components in the IR basis.

In the treatment of large-scale data, it is common to truncate bases according to the singular values, a process called \emph{dimensionality reduction}.
Here, it is important to emphasize the difference between ordinary dimensionality reduction and the problem of analytical continuation.
Normally, as the number of retained bases increases, the accuracy of the reduced representation improves.
A typical example in quantum many-body physics is the density matrix renormalization group method, in which the density matrix is approximated with this technique.
In the present case of analytical continuation, using more bases results in a worse spectrum because the noise is enlarged by small $s_l$.
Therefore, we need to \emph{select} appropriate bases rather than perform truncation.

\subsection{Selection of bases}

We now select bases in the IR representation to remove the influence of noise. 
In other words, we need to find a sparse solution in this representation and
thus, sparse modeling in Sect.~\ref{sec:spm} applies to the present problem.

We first define the squared error $\chi^2$ of the target equation [Eq.~(\ref{eq:G_K_rho})] as
\begin{align}
\chi^2 (\brho | \bG) = \frac12 \| \bG - K \brho \|_2^2.
\label{eq:chi-1}
\end{align}
Here, the argument on the left-hand side ($\brho$) indicates a quantity to be varied, and that on the right-hand side of the bar ($\bG$) indicates a fixed quantity.
We can represent $\chi^2$ using the IR as
\begin{align}
\chi^2 (\brho' | \bG') = \frac12 \| \bG' - S \brho' \|_2^2.
\label{eq:chi-2}
\end{align}
The expressions in Eqs.~(\ref{eq:chi-1}) and (\ref{eq:chi-2}) are mathematically equivalent, but may give different values in numerical calculations because of the ill-conditioned nature of $K$.

To enforce sparseness on a solution in the IR domain, we introduce the $L_1$ regularization term and consider a LASSO-type minimization problem of the form\cite{Otsuki17,Yoshimi19}
\begin{align}
\label{eq:F}
F(\brho'|\bG',\lambda) = \chi^2 (\brho' | \bG') + \lambda \| \bm{\rho}' \|_1.
\end{align}
The second term is the $L_1$ regularization, which makes the solution sparse to remove irrelevant bases.
The parameter $\lambda$ controls the extent to which sparseness is enforced.
We can determine an optimal value automatically, as discussed later.

In practical situations, the input $G(\tau_i)$ computed in QMC calculations may be associated with statistical errors (error bars) $\Delta G_i$.
More generally, the statistical errors are expressed by the covariance matrix $C$, whose diagonal part corresponds to $C_{ii}=(\Delta G_i)^2$.
In this case, we could extend the squared error in Eq.~(\ref{eq:chi-1}) in a form that takes $C$ into account\cite{MaxEnt}:
\begin{align}
\chi^2 (\brho|\bG,C) = \frac12 \sum_{ij} (\bG - K \brho )_i^\mathrm{T} (C^{-1})_{ij} (\bG - K \brho )_j.
\label{eq:chi-covariance}
\end{align}
The role of $C$ can be understood by considering that the diagonal components, $(\bG - K \brho )_i^2$, are weighted by $1/(\Delta G_i)^2$ in the summation and thus more accurate components have stronger influence on $\chi^2$.
It has been shown that for the MEM and the stochastic method, the inclusion of $\Delta G_i$ or $C$ improves the results of analytical continuation\cite{MaxEnt,Sandvik98,Shao17}.
Using the IR basis, Eq.~(\ref{eq:chi-covariance}) is rewritten as
\begin{align}
\chi^2 (\brho'|\bG',C) = \frac12 (\bm{G}' - S \bm{\rho}')^{\mathrm{T}} W (\bm{G}' - S \bm{\rho}'),
\end{align}
where the matrix $W$ is defined as $W=U^{\mathrm{T}}C^{-1}U$.
The function to be minimized is
\begin{align}
\label{eq:F2}
F(\bm{\rho}'|\bG',C,\lambda) = \chi^2 (\brho'|\bG',C)  + \lambda \| \bm{\rho}' \|_1.
\end{align}
This expression is reduced to Eq.~(\ref{eq:F}) by replacing $C$ with a unit matrix, i.e.,
$F(\bm{\rho}'|\bG',\bm{1},\lambda) = F(\bm{\rho}'|\bG',\lambda)$.

Our task now is to find $\brho'$ that minimizes $F(\brho')$ in Eq.~(\ref{eq:F}) or Eq.~(\ref{eq:F2}) subject to 
two constraints, namely non-negativity and the sum rule:
\begin{align}
&\rho(\omega) \geq 0,
\\
&\int_{-\infty}^{\infty} d\omega \rho(\omega)=c,
\end{align}
where $c=1$ for ordinary (spin- and orbital-) diagonal fermionic Green's function, and $c=0$ for off-diagonal components.
In general, one can determine the value of $c$ as
\begin{align}
c = \begin{cases}
\displaystyle \left. G^\mathrm{F}(\tau) \right|_{\tau=+0} + \left. G^\mathrm{F}(\tau)\right|_{\tau=\beta-0} & (\alpha=\mathrm{F}) \\
\displaystyle \int_0^{\beta} G^\mathrm{B}(\tau) d\tau & (\alpha=\mathrm{B})
\end{cases}.
\end{align}
The fermionic expression corresponds to the coefficient of the high-frequency asymptotics, $G^\mathrm{F}(i\omega_n) \sim -c/i\omega_n$, and the bosonic one corresponds to the static susceptibility, $G^\mathrm{B}(i\omega_n=0)$.
Even with these additional constraints, the minimization problem in Eq.~(\ref{eq:F2}) can be solved using the ADMM algorithm presented in Sect.~\ref{subsec:admm}.
For details, see Appendix\ref{app:ADMM}.

\subsection{Example}
\label{subsec:spm_example}
\begin{figure}
    \centering
    \includegraphics[width=\linewidth]{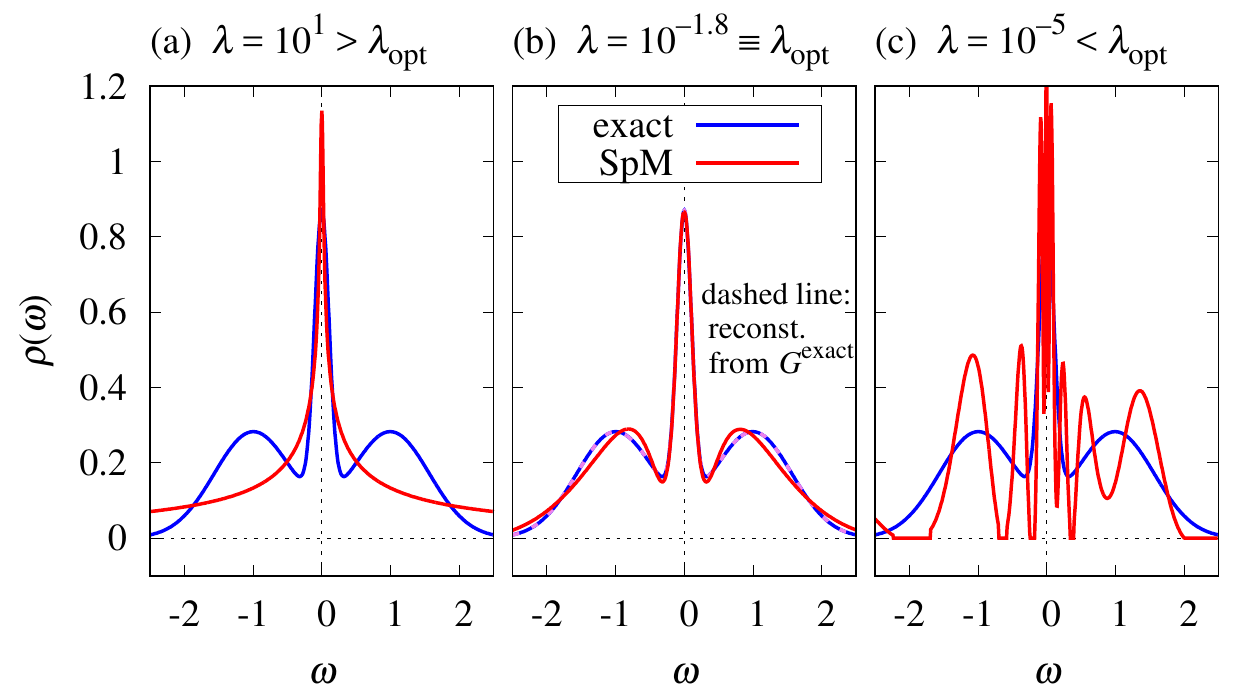}
    \caption{(Color online) Spectrum computed using analytical continuation $\rho(\omega)$ (red) and the exact spectrum (blue). The three subfigures correspond to cases where the regularization parameter $\lambda$ is (a) large, (b) optimal, and (c) small.
    Reprinted with permission from Ref.~\citen{Otsuki17} \copyright 2017 the American Physical Society.
    }
    \label{fig:results_rho}
\end{figure}

We now review the results in Ref.~\citen{Otsuki17} and how $L_1$-norm regularization is applied to analytical continuation.
Here, the covariance matrix is not taken into account, i.e., the optimization problem in Eq.~(\ref{eq:F}) is solved.
Figure~\ref{fig:results_rho} shows the spectrum computed for various values of $\lambda$.
A reasonable spectrum was obtained by taking a moderate value of $\lambda$ as shown in Fig.~\ref{fig:results_rho}(b).
Inappropriate choices of $\lambda$ result in a featureless or a spiky spectrum as shown in Figs.~\ref{fig:results_rho}(a) or \ref{fig:results_rho}(c), respectively.

\begin{figure}
    \centering
    \includegraphics[width=\linewidth]{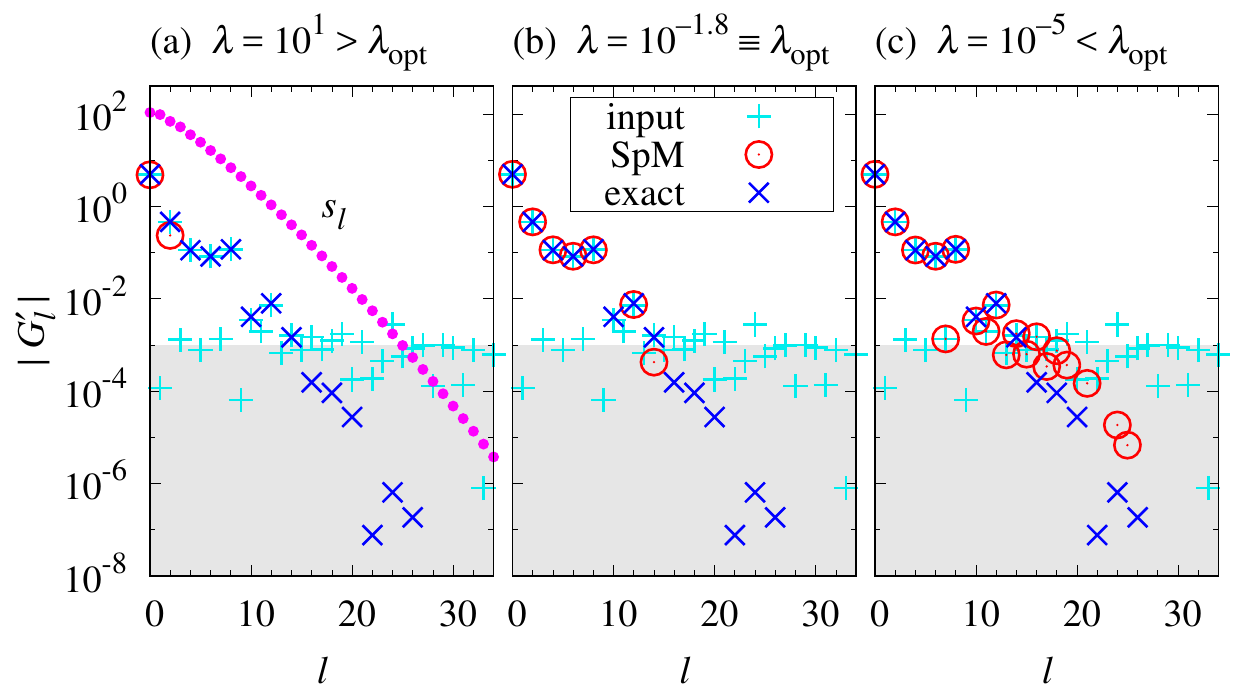}
    \caption{(Color online) $G'_l$ constructed after analytical continuation (red circles), input data (light blue $+$ symbols), and exact data (blue $\times$ symbols) for $\lambda$ values of (a) $10^1 > \lambda_{\rm opt}$, (b) $10^{-1.8} \equiv \lambda_{\rm opt}$, and (c) $10^{-5} < \lambda_{\rm opt}$ (see the caption of Fig.~\ref{fig:results_rho}).
    Reprinted with permission from Ref.~\citen{Otsuki17} \copyright 2017 the American Physical Society.
    }
    \label{fig:results_g}
\end{figure}

Let us consider how different spectra were derived depending on the value of $\lambda$.
Figure~\ref{fig:results_g} shows the IR representation, $G'_l=s_l \rho'_l$, computed after analytical continuation (red circles).
For comparison, the input data (light blue crosses) and the exact data (blue crosses) are also plotted [the same data as in Fig.~\ref{fig:transform}(d)].
The shaded region indicates $|G'_l|<\sigma=10^{-3}$, where the noise may dominate the exact value.
By comparing the SpM result and the input data, we can find which bases are retained by $L_1$-norm regularization (the deviation between them is due to the sum rule and non negativity constraints enforced in the SpM process).
Note that the exact data are \emph{not} used in deriving the SpM result, i.e. SpM does not ``know'' the exact data in choosing which bases should be retained.

When $\lambda$ is too large [Fig.~\ref{fig:results_g}(a)], SpM retains only two bases, which are incapable of reproducing the three-peak structure of the exact spectrum.
On the opposite side [Fig.~\ref{fig:results_g}(c)], the spectrum is constructed using many bases, including those that contain no relevant information.
Such redundant bases result in spiky and oscillatory spectra, which lack reproducibility.
When $\lambda$ is properly optimized, SpM selects relevant components that are not influenced by noise.

\begin{figure}
    \centering
    \includegraphics[width=0.8\linewidth]{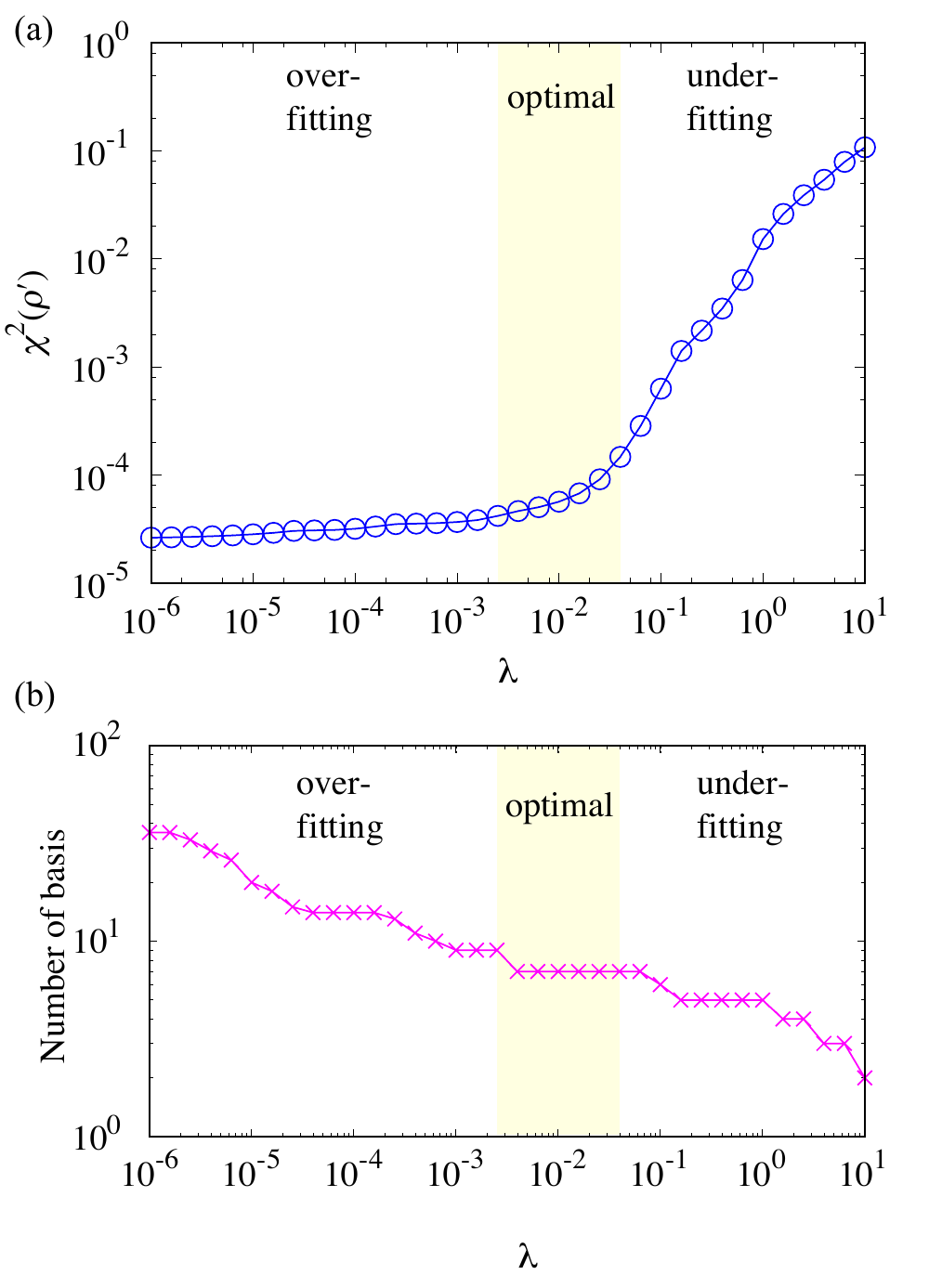}
    \caption{(Color online) $\lambda$-dependence of (a) the squared error $\chi^2(\brho'(\lambda) | \bG')$ and (b)  number of bases which are retained in $\brho'$. Data in (a) are taken from Ref.~\citen{Otsuki17}. }
    \label{fig:lambda}
\end{figure}

The optimal value of $\lambda$ was determined using the elbow method (Sect.~\ref{subsubsec:elbow}).
Figure~\ref{fig:lambda}(a) shows the $\lambda$-dependence of the squared error $\chi^2(\brho'(\lambda) | \bG')$.
The behavior is similar to that for the random matrix model in Fig.~\ref{fig:lasso_example_score}(a), where $\chi^2$ is insensitive to the variation of $\lambda$ in the small-$\lambda$ region (overfitting), and increases steeply in the large-$\lambda$ region (underfitting).
The reasonable spectrum in Fig.~\ref{fig:results_rho}(b) was obtained around the elbow at $\lambda=10^{-1.8} \equiv \lambda_\mathrm{opt}$.
Figure~\ref{fig:lambda}(b) shows the number of bases that are retained in the converged solution $\brho'$ (number of red circles in Fig.~\ref{fig:results_g}) exhibits a plateau around $\lambda=\lambda_\mathrm{opt}$,
indicating the stability of the solution against an order of magnitude change in $\lambda$.

\subsection{Robustness against noise}
\begin{figure}
    \centering
    \includegraphics[width=\linewidth]{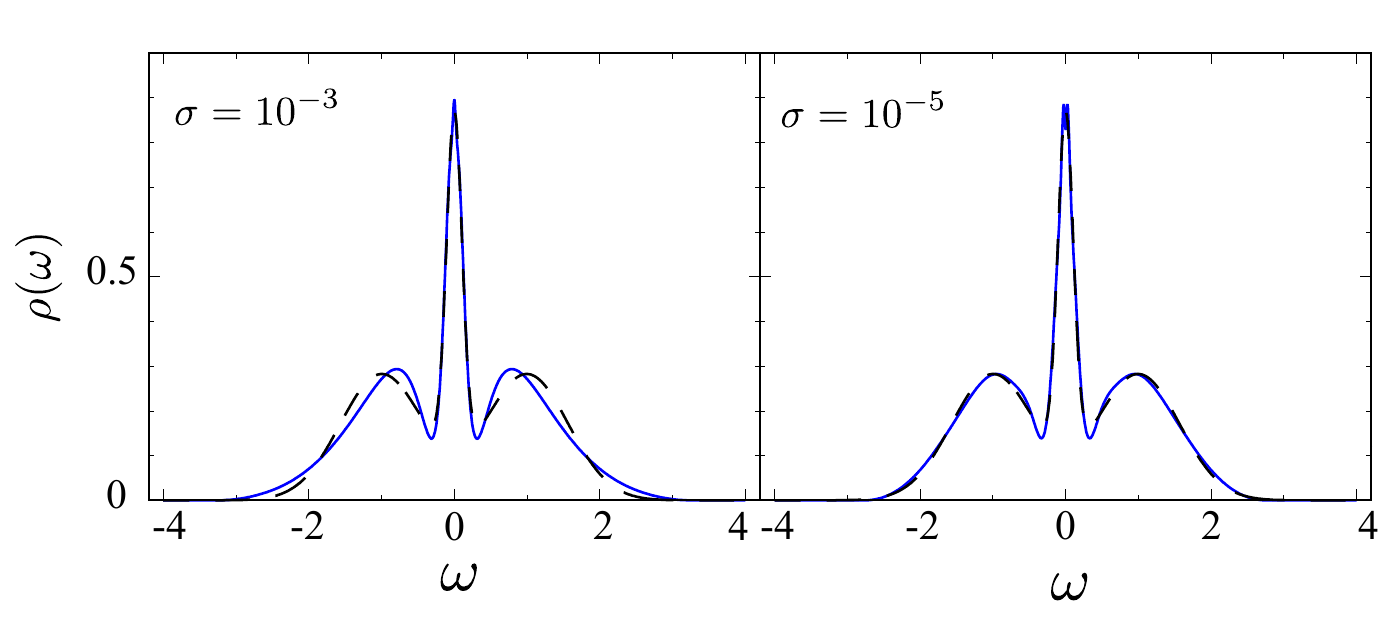}
    \caption{(Color online) Spectrum $\rho(\omega)$ computed using the SpM analytical continuation method.
    See the caption of Fig.~\ref{fig:results_pade} for an explanation of the plot.
    The standard deviations of the SpM results are so small that the shaded area is invisible (narrower than the line width).
    Reprinted with permission from Ref.~\citen{Yoshimi19} \copyright 2019 Elsevier.}
    \label{fig:results_spm}
\end{figure}

The results in Sect.~\ref{subsec:spm_example} demonstrate that the SpM method automatically removes irrelevant components that are strongly affected by noise [Fig.~\ref{fig:results_g}(b)].
Therefore, $\rho(\omega)$ evaluated using SpM is expected to be robust against noise.
This property was verified in Ref.~\citen{Yoshimi19}, where SpM analytical continuation was performed using 30 datasets of $G(\tau)$ (different noise configurations) and the mean value and the standard deviation were estimated at each $\omega$.
Figure~\ref{fig:results_spm} shows the results for the SpM method.
It is clear that SpM yields quite robust spectra even for $\sigma=10^{-3}$, whereas the Pad\'e results (Fig.~\ref{fig:results_pade}) show severe noise dependence.

\subsection{Extent to which a spectrum is reconstructable}

We have shown that the sparse modeling technique allows stable analytical continuation even in the presence of noise.
We can obtain almost the same results if the noise level, or statistical errors in QMC calculations, is of a given order.
Note, however, that this does \emph{not} mean that the obtained spectrum is correct.
SpM fully uses relevant information contained in the input, but the information lost due to noise is not regenerated.

\begin{figure}
    \centering
    \includegraphics[width=0.98\linewidth]{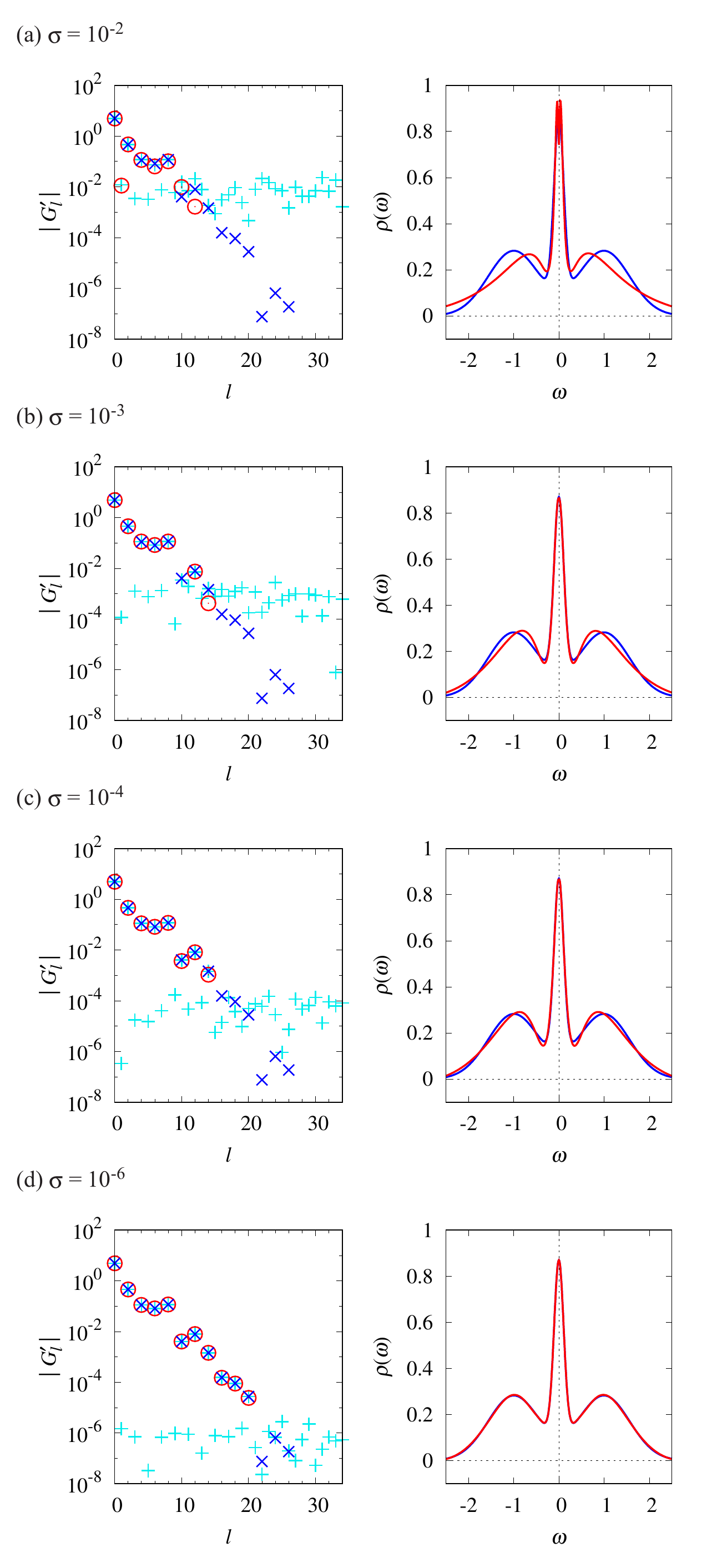}
    \caption{(Color online) $G'_l$ (left) and $\rho(\omega)$ (right) for various noise levels. See the captions of Fig.~\ref{fig:results_g} and Fig.~\ref{fig:results_rho} for explanations of the plots. (a)--(d) correspond to $\sigma=10^{-2}$, $10^{-3}$, $10^{-4}$, and $10^{-6}$, respectively.}
    \label{fig:noise_level}
\end{figure}

Let us discuss in more detail the extent to which the true spectrum is reconstructable.
Figure~\ref{fig:noise_level} shows spectra obtained using SpM analytical continuation from several input datasets with different noise levels $\sigma$.
The variation of the reconstructed $\rho(\omega)$ (red curves) signifies how much relevant information remains against noise.
The result for $\sigma=10^{-2}$ shows a large deviation because only five data points of $G'_l$ are retained above the noise level.
For smaller noise levels, namely $\sigma=10^{-3}$, $\sigma=10^{-4}$, and $\sigma=10^{-6}$, 6, 8, and 11 relevant points of $G_l$ are retained, respectively, and thus a better $\rho(\omega)$ can be reconstructed. 
The result for $\sigma=10^{-6}$ shows perfect agreement with the exact $\rho(\omega)$.
These results demonstrate the limitation of analytical continuation in the presence of noise.

\section{Application of Intermediate Representation of Green's Functions to Many-Body Calculations}
\label{sec:compressed_sampling}
The IR is a compact representation of the imaginary-time dependence of Green's functions.\cite{Shinaoka17}
As shown in previous sections, the IR plays a substantial role in SpM analytical continuation.
The compactness of the IR may make it useful for reducing the computation time and memory consumption of quantum many-body simulations.
In Ref.~\citen{Shinaoka17}, the present authors proposed the use of the IR basis for efficient and compact many-body calculations, such as the measurement of Green's function in QMC calculations.
In Ref.~\citen{Chikano18PRB}, some of the present authors proposed a numerical algorithm for computing the IR basis functions precisely, and investigated
the properties of the IR in greater depth.
Numerical data of the IR basis functions are available online.~\cite{Chikano18CPC,irbasis}
In this section, we describe the properties of the IR basis functions in more detail and show some applications to many-body calculations.

\subsection{Mathematical properties of IR basis functions}
We now formulate the IR basis functions in the continuous limit~\cite{Shinaoka17,Chikano18PRB}.
The spectral (Lehmann) representation of the single-particle Green's function is
\begin{align}
	G^{\alpha}(\tau) &= -\int_{-\wmax}^{\wmax} d\omega K^{\alpha}(\tau, \omega) \rho^{\alpha}(\omega),\label{eq:fwd}
\end{align}
where we introduce the cutoff frequency $\wmax$.
Here, we assume that the spectrum is bounded in the interval $[-\wmax, \wmax]$.
The superscript $\alpha$ specifies statistics: $\alpha$ = F for fermions and $\alpha$ = B for bosons. 

The two complete orthonormal basis sets for IR, $\{U^\alpha_l(\tau)\}$ and $\{V^\alpha_l(\omega)\}$, are defined through the decomposition
\begin{align}
	K^{\alpha}(\tau, \omega) &= \sum_{l=0}^{\infty} S^{\alpha}_l U^{\alpha}_l(\tau) V^{\alpha}_l(\omega) \label{eq:kernel-exp}
\end{align}
for $\tau\in[0,\beta]$ and $\omega \in [-\wmax,\wmax]$.
These basis sets are orthogonalized as  $\int_0^\beta d \tau U_l^\alpha(\tau)U_{l^\prime}^\alpha(\tau) =\int_{-\wmax}^\wmax d \omega V_l^\alpha(\omega)V_{l^\prime}^\alpha(\omega) = \delta_{ll^\prime}$.
This decomposition corresponds to the continuous limit of the SVD of the kernel discretized on a discrete and uniform mesh of $\tau$ and $\omega$ (refer to Sect.~\ref{sec:exact_relations}).
In practice, the basis functions can be computed by solving the integral equation
\begin{align}
 S^\alpha_l U^{\alpha}_l(\tau) &= \int_{-\wmax}^{\wmax} d\omega K^{\alpha}(\tau, \omega) V_l^{\alpha}(\omega)\label{eq:integral-eq}
\end{align}
under the orthonormal conditions.
The basis $\{U_l^\alpha(\tau)\}$ also satisfies the inhomogeneous Fredholm equation of the second kind  
\begin{align}
\int_0^\beta \tilde{K}(\tau, \tau^\prime)U^{\alpha}_l(\tau^\prime)d\tau^\prime &= (S_l^\alpha)^2 U_l^\alpha(\tau),
\end{align}
where
\begin{align}
\tilde{K}(\tau, \tau^\prime) &\equiv \int_{-\wmax}^{\wmax} d\omega K^\alpha(\tau,\omega) K^\alpha(\tau^\prime,\omega).
\end{align}

The integral equation (\ref{eq:integral-eq}) can be recast into a dimensionless form by using the change of variables ($x\equiv 2\tau/\beta -1$ and $y\equiv \omega/\wmax$).~\cite{Shinaoka17}
This explicitly shows that the singular values depend on only the statistics and a dimensionless parameter $\Lambda \equiv \beta\wmax$ up to a constant.

Recently, some of the present authors and co-workers developed an efficient numerical algorithm for solving the integral equation [Eq. (\ref{eq:integral-eq})].\cite{Chikano18PRB}
Although the analytic form of the solution is unknown,
numerical studies have revealed some  interesting properties.
When the singular values are ordered in descending order,
for even (odd) values of $l$, $U^\alpha_l(\tau)$ and $V^\alpha_l(\omega)$ are even (odd) functions with respect to the center of the domain, i.e., $\tau=\beta/2$ or $\omega=0$.
More interestingly, $U_l^{\alpha}(\tau)$ and $V_l^{\alpha}(\omega)$ are reduced to the Legendre polynomials in the limit $\Lambda \to 0$ (if the ranges of $\tau$ and $\omega$ are scaled properly).\cite{Shinaoka17}
Similarly to classical orthogonal polynomials such as Legendre and Chebyshev polynomials,
all the available numerical data indicate that $U^\alpha_l(\tau)$ and $V^\alpha_l(\omega)$ have $l$ zeros in their domains.

Figure~\ref{fig:basis} shows the IR basis functions computed for $\Lambda=100$ ($\beta=10$ and $\wmax=10$).
One can clearly see the interesting properties discussed above.
We refer readers to Ref.~\citen{Chikano18PRB} for more details on the properties of the IR basis. 

The integral equation [Eq. (\ref{eq:integral-eq})] is ill-conditioned as the singular values decay exponentially.
Thus, solving it numerically requires arbitrary-precision arithmetic, which is computationally expensive.
A library is provided with precomputed numerical data of the basis functions.~\cite{Chikano18CPC,irbasis}
It allows us to use the IR basis as easily as classical orthogonal polynomials. 

\subsection{Convergence properties of IR}
\begin{figure}
	\centering
	\includegraphics[width=.4\textwidth,clip]{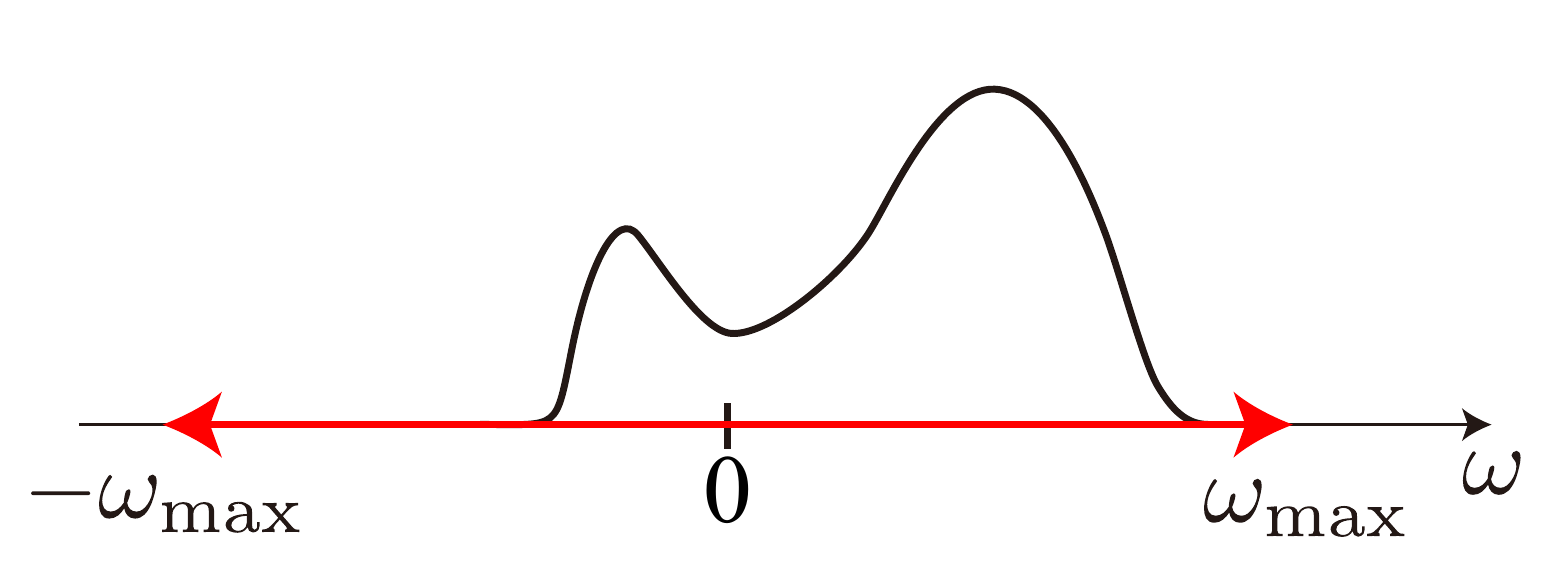}
	\caption{
    Compactness of IR is guaranteed when the frequency window $[-\wmax,\wmax]$ covers the whole spectrum.}
	\label{fig:bounded}
\end{figure}

\begin{figure}
	\centering
    \begin{flushleft}\hspace{4em}(a)\end{flushleft}
	\includegraphics[width=.4\textwidth,clip]{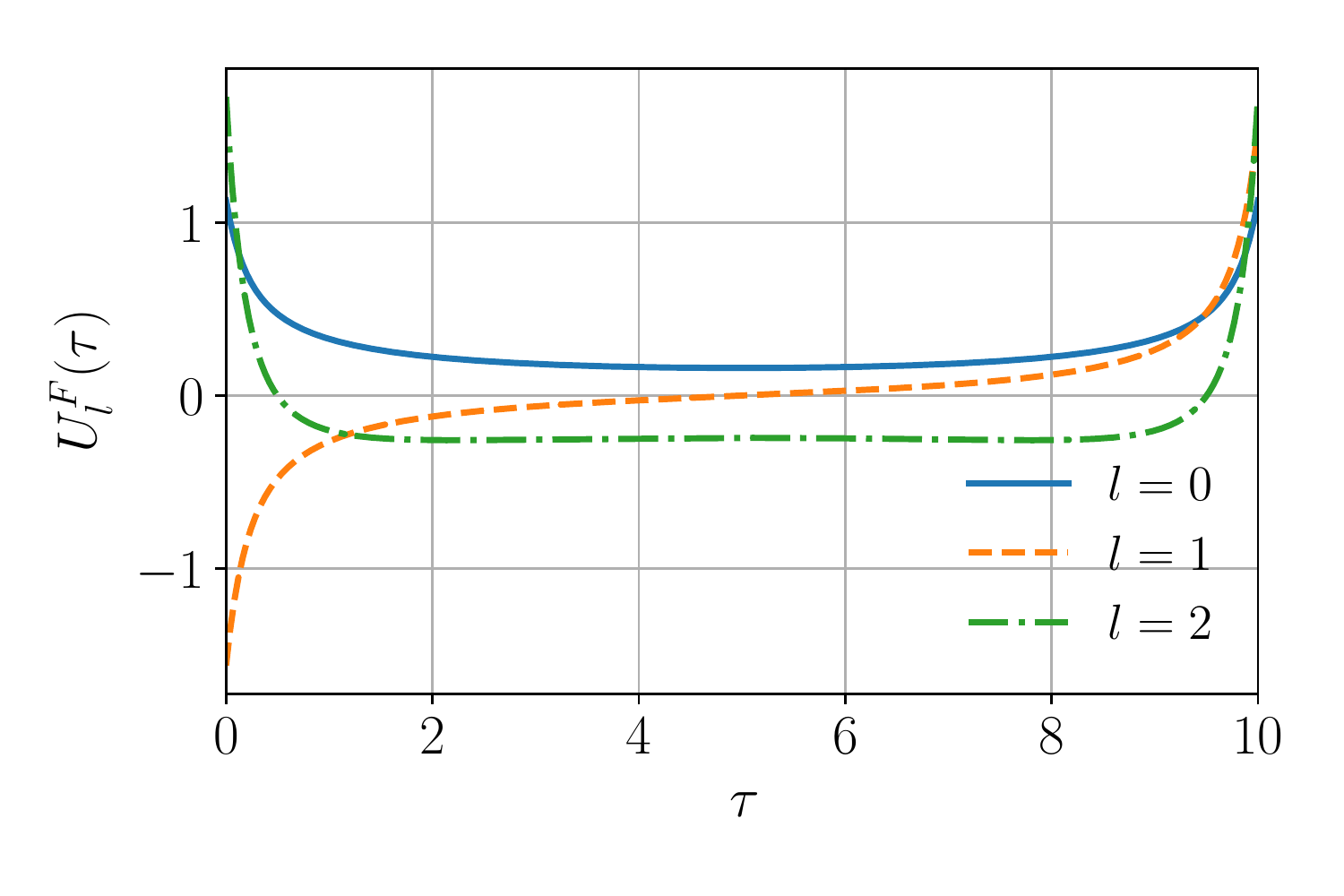}
	\begin{flushleft}\hspace{4em}(b)\end{flushleft}
	\includegraphics[width=.4\textwidth,clip]{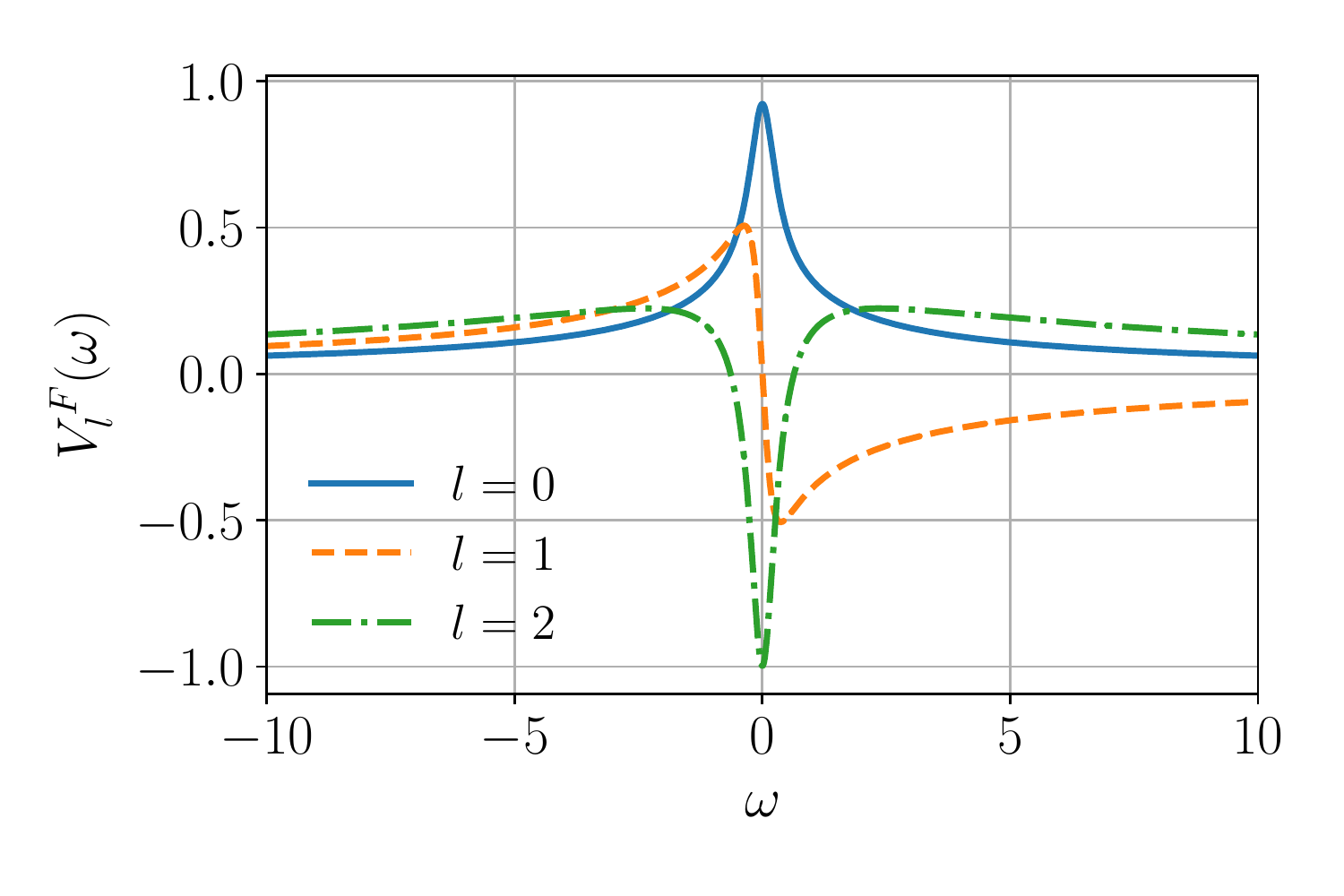}
	\caption{(Color online) IR basis functions computed for 
	$\Lambda=100$ ($\wmax=10$ and $\beta=10$).
	}
	\label{fig:basis}
\end{figure}

We expand $G^\alpha(\tau)$ using the complete basis $\{U^\alpha(\tau)\}$ as follows:
\begin{align}
	G^{\alpha}(\tau) &= \sum_{l=0}^\infty  G_l^{\alpha} U_l^{\alpha}(\tau)\label{eq:IR-decomp}.
\end{align}
If the spectrum of $G^{\alpha}(\tau)$ is bounded in $[-\wmax,\wmax]$ (see Fig.~\ref{fig:bounded}),
substituting Eq.~(\ref{eq:kernel-exp}) into Eq.~(\ref{eq:fwd}) and comparing with the above equation, we obtain
\begin{align}
	G^{\alpha}_l = - S^{\alpha}_l \rho^{\alpha}_l,\label{eq:gl}
\end{align}
where $\rho^{\alpha}_l$ is given by
\begin{align}
	\rho_l^{\alpha} = \int_{-\wmax}^{\wmax} d \omega \rho^\alpha(\omega) V^\alpha_l(\omega).\label{eq:rhol}
\end{align}
Equation~(\ref{eq:gl}) shows that the expansion coefficients $G_l^{\alpha}$ decay at least as fast as $S_l^{\alpha}$.
Because the singular values decay exponentially with $l$,
one may not need basis functions that correspond to small singular values below $S_l^\alpha/S_0^\alpha < \delta$  to express Green's function in practical calculations (i.e., $\delta=10^{-5}$ may be sufficient for typical noisy Monte Carlo data).

\begin{figure}
	\centering
	\includegraphics[width=.45\textwidth,clip]{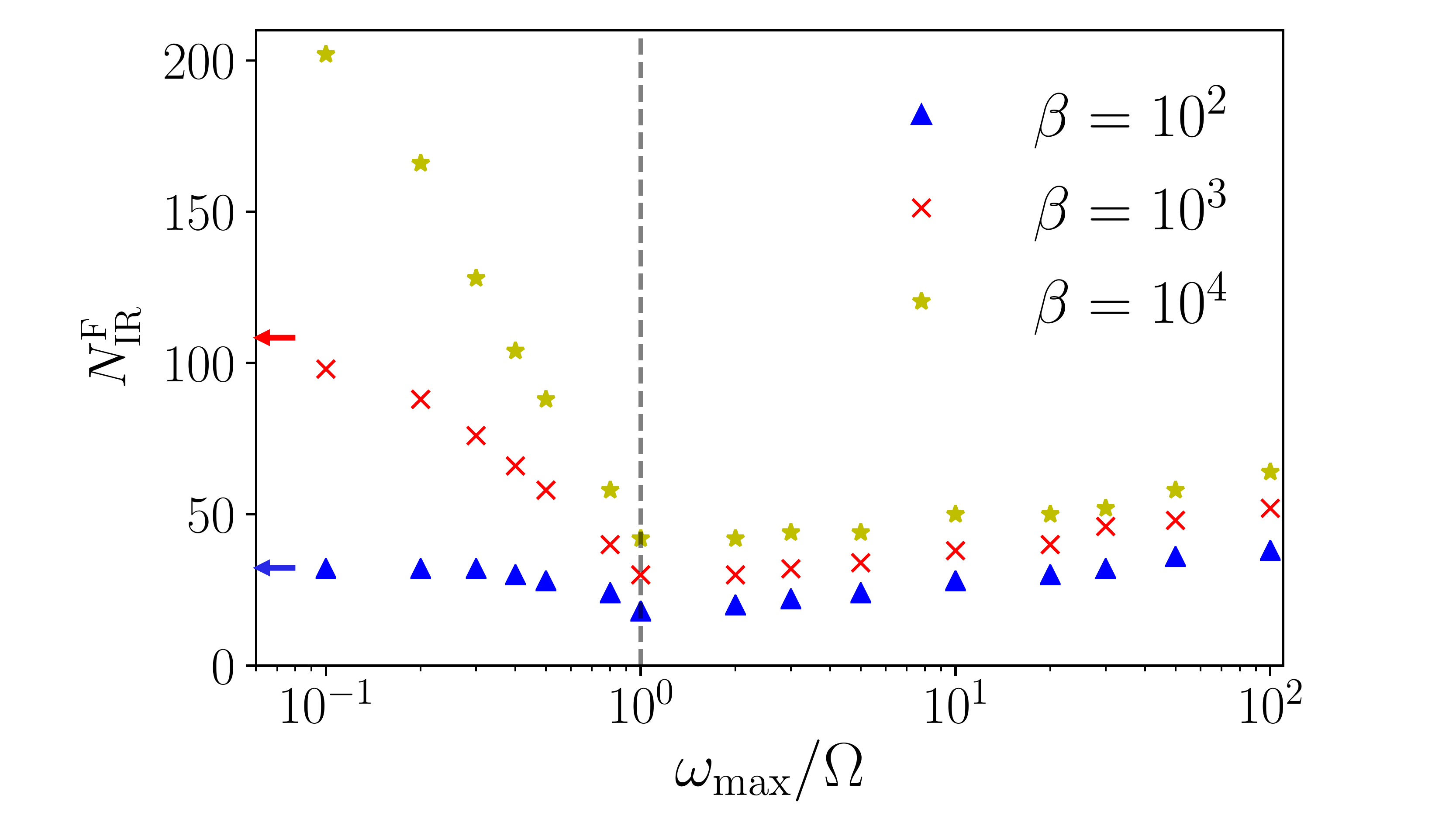}
	\includegraphics[width=.45\textwidth,clip]{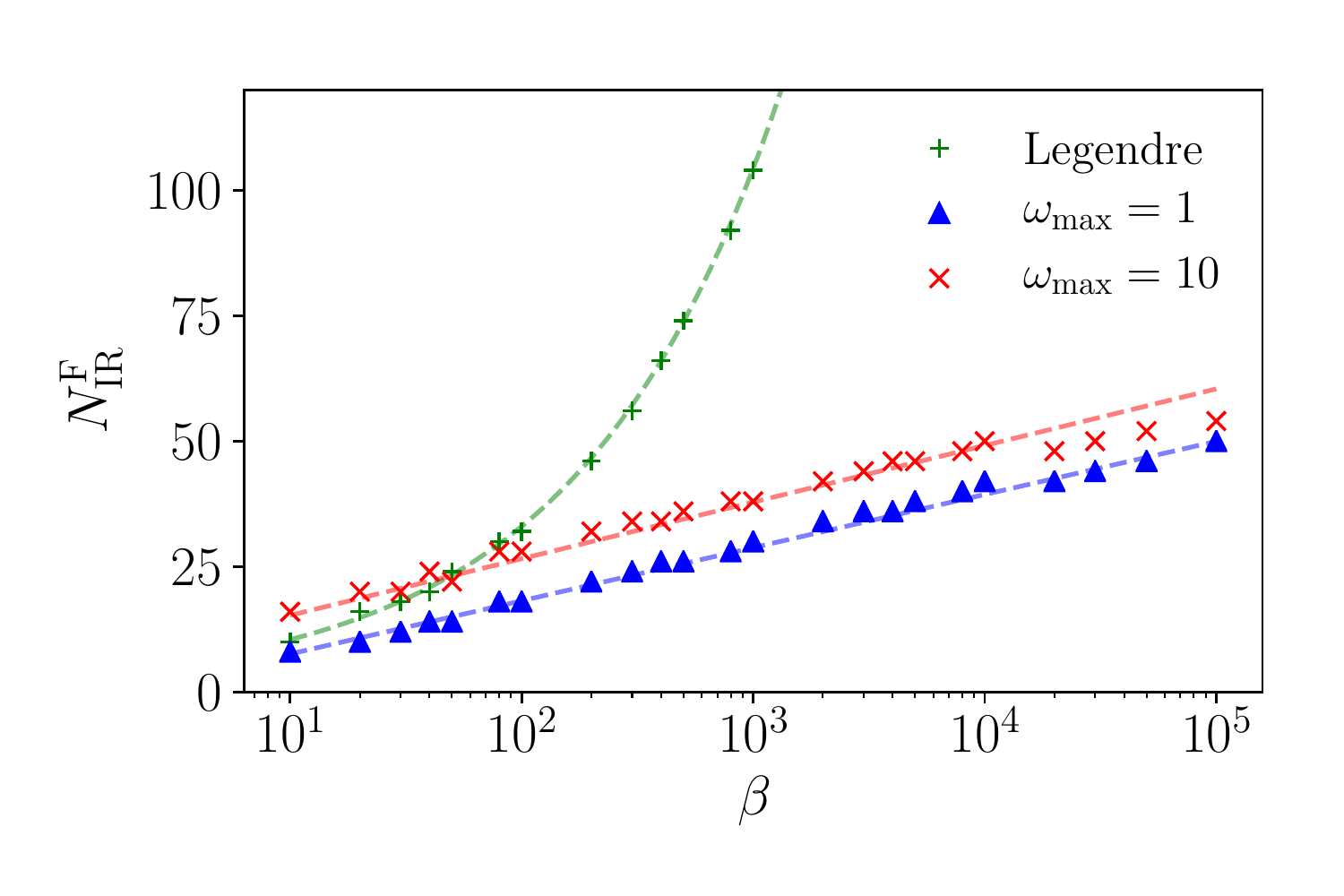}
	\caption{
		(Color online) Minimum number of basis functions required to express $G(\tau)$ with a precision of $10^{-5}$ at arbitrary $\tau$ for fermions.
		The data were computed for the Green's function for the spectral function consisting of two poles at $\omega=\pm \Omega = \pm 1$.
		In the upper and lower panels, the data are plotted with respect to $\wmax$ and $\beta$, respectively.
		Reprinted with permission from Ref.~\citen{Chikano18PRB} \copyright 2018 the American Physical Society.
	}
	\label{fig:Lambdadep}
\end{figure}

The upper panel of Fig.~\ref{fig:Lambdadep} shows the number of basis functions required for representing a typical model of fermionic $G(\tau)$ with a certain precision.
The data obtained for two choices of $\wmax$ are plotted against $\beta$.
The dimension of the basis increases logarithmically with $\Lambda$.
Surprisingly, for bosons, it becomes saturated (see Fig. 4 in Ref.~\citen{Chikano18PRB}).
These behaviors are in contrast to the power-law increase $\propto \beta^{1/2}$ observed for the Legendre basis~\cite{Boehnke11,Chikano18PRB} and the Chebyshev polynomial basis.~\cite{EGull2018}
These results indicate that only a constant number of IR basis functions will suffice in many-body calculations based on the imaginary-time Green's function at low $T$.

In practical calculations, we set $\wmax$ to a value much larger than the spectral width of the system
to ensure that the spectrum is bounded in $[-\wmax,\wmax]$.
The speed of convergence of $G^\alpha_l$ depends on the choice of $\wmax$.
The choice of cutoff value only slightly influences convergence (only logarithmically).~\cite{Chikano18PRB}
This is demonstrated in the lower panel of Fig.~\ref{fig:Lambdadep}.
The compactness of the data is not affected by the choice of $\Lambda$ ($\wmax$)
as long as the spectrum is bounded in $[-\wmax,\wmax]$.

\subsection{Application of IR basis functions}
\subsubsection{Efficient quantum Monte Carlo sampling}
Continuous-time QMC methods are widely used in the field of condensed matter physics.
As demonstrated by the present authors in Ref.~\citen{Shinaoka18},
the single-particle Green's function can be accumulated directly in terms of the IR basis.
The authors considered the particle-hole symmetric single-site Anderson impurity model defined by the Hamiltonian
\begin{eqnarray}
	\mathcal{H} &=& - \mu  \sum_\sigma c^\dagger_{\sigma} c_{\sigma} + U n_{\uparrow} n_{\downarrow} + \sum_{k\sigma} (c^\dagger_{\sigma} a_{k \sigma} +a^\dagger_{k \sigma} c_{\sigma} )\nonumber\\
	&& + \sum_{\alpha} \sum_{k \sigma} \epsilon_k a^\dagger_{k \sigma} a_{k \sigma}\label{eq:model}
\end{eqnarray} 
where $\mu=U/2$ and $\sigma$ is the spin index.
$c_\sigma$ and $c^\dagger_\sigma$ are annihilation and creation operators at the impurity site, respectively, and $a_{k\sigma}$ and $a^\dagger_{k\sigma}$ are those at the bath sites ($k$ is the internal degree of freedom of the bath), respectively.
The distribution of $\epsilon_k$ is a semi-circular density of states of width $4$.

\begin{figure}
	\centering
	\includegraphics[width=.45\textwidth,clip]{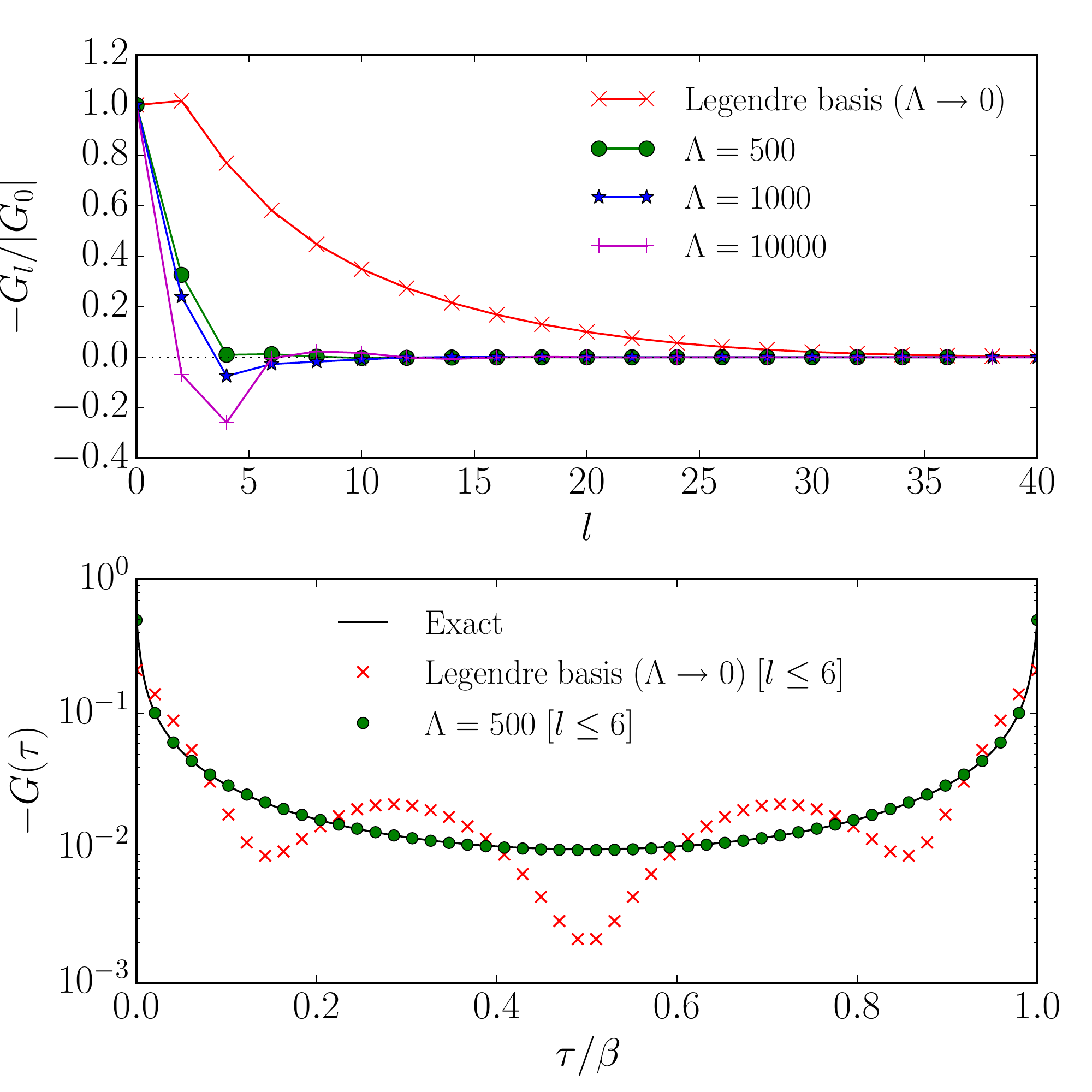}
	\caption{(Color online) Expansion coefficients of QMC data for the single-particle Green's function in terms of the IR basis computed for the Anderson impurity model with $U=4$ and $\beta=100$.
	Reprinted with permission from Ref.~\citen{Shinaoka17} \copyright 2017 the American Physical Society.
	}
	\label{fig:gf}
\end{figure}

Figure~\ref{fig:gf} shows the coefficients of the single-particle Green's function computed for $U=4$ and $\beta=100$.
Note that the data were accumulated directly in the IR basis (and the Legendre basis as a reference).
The model was solved using the hybridization expansion continuous-time Monte Carlo technique~\cite{Werner:2006ko}.
In the upper panel of Fig.~\ref{fig:gf}, one can clearly see that the IR yields coefficients that decay even faster than those for the Legendre basis.
One can also see that the most compact representation is obtained when $\wmax = \Lambda/\beta$ matches the actual width of the spectrum.
The optimal value obtained is $\Lambda\simeq 1000$ for $\beta=100$, which is consistent with the largest dimensionless energy scale of the system, i.e., $\beta U$, $\beta W = 400$.
As $\Lambda$ exceeds the optimal value, the efficiency only slowly decreases.
The direct measurement of Green's function will reduce memory consumption and computational time.

The lower panel of Fig.~\ref{fig:gf} shows $G(\tau)$ reconstructed from the coefficients for $l\le 6$.
The data obtained for the IR ($\Lambda=500$) shows perfect agreement with the numerically exact data. The truncation in the Legendre representation results in large Gibbs oscillations.

\subsubsection{Noise filtering to  finite-size effects}
The projection of a single-particle object is useful not only for QMC data but also for those without statistical errors.
This was demonstrated by Nagai and one of the present authors in Ref.~\citen{Nagai:2019de} in the context of 
dynamical mean-field calculations using an exact-diagonalization impurity solver (DMFT+ED) at finite temperature.

\begin{figure}
	\centering
	\includegraphics[width=.45\textwidth,clip]{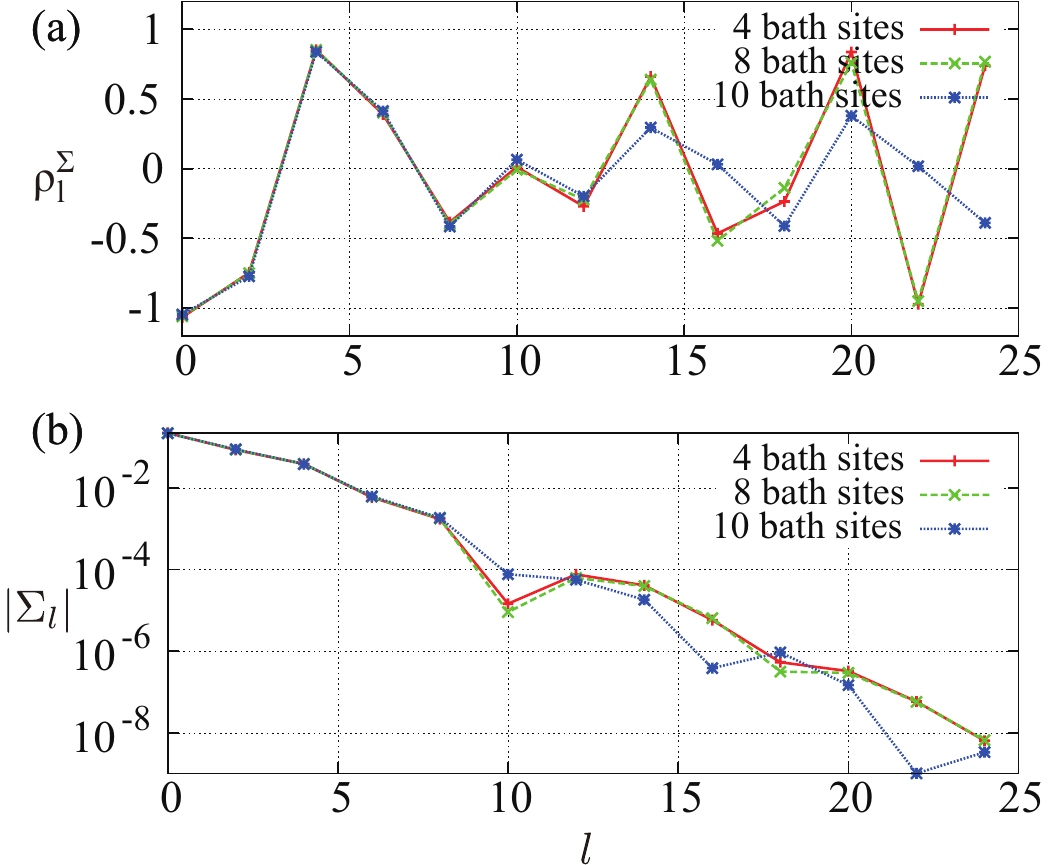}
	\caption{(Color online)
		(a) Expansion coefficients of the imaginary part of the real-frequency self-energy and the imaginary-time self-energy.
		(b) Imaginary part of the real-frequency self-energy before and after filtering in terms of the IR basis.
		Reprinted with permission from Ref.~\citen{Nagai:2019de} \copyright 2019 The Physical Society of Japan.
		}
	\label{fig:nagai}
\end{figure}

In the finite-$T$ DMFT+ED method, the self-energy $\Sigma(\omega)$, which is assumed to be local in space, is determined self-consistently in the procedure illustrated in Fig.~\ref{fig:nagai}.
The effective impurity model is solved using the exact diagonalization method after the bath has been discretized.
Although one can compute the real-frequency self-energy in the finite-$T$ DMFT+ED method,
its imaginary part, Im $\Sigma(\omega+i\delta)$, is usually spiky because the bath is approximated by a finite number of bath sites.

A fundamental question is how much information is included in the self-energy.
The imaginary part of the real-frequency self-energy can be regarded as the spectral function of the self-energy.
Thus, the IR of the self-energy is
\begin{align}
\Sigma(i\omega_n) - \Sigma_\mathrm{const} &= \sum_{l=0}^\infty  \Sigma_{l} U^\mathrm{F}_l(i \omega_n), \label{eq:sigmal}\\
\rho^{\Sigma}(\omega) &\equiv  -\frac{1}{\pi} \mathrm{Im} \Sigma^R(\omega) = \sum_{l=0}^\infty  \rho^{\Sigma}_{l} V^\mathrm{F}_l(\omega),\label{eq:sigma-rhol}
\end{align}
where $\Sigma_\mathrm{const}$ is a frequency-independent term.

Nagai and Shinaoka investigated how these expansion coefficients depend on the number of bath sites for the single-orbital Hubbard model with a semi-circular non-interacting density of states of width $D$.
Figure~\ref{fig:nagai}(a) shows the results computed for a paramagnetic metallic solution at $U= 2D$ and $\beta = 20$.
As can be clearly seen, only the first few coefficients $\rho^\Sigma_l$ converge with respect to the number of bath sites.
Larger $l$ components are affected by finite-bath-size effects.
This clearly indicates that only the first few components carry relevant information that converges with the number of bath sites.
Correspondingly, as shown in Fig.~\ref{fig:nagai}(b), $\Sigma_l$ for $l \ge 10$ depends on the number of bath sites.

\begin{figure}
	\centering
	\includegraphics[width=.45\textwidth,clip]{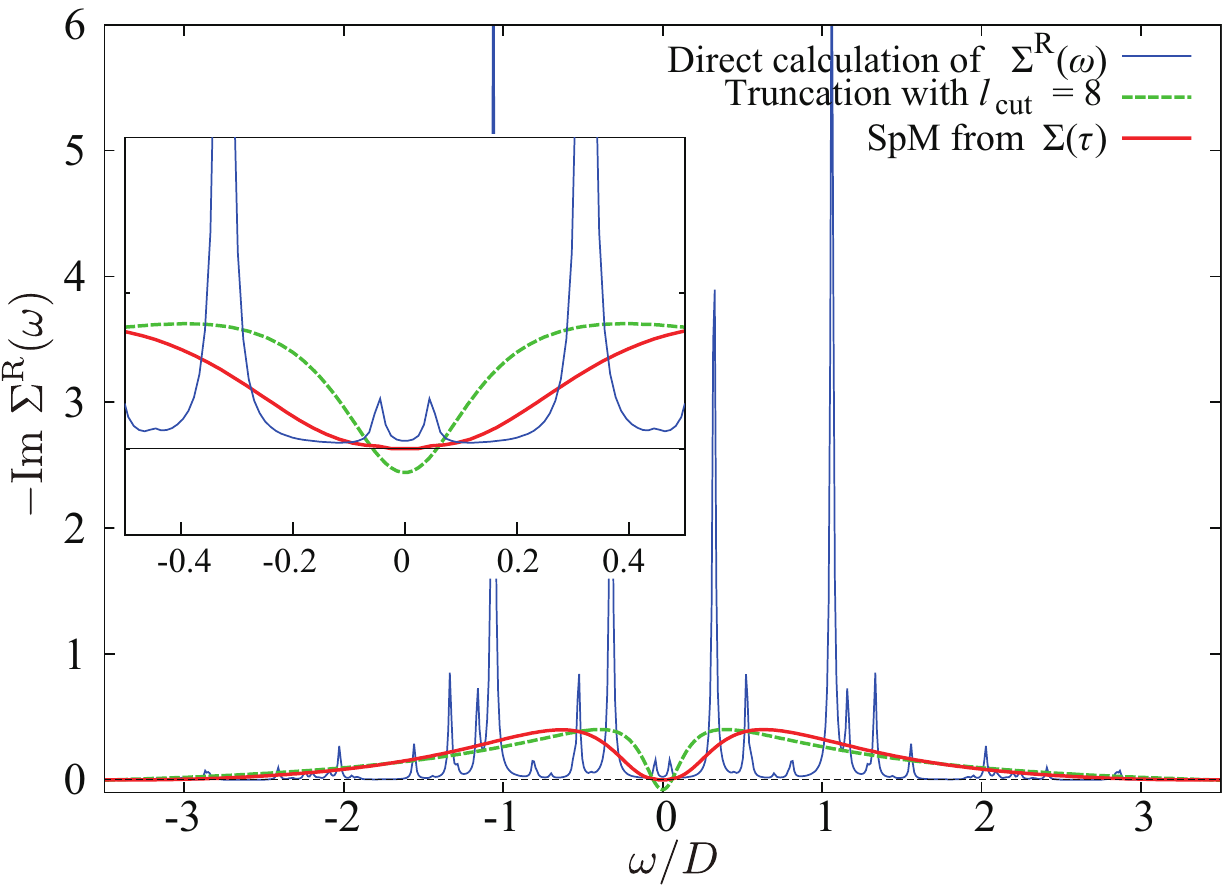}
	\caption{(Color online)
		Imaginary part calculated using the direct calculation in the DMFT+ED method, truncation with $l=8$, and SpM analytic continuation for $U=2D$. 
		Reprinted with permission from Ref.~\citen{Nagai:2019de} \copyright 2019 The Physical Society of Japan.
	}
	\label{fig:nagai2}
\end{figure}

Nagai and Shinaoka also computed the physically relevant smooth spectral function, i.e., $\rho^\Sigma(\omega)$, by truncating the expansion of the self-energy at $l=8$ up to which the coefficients converge~~\cite{Nagai:2019de}.
As clearly demonstrated in Fig.~\ref{fig:nagai2}, the spiky components, arising from the finite-size effects, are removed.
However, this native truncation breaks the causality of the self-energy.
As shown in Fig.~\ref{fig:nagai2}, this can be remedied by using SpM analytic continuation techniques.

\subsubsection{IR approach for two-particle Green's functions}
The concept of the IR was first proposed in the context of the single-particle Green's function.
The present authors and co-workers extended the IR to two-particle Green's functions.\cite{Shinaoka18}
Two-particle Green's functions and vertex functions play a critical role in theoretical frameworks for describing strongly correlated electron systems.
However, numerical calculations at the two-particle level often suffer from large computation time and massive memory consumption because these objects depend on multiple Matsubara frequencies and have a high-frequency and long-tail structure in the Matsubara frequency domain, which requires an elaborate treatment in practical applications~\cite{Li:2016bn,Kunes:2011is,Rohringer:2012cc,Kaufmann:2017hw,Wentzell-arXiv}.

In Ref.~\citen{Shinaoka18}, the present authors and co-workers derived a general expansion formula for two-particle Green's functions in terms of an overcomplete representation based on the IR basis functions.
The expansion formula was obtained by decomposing the spectral representation of the two-particle Green's function.
It was rigorously shown that the expansion coefficients decay exponentially (the upper bound is also given by singular values $S_l^\alpha$) while all high-frequency and long-tail structures in the Matsubara frequency domain are retained.

\begin{figure}
	\centering
	\includegraphics[width=.5\textwidth,clip]{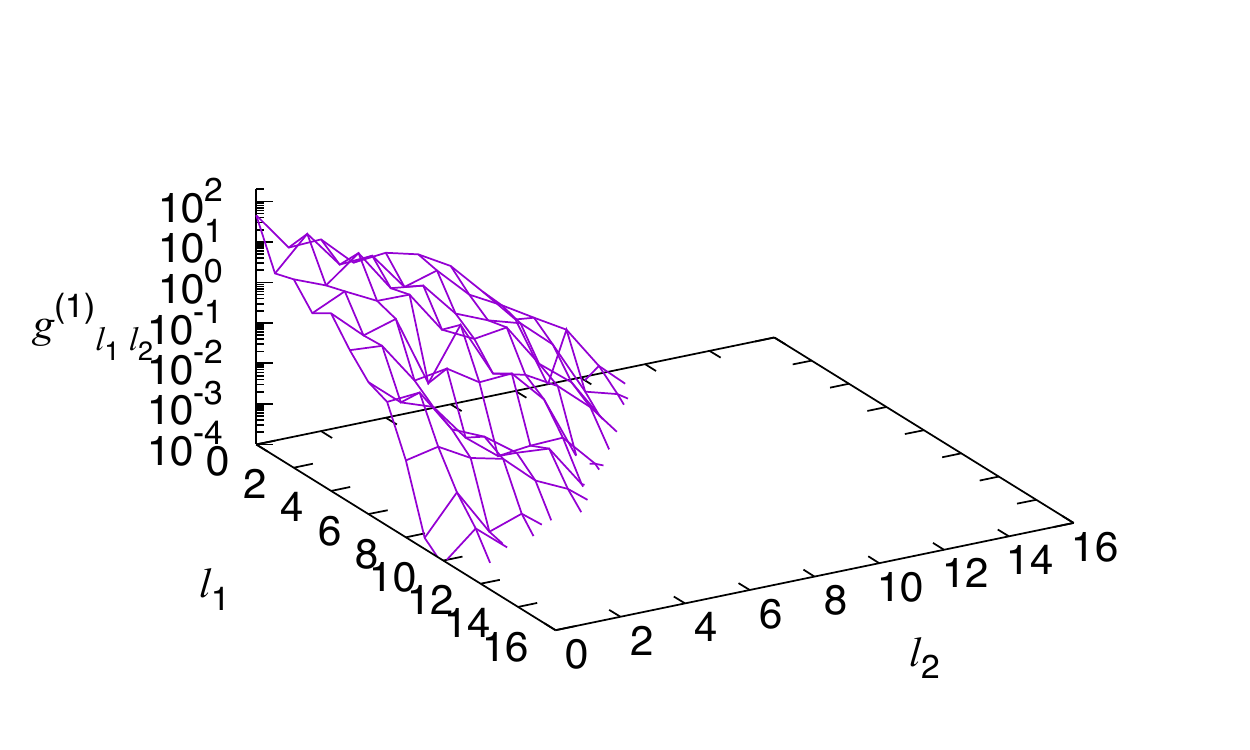}
	\caption{(Color online)
		One of three expansion coefficients, $g^{(i)}_{l_1 l_2}$ $(i=1,2,3)$, obtained from fitting the QMC data of the three-point Green's function in Eq.~(\ref{eq:Gph}). The model is the Hubbard model with a semi-circular density of states of bandwidth 2 at half filling for $U=2$ and $\beta=20$.
		Reprinted with permission from Ref.~\citen{Shinaoka18} \copyright 2018 the American Physical Society.
	}
	\label{fig:two-particle}
\end{figure}

Because the expansion formula is rather complicated, we only present some numerical results here.
The present authors and co-workers solved the Hubbard model 
using the dynamical mean-field theory combined with the continuous-time hybridization expansion QMC method.\cite{Werner:2006ko}
We measured the three-point Green's function in the particle-hole channel:
\begin{equation}\begin{split}
&G^{\mathrm{ph}}_{\uparrow\uparrow}(i\nu_n, i\omega_{n^\prime}) \\
&\quad=\int_0^\beta d\tau_{12}\ d\tau_{23}\ e^{i\nu_n \tau_{12} + i\omega_{n^\prime} \tau_{23}}\ G^{\mathrm{ph}}_{\uparrow\uparrow}(\tau_1, \tau_2, \tau_3),
\end{split}\label{eq:Gph}
\end{equation}
where
\begin{align}
G^{\mathrm{ph}}_{\uparrow\uparrow}(\tau_1, \tau_2, \tau_3) &= \braket{T_\tau c_\uparrow(\tau_1) c^\dagger_\uparrow(\tau_2) c_\uparrow(\tau_3) c_\uparrow^\dagger(\tau_3)}.\label{eq:Gph2}
\end{align}
Here, $\nu_n$ and $\omega_{n^\prime}$ are fermionic and bosonic Matsubara frequencies, respectively.
We measured $G^{\mathrm{ph}}_{\uparrow\uparrow}(i\nu_n, i\omega_{n^\prime})$ in the rectangular Matsubara frequency domain of $-100 \le n \le 99$ and $-100 \le n^\prime \le 100$.
Then, the data were transformed into the IR basis.
Because the IR for the two-particle Green's function is overcomplete,
$G^{\mathrm{ph}}_{\uparrow\uparrow}(i\nu_n, i\omega_{n^\prime})$ is expanded as a combination of three sets of expansion coefficients, $g^{(i)}_{l_1 l_2}$ $(i=1,2,3)$.
Figure~\ref{fig:two-particle} shows one of the coefficients, $g^{(1)}_{l_1 l_2}$ (other coefficients show similar behavior).
As can be clearly seen, $g^{(1)}_{l_1 l_2}$ decays (super-) exponentially in both directions.

This representation thus enables an efficient treatment of two-particle quantities and opens a route for the application of modern many-body theories to realistic strongly correlated electron systems.

\section{Concluding Remarks}
\label{sec:summary}

We reviewed the theoretical background and applications of sparse modeling.
Sparsity has been recognized as a useful criterion that selects a reasonable solution of an ill-conditioned inverse problem.
Compressed sensing fully utilizes the sparsity of the expected solution, reducing measurement time.
We emphasize that the choice of the basis (representation) is crucial for successful application.
One needs to design a sparse representation based on experience.
If a sparse representation is available for a problem, it is worth considering to apply sparse modeling.
Sparse modeling will reveal information that is covered by noise in ordinary data analysis.

It has been demonstrated that sparsity is useful also in quantum many-body theory.
The kernel $K$ in the relation $\bG=K\brho$ is ill-conditioned, meaning that the Matsubara Green's function $\bG$ has little information on the physical spectrum $\rho$ in the presence of noise.
The sparse-modeling technique offers a way to single out relevant information, allowing $\bG$ to be compressed essentially without loss of information.
The resultant compact representation is called the IR.
Once $\bG$ is represented in the IR basis, one can perform analytical continuation to obtain $\rho$ or carry out many-body calculations within the IR basis.
Methods based on Matsubara Green's functions can be reformulated using the IR basis, allowing efficient calculations of high-dimensional quantities in systems with multiple degrees of freedom.


\begin{acknowledgments}
We would like to thank 
K. Hukushima, Y. Nakanishi-Ohno, H. Tsunetsugu, Y. Motoyama, and N. Chikano
for useful comments and discussions.
This work was supported by JSPS KAKENHI Grant No. 18H01158.
J.O. was supported by JSPS KAKENHI Grant No. 18H04301 (J-Physics).
M.O. was supported by MEXT KAKENHI Grant No. 25120008, JST CREST, and JSPS KAKENHI No. 16H04382.
H.S. was supported by JSPS KAKENHI Grant Nos. 16H01064 (J-Physics) and 16K17735.
K.Y. was supported by JSPS KAKENHI Grant No.19K03649, and the Building of Consortia for the Development of Human Resources in Science and Technology, MEXT, Japan.
\end{acknowledgments}

\appendix
\section{Lipschitz Continuity}
\label{app:lipschitz}

A function $f(\bx)$ is said to be \emph{Lipschitz continuous} or \emph{$L$-smooth} between two points $\bx=\bm{a}$ and $\bx=\bm{b}$ if it satisfies the inequality
\begin{equation}
\left\| f(\bm{a}) - f(\bm{b}) \right\|_2 \le L \left\| \bm{a} -\bm{b} \right\|_2.
\end{equation}
Here, $L$ is called a Lipschitz constant.
This inequality quantifies the continuity of $f(\bx)$:
The variation of $f(\bm{x})$ is no more than a certain value proportional to the ``distance'' $\left\| \bm{a} -\bm{b} \right\|_2$.

\section{ADMM Algorithm for SpM Analytical Continuation}
\label{app:ADMM}

In Sect.~\ref{sec:analytical_continuation}, analytical continuation has been formulated as an optimization problem. The function to be minimized is given by Eq.~(\ref{eq:F2}) if the covariant matrix $C$ is taken into account, and Eq.~(\ref{eq:F}) if not.
The algorithm for the case with Eq.~(\ref{eq:F}) is presented in Ref.~\citen{Otsuki17}.
Here, we extend it to the case with a covariance matrix, namely the optimization problem in Eq.~(\ref{eq:F2}).

For consistency with Sect.~\ref{sec:spm}, we change the notation from \{$\brho$, $\bG$\} to \{$\bm{x}$, $\bm{y}$\}.
The function in Eq.~(\ref{eq:F2}) is then
\begin{align}
\label{S:eq:F}
F(\bm{x}' |\bm{y}',W,\lambda) = \frac12 \| W^{1/2} ( \bm{y}' - S \bm{x}' ) \|_2^2 + \lambda \| \bm{x}' \|_1,
\end{align}
and the constraints are expressed as
\begin{align}
\label{S:eq:constraint_x}
(V\bm{x}')_j \geq 0,\quad \langle V\bm{x}' \rangle \equiv \sum_j (V\bm{x}')_j = c.
\end{align}
We have used the relation $W=W^\mathrm{T}$ to express the first term in Eq.~(\ref{S:eq:F}) in the form of the $L_2$ norm.

Following the ADMM procedure presented in Sect.~\ref{subsec:admm},
we introduce auxiliary vector $\bm{z}'$ to separate the $L_1$ regularization term.
We further introduce $\bm{z}$, which imposes the non-negative constraint in Eq.~(\ref{S:eq:constraint_x}).
Thus, the alternative function to be minimized is
\begin{align}
\widetilde{F}(\bm{x}', \bm{z}', \bm{z})
&=
\frac{1}{2\lambda} \| W^{1/2} (\bm{y}' - S \bm{x}' ) \|_2^2
-\nu(\langle V\bm{x}' \rangle -1)
\nonumber \\
&+ \| \bm{z}' \|_1
+ \lim_{\gamma\to\infty} \gamma \sum_j \Theta(-z_j),
\label{S:eq:ADMM_cost}
\end{align}
subject to
\begin{align}
\bm{z}'=\bm{x}',
\quad
\bm{z}=V\bm{x}'.
\label{S:eq:ADMM_constraint}
\end{align}
The sum rule is imposed by the Lagrange multiplier $\nu$,
and non-negativity is expressed by an infinite potential $\gamma$ that acts on negative elements ($\Theta$ is the Heaviside step function).

The constraints in Eq.~(\ref{S:eq:ADMM_constraint}) are treated using the augmented Lagrange multiplier method. Parameters $\mu'$ and $\mu$ are introduced for the first and second constraints, respectively.
The update formulas are given by
\begin{subequations}
\label{S:eq:update}
\begin{align}
\bm{x}'
\leftarrow&
\left( \frac{1}{\lambda} S^\mathrm{T} WS
 + (\mu'+\mu) \bm{1} \right)^{-1}
\nonumber\\
&\times
\left( \frac{1}{\lambda} S^\mathrm{T} W\bm{y}'
+ \mu' (\bm{z}' - \bm{u}')
+ \mu V^\mathrm{T}(\bm{z} - \bm{u})
+ \nu V^\mathrm{T}\bm{e} \right)
\nonumber \\
&\equiv
\bm{\xi}_1 + \nu\bm{\xi}_2,
\label{S:eq:update-x'}
\\
\bm{z}' \leftarrow&\ {\cal S}_{1/\mu'} (\bm{x}' + \bm{u}'),
\\
\bm{u}' \leftarrow&\ \bm{u}' + \bm{x}' - \bm{z}',
\\
\bm{z} \leftarrow&\ {\cal P}_+
(V\bm{x}' + \bm{u}),
\\
\bm{u} \leftarrow&\ \bm{u} + V\bm{x}' - \bm{z},
\end{align}
\end{subequations}
where $e_i=1$ and
\begin{align}
\nu= \frac{c-\langle V\bm{\xi}_1 \rangle}{\langle V\bm{\xi}_2 \rangle}.
\end{align}
${\cal P}_+$ denotes a projection onto the non-negative quadrant, i.e., ${\cal P}_+ z_j = \max(z_j, 0)$ for each element.
${\cal S}_{\alpha}(\bm{x})$ is the element-wise soft threshold function defined in Eq.~(\ref{eq:soft_threshold}).
Starting with the initial condition, $\bm{x}'=\bm{z}'=\bm{u}'=\bm{0}$ and $\bm{z}=\bm{u}=\bm{0}$,
the updates in Eqs.~(\ref{S:eq:update}) are repeated until convergence is reached.
The inverse of the matrix in Eq.~(\ref{S:eq:update-x'}) is performed only once before the  iteration.
Then, the updates include only matrix-vector products, and thus the computational cost is quite low.

\bibliographystyle{jpsj}
\bibliography{bib/refs,bib/unpublished,bib/footnote,bib/ref_shinaoka}

\begin{thebibliography}{100}

\bibitem{Candes06a}
E.~J. {Cand\`es}, J.~{Romberg}, and T.~{Tao}: IEEE Trans. Inf. Theory
  {\bfseries 52} (2006) 489.

\bibitem{Candes06b}
E.~J. {Cand\`es}, J.~{Romberg}, and T.~{Tao}: Commun. Pure Appl. Math.
  {\bfseries 59} (2006) 1207.

\bibitem{Donoho06}
D.~L. {Donoho}: IEEE Trans. Inf. Theory {\bfseries 52} (2006) 1289.

\bibitem{Lustig07}
M.~Lustig, D.~Donoho, and J.~M. Pauly: Magnetic Resonance in Medicine
  {\bfseries 58} (2007) 1182.

\bibitem{Lustig08}
M.~{Lustig}, D.~L. {Donoho}, J.~M. {Santos}, and J.~M. {Pauly}: IEEE Signal
  Processing Magazine {\bfseries 25} (2008) 72.

\bibitem{EventHorizonTelescope19a}
{The Event Horizon Telescope Collaboration}: Astrophys. J. Lett. {\bfseries
  875} (2019) L1.

\bibitem{EventHorizonTelescope19b}
{The Event Horizon Telescope Collaboration}: Astrophys. J. Lett. {\bfseries
  875} (2019) L2.

\bibitem{EventHorizonTelescope19c}
{The Event Horizon Telescope Collaboration}: Astrophys. J. Lett. {\bfseries
  875} (2019) L3.

\bibitem{EventHorizonTelescope19d}
{The Event Horizon Telescope Collaboration}: Astrophys. J. Lett. {\bfseries
  875} (2019) L4.

\bibitem{EventHorizonTelescope19e}
{The Event Horizon Telescope Collaboration}: Astrophys. J. Lett. {\bfseries
  875} (2019) L5.

\bibitem{EventHorizonTelescope19f}
{The Event Horizon Telescope Collaboration}: Astrophys. J. Lett. {\bfseries
  875} (2019) L6.

\bibitem{footnote_minL2}
For the opposite case with $M>N$ (overdetermined systems), the solution is
  given by $\bx^* = (A^{\rm T}A)^{-1}A^{\rm T}\by$, which corresponds to the
  least-square solution.

\bibitem{Donoho05}
D.~L. Donoho and J.~Tanner: Proc. Nat. Acad. Sci. {\bfseries 102} (2005) 9452.

\bibitem{Donoho06b}
D.~L. Donoho: Discrete {\&} Computational Geometry {\bfseries 35} (2006) 617.

\bibitem{Kabashima09}
Y.~Kabashima, T.~Wadayama, and T.~Tanaka: J. Stat. Mech. Theory Exp. {\bfseries
  2009} (2009) L09003.

\bibitem{Donoho09}
D.~L. Donoho, A.~Maleki, and A.~Montanari: Proc. Natl. Acad. Sci. U.S.A.
  {\bfseries 106} (2009) 18914.

\bibitem{footnote_sample_script}
We solved the example problems using {\tt scikit-learn}
  package\cite{scikit-learn} of Python. Our scripts used are available in the
  GitHub repository. To download them, go to
  \url{https://github.com/SpM-lab/CS-tools} and follow the instruction there.

\bibitem{Tibshirani96}
R.~Tibshirani: J. R. Stat. Soc. B {\bfseries 58} (1996) 267.

\bibitem{MaxEnt}
M.~Jarrell and J.~Gubernatis: Physics Reports {\bfseries 269} (1996) 133.

\bibitem{Kullback-book}
S.~Kullback: {\em Information Theory and Statistics} (Dover Publications,
  1997).

\bibitem{Gull-Skilling84}
S.~Gull and J.~Skilling: IEE Proceedings F (Communications, Radar and Signal
  Processing) {\bfseries 131} (1984) 646.

\bibitem{Carcamo18}
M.~C\'arcamo, P.~Rom\'an, S.~Casassus, V.~Moral, and F.~Rannou: Astronomy and
  Computing {\bfseries 22} (2018) 16 .

\bibitem{footnote_smooth}
More precisely, ``smooth'' means $\nabla f(\bx)$ is $1/\eta$-smooth between
  $\bx$ and $\bx_{t}$ in the sense of Lipschitz continuity. See
  Appendix~\ref{app:lipschitz} for the definition of Lipschitz continuity.

\bibitem{Lange-book}
K.~Lange: {\em MM Optimization Algorithms} (Society for Industrial and Applied
  Mathematics, 2016).

\bibitem{Nocedal-Wright-book}
J.~Nocedal and S.~J. Wright: {\em Numerical Optimization} (Springer, New York,
  2006).

\bibitem{NAG}
Y.~E. Nesterov: Dokl. Akad. Nauk SSSR {\bfseries 269} (1983) 543.

\bibitem{FISTA}
A.~Beck and M.~Teboulle: SIAM Journal on Imaging Sciences {\bfseries 2} (2009)
  183.

\bibitem{Boyd11}
S.~Boyd, N.~Parikh, E.~Chu, B.~Peleato, and J.~Eckstein: Foundations and
  Trends\textsuperscript{\textregistered} in Machine Learning {\bfseries 3}
  (2011) 1.

\bibitem{Beck2009}
A.~Beck and M.~Teboulle: IEEE Trans. Image Process. {\bfseries 18} (2009) 2419.

\bibitem{Wahlberg2012}
B.~Wahlberg, S.~Boyd, M.~Annergren, and Y.~Wang: IFAC Proceedings Volumes
  {\bfseries 45} (2012) 83 .

\bibitem{Bertsekas1996}
D.~P. Bertsekas: {\em Constrained Optimization and Lagrange Multiplier Methods
  (Optimization and Neural Computation Series)} (Athena Scientific, 1996).

\bibitem{Thorndike53}
R.~L. Thorndike: Psychometrika {\bfseries 18} (1953) 267.

\bibitem{Decelle14}
A.~Decelle and F.~Ricci-Tersenghi: Phys. Rev. Lett. {\bfseries 112} (2014)
  070603.

\bibitem{Yamanaka15}
S.~Yamanaka, M.~Ohzeki, and A.~Decelle: J. Phys. Soc. Jpn. {\bfseries 84}
  (2015) 024801.

\bibitem{Kohavi95}
R.~Kohavi: Proceedings of the Fourteenth International Joint Conference on
  Artificial Intelligence  (1995) 1137.

\bibitem{Obuchi16}
T.~Obuchi and Y.~Kabashima: J. Stat. Mech.: Theory and Experiment {\bfseries
  2016} (2016) 053304.

\bibitem{Claerbout73}
J.~F. Claerbout and F.~Muir: GEOPHYSICS {\bfseries 38} (1973) 826.

\bibitem{Eldar-book}
Y.~Eldar and G.~Kutyniok: {\em Compressed Sensing: Theory and Applications}
  (Cambridge University Press, 2012).

\bibitem{Candes08}
E.~J. {Cand\`es} and M.~B. {Wakin}: IEEE Signal Processing Magazine {\bfseries
  25} (2008) 21.

\bibitem{Elad-book}
M.~Elad: {\em Sparse and Redundant Representations: From Theory to Applications
  in Signal and Image Processing} (Springer New York, 2010).

\bibitem{Sakata90}
M.~Sakata and M.~Sato: Acta Crystallographica Section A {\bfseries 46} (1990)
  263.

\bibitem{Sakata93}
M.~Sakata, T.~Uno, M.~Takata, and C.~J. Howard: Journal of Applied
  Crystallography {\bfseries 26} (1993) 159.

\bibitem{Tanaka19}
H.~Tanaka, M.~Oie, and K.~Oko: J. Phys. Soc. Jpn. {\bfseries 88} (2019) 053501.

\bibitem{Kazimierczuk11}
K.~Kazimierczuk and V.~Y. Orekhov: Angewandte Chemie International Edition
  {\bfseries 50} (2011) 5556.

\bibitem{Holland11}
D.~J. Holland, M.~J. Bostock, L.~F. Gladden, and D.~Nietlispach: Angewandte
  Chemie International Edition {\bfseries 50} (2011) 6548.

\bibitem{Nakanishi16}
Y.~Nakanishi-Ohno, M.~Haze, Y.~Yoshida, K.~Hukushima, Y.~Hasegawa, and
  M.~Okada: J. Phys. Soc. Jpn. {\bfseries 85} (2016) 093702.

\bibitem{Matsushita16}
T.~Matsushita: e-Journal of Surface Science and Nanotechnology {\bfseries 14}
  (2016) 158.

\bibitem{Akai18}
I.~Akai, K.~Iwamitsu, Y.~Igarashi, M.~Okada, H.~Setoyama, T.~Okajima, and
  Y.~Hirai: J. Phys. Soc. Jpn. {\bfseries 87} (2018) 074003.

\bibitem{Honma14}
M.~Honma, K.~Akiyama, M.~Uemura, and S.~Ikeda: Publ. Astron. Soc. Jpn.
  {\bfseries 66} (2014) 95.

\bibitem{Fienup78}
J.~R. Fienup: Opt. Lett. {\bfseries 3} (1978) 27.

\bibitem{Fienup82}
J.~R. Fienup: Appl. Opt. {\bfseries 21} (1982) 2758.

\bibitem{Matthew07}
R.~G.~B. Matthew L.~Moravec, Justin K.~Romberg.
\newblock Compressive phase retrieval, 2007.

\bibitem{Chan08}
W.~L. Chan, M.~L. Moravec, R.~G. Baraniuk, and D.~M. Mittleman: Opt. Lett.
  {\bfseries 33} (2008) 974.

\bibitem{Newton12}
M.~C. Newton: Phys. Rev. E {\bfseries 85} (2012) 056706.

\bibitem{Yokoyama19}
Y.~Yokoyama, T.-h. Arima, M.~Okada, and Y.~Yamasaki: J. Phys. Soc. Jpn.
  {\bfseries 88} (2019) 024009.

\bibitem{Miyama18}
M.~J. Miyama and K.~Hukushima: J. Phys. Soc. Jpn. {\bfseries 87} (2018) 044801.

\bibitem{Nagata12}
K.~Nagata, S.~Sugita, and M.~Okada: Neural Networks {\bfseries 28} (2012) 82 .

\bibitem{Igarashi16}
Y.~Igarashi, K.~Nagata, T.~Kuwatani, T.~Omori, Y.~Nakanishi-Ohno, and M.~Okada:
  J. Phys. Conf. Ser. {\bfseries 699} (2016) 012001.

\bibitem{Tokuda17}
S.~Tokuda, K.~Nagata, and M.~Okada: J. Phys. Soc. Jpn. {\bfseries 86} (2017)
  024001.

\bibitem{Nelson13a}
L.~J. Nelson, G.~L.~W. Hart, F.~Zhou, and V.~Ozoli\ifmmode \mbox{\c{n}}\else
  \c{n}\fi{}\ifmmode~\check{s}\else \v{s}\fi{}: Phys. Rev. B {\bfseries 87}
  (2013) 035125.

\bibitem{Nelson13b}
L.~J. Nelson, V.~Ozoli\ifmmode \mbox{\c{n}}\else
  \c{n}\fi{}\ifmmode~\check{s}\else \v{s}\fi{}, C.~S. Reese, F.~Zhou, and
  G.~L.~W. Hart: Phys. Rev. B {\bfseries 88} (2013) 155105.

\bibitem{Ghiringhelli15}
L.~M. Ghiringhelli, J.~Vybiral, S.~V. Levchenko, C.~Draxl, and M.~Scheffler:
  Phys. Rev. Lett. {\bfseries 114} (2015) 105503.

\bibitem{Ghiringhelli17}
L.~M. Ghiringhelli, J.~Vybiral, E.~Ahmetcik, R.~Ouyang, S.~V. Levchenko,
  C.~Draxl, and M.~Scheffler: New J. Phys. {\bfseries 19} (2017) 023017.

\bibitem{Seko14}
A.~Seko, A.~Takahashi, and I.~Tanaka: Phys. Rev. B {\bfseries 90} (2014)
  024101.

\bibitem{Seko15}
A.~Seko, A.~Takahashi, and I.~Tanaka: Phys. Rev. B {\bfseries 92} (2015)
  054113.

\bibitem{Tadano15}
T.~Tadano and S.~Tsuneyuki: Phys. Rev. B {\bfseries 92} (2015) 054301.

\bibitem{Zhou14}
F.~Zhou, W.~Nielson, Y.~Xia, and V.~Ozoli\ifmmode \mbox{\c{n}}\else
  \c{n}\fi{}\ifmmode~\check{s}\else \v{s}\fi{}: Phys. Rev. Lett. {\bfseries
  113} (2014) 185501.

\bibitem{Ozolins:2013ig}
V.~Ozolins, R.~Lai, R.~Caflisch, and S.~Osher: Proc. Natl. Acad. Sci. U.S.A.
  {\bfseries 110} (2013) 18368.

\bibitem{Budich14}
J.~C. Budich, J.~Eisert, E.~J. Bergholtz, S.~Diehl, and P.~Zoller: Phys. Rev. B
  {\bfseries 90} (2014) 115110.

\bibitem{Tamura17}
R.~Tamura and K.~Hukushima: Phys. Rev. B {\bfseries 95} (2017) 064407.

\bibitem{Mototake19}
Y.-i. Mototake, M.~Mizumaki, I.~Akai, and M.~Okada: J. Phys. Soc. Jpn.
  {\bfseries 88} (2019) 034004.

\bibitem{Fujita18}
H.~Fujita, Y.~O. Nakagawa, S.~Sugiura, and M.~Oshikawa: Phys. Rev. B {\bfseries
  97} (2018) 075114.

\bibitem{Nakanishi18b}
Y.~Nakanishi-Ohno and K.~Hukushima: Phys. Rev. E {\bfseries 98} (2018) 052120.

\bibitem{Nakanishi18a}
Y.~Nakanishi-Ohno and K.~Hukushima: J. Phys. Conf. Ser. {\bfseries 1036} (2018)
  012014.

\bibitem{Chen98}
S.~Chen, D.~Donoho, and M.~Saunders: SIAM Journal on Scientific Computing
  {\bfseries 20} (1998) 33.

\bibitem{Elad06}
M.~{Elad} and M.~{Aharon}: IEEE Trans. Image Process. {\bfseries 15} (2006)
  3736.

\bibitem{Aharon06}
M.~{Aharon}, M.~{Elad}, and A.~{Bruckstein}: IEEE Trans. Signal Process.
  {\bfseries 54} (2006) 4311.

\bibitem{Mairal08}
J.~{Mairal}, M.~{Elad}, and G.~{Sapiro}: IEEE Trans. Image Process. {\bfseries
  17} (2008) 53.

\bibitem{Mairal10}
J.~Mairal, F.~Bach, J.~Ponce, and G.~Sapiro: {J. Mach. Learn. Res.} {\bfseries
  11} (2010) 19.

\bibitem{Candes09}
E.~J. Cand{\`e}s and B.~Recht: Foundations of Computational Mathematics
  {\bfseries 9} (2009) 717.

\bibitem{Cai10}
J.~Cai, E.~Cand\`es, and Z.~Shen: SIAM Journal on Optimization {\bfseries 20}
  (2010) 1956.

\bibitem{Candes10}
E.~J. {Cand\`es} and Y.~{Plan}: Proc. IEEE {\bfseries 98} (2010) 925.

\bibitem{AGD}
A.~A. Abrikosov, L.~P. Gorkov, and I.~E. Dzyaloshinski: {\em Methods of quantum
  field theory in statistical physics} (Dover, 1963).

\bibitem{Chikano18PRB}
N.~Chikano, J.~Otsuki, and H.~Shinaoka: Phys. Rev. B {\bfseries 98} (2018)
  035104.

\bibitem{Shinaoka17}
H.~Shinaoka, J.~Otsuki, M.~Ohzeki, and K.~Yoshimi: Phys. Rev. B {\bfseries 96}
  (2017) 035147.

\bibitem{Otsuki17}
J.~Otsuki, M.~Ohzeki, H.~Shinaoka, and K.~Yoshimi: Phys. Rev. E {\bfseries 95}
  (2017) 061302.

\bibitem{Vidberg77}
H.~J. Vidberg and J.~W. Serene: J. Low Temp. Phys. {\bfseries 29} (1977) 179.

\bibitem{Yoshimi19}
K.~Yoshimi, J.~Otsuki, Y.~Motoyama, M.~Ohzeki, and H.~Shinaoka: Comput. Phys.
  Commun. {\bfseries 244} (2019) 319 .

\bibitem{Gunnarsson10b}
O.~Gunnarsson, M.~W. Haverkort, and G.~Sangiovanni: Phys. Rev. B {\bfseries 81}
  (2010) 155107.

\bibitem{Bergeron16}
D.~Bergeron and A.-M.~S. Tremblay: Phys. Rev. E {\bfseries 94} (2016) 023303.

\bibitem{Levy17}
R.~Levy, J.~LeBlanc, and E.~Gull: Comput. Phys. Commun. {\bfseries 215} (2017)
  149 .

\bibitem{Sim-arXiv}
J.-H. Sim and M.~J. Han: arXiv:1804.01683 .

\bibitem{Sandvik98}
A.~W. Sandvik: Phys. Rev. B {\bfseries 57} (1998) 10287.

\bibitem{Mishchenko00}
A.~S. Mishchenko, N.~V. Prokof'ev, A.~Sakamoto, and B.~V. Svistunov: Phys. Rev.
  B {\bfseries 62} (2000) 6317.

\bibitem{Fuchs10}
S.~Fuchs, T.~Pruschke, and M.~Jarrell: Phys. Rev. E {\bfseries 81} (2010)
  056701.

\bibitem{Beach-arXiv}
K.~S.~D. Beach: arXiv:cond-mat/0403055 .

\bibitem{Sandvik16}
A.~W. Sandvik: Phys. Rev. E {\bfseries 94} (2016) 063308.

\bibitem{Bao16}
F.~Bao, Y.~Tang, M.~Summers, G.~Zhang, C.~Webster, V.~Scarola, and T.~A. Maier:
  Phys. Rev. B {\bfseries 94} (2016) 125149.

\bibitem{Krivenko-arXiv}
I.~Krivenko and M.~Harland: arXiv:1808.00603 .

\bibitem{Arsenault17}
L.-F. Arsenault, R.~Neuberg, L.~A. Hannah, and A.~J. Millis: Inverse Problems
  {\bfseries 33} (2017) 115007.

\bibitem{Yoon-arXiv}
H.~Yoon, J.-H. Sim, and M.~J. Han: arXiv:1806.03841 .

\bibitem{Beach00}
K.~S.~D. Beach, R.~J. Gooding, and F.~Marsiglio: Phys. Rev. B {\bfseries 61}
  (2000) 5147.

\bibitem{Oestlin12}
A.~\"Ostlin, L.~Chioncel, and L.~Vitos: Phys. Rev. B {\bfseries 86} (2012)
  235107.

\bibitem{Dirks13}
A.~Dirks, M.~Eckstein, T.~Pruschke, and P.~Werner: Phys. Rev. E {\bfseries 87}
  (2013) 023305.

\bibitem{Schoett16}
J.~Sch\"ott, I.~L.~M. Locht, E.~Lundin, O.~Gr\aa{}n\"as, O.~Eriksson, and
  I.~Di~Marco: Phys. Rev. B {\bfseries 93} (2016) 075104.

\bibitem{Bertaina16}
G.~Bertaina, D.~E. Galli, and E.~Vitali: Adv. Phys.: X {\bfseries 2} (2017)
  302.

\bibitem{Xuping-arXiv}
X.~Xie, F.~Bao, T.~Maier, and C.~Webster: arXiv:1905.10430 .

\bibitem{Goulko17}
O.~Goulko, A.~S. Mishchenko, L.~Pollet, N.~Prokof'ev, and B.~Svistunov: Phys.
  Rev. B {\bfseries 95} (2017) 014102.

\bibitem{Shao17}
H.~Shao, Y.~Q. Qin, S.~Capponi, S.~Chesi, Z.~Y. Meng, and A.~W. Sandvik: Phys.
  Rev. X {\bfseries 7} (2017) 041072.

\bibitem{Chikano18CPC}
N.~Chikano, K.~Yoshimi, J.~Otsuki, and H.~Shinaoka: Comput. Phys. Commun.
  {\bfseries 240} (2019) 181 .

\bibitem{irbasis}
\url{https://github.com/SpM-lab/irbasis}.

\bibitem{Boehnke11}
L.~Boehnke, H.~Hafermann, M.~Ferrero, F.~Lechermann, and O.~Parcollet: Phys.
  Rev. B {\bfseries 84} (2011) 075145.

\bibitem{EGull2018}
E.~Gull, S.~Iskakov, I.~Krivenko, A.~A. Rusakov, and D.~Zgid: Phys. Rev. B
  {\bfseries 98} (2018) 075127.

\bibitem{Shinaoka18}
H.~Shinaoka, J.~Otsuki, K.~Haule, M.~Wallerberger, E.~Gull, K.~Yoshimi, and
  M.~Ohzeki: Phys. Rev. B {\bfseries 97} (2018) 205111.

\bibitem{Werner:2006ko}
P.~Werner, A.~Comanac, L.~de{\textquoteright} Medici, M.~Troyer, and A.~Millis:
  Phys. Rev. Lett. {\bfseries 97} (2006) 076405.

\bibitem{Nagai:2019de}
Y.~Nagai and H.~Shinaoka: J. Phys. Soc. Jpn. {\bfseries 88} (2019) 064004.

\bibitem{Li:2016bn}
G.~Li, N.~Wentzell, P.~Pudleiner, P.~Thunstr{\"o}m, and K.~Held: Phys. Rev. B
  {\bfseries 93} (2016) 165103.

\bibitem{Kunes:2011is}
J.~Kune{\v s}: Phys. Rev. B {\bfseries 83} (2011) 085102.

\bibitem{Rohringer:2012cc}
G.~Rohringer, A.~Valli, and A.~Toschi: Phys. Rev. B {\bfseries 86} (2012)
  125114.

\bibitem{Kaufmann:2017hw}
J.~Kaufmann, P.~Gunacker, and K.~Held: Phys. Rev. B {\bfseries 96} (2017)
  035114.

\bibitem{Wentzell-arXiv}
N.~Wentzell, G.~Li, A.~Tagliavini, C.~Taranto, G.~Rohringer, K.~Held,
  A.~Toschi, and S.~Andergassen: arXiv:1610.06520 .

\bibitem{scikit-learn}
F.~Pedregosa, G.~Varoquaux, A.~Gramfort, V.~Michel, B.~Thirion, O.~Grisel,
  M.~Blondel, P.~Prettenhofer, R.~Weiss, V.~Dubourg, J.~Vanderplas, A.~Passos,
  D.~Cournapeau, M.~Brucher, M.~Perrot, and E.~Duchesnay: J. Mach. Learn. Res.
  {\bfseries 12} (2011) 2825.

\end{thebibliography}

\end{document}